%% file: main_sts.tex
\documentclass[a4paper,12pt]{article}
\usepackage{array,graphicx,amsmath,amssymb,dcolumn,float,tabu,booktabs,framed,multirow,tikz,subcaption,pdflscape,bm,rotating,nicefrac,fancyhdr,lastpage,centernot,url,colortbl,mathdots,subdepth,longtable,lipsum,mathtools,extarrows,mdframed,algpseudocode,setspace}
\usepackage[vmargin=.9in,hmargin=.5in,footskip=.4in]{geometry}
\usetikzlibrary{calc,arrows.meta,shapes,decorations,patterns,decorations.text,shadows,matrix,arrows,decorations.pathreplacing,calligraphy}

\usepackage{bm,fouriernc} %
\usepackage[T1]{fontenc}
\usepackage[mathcal]{euscript}
\usepackage[dvipsnames]{xcolor}
\usepackage{bibunits}

\defaultbibliographystyle{unsrt}
\defaultbibliography{rci}

\input{notation.tex}

\title{Reframing cross-world independence in~multimediator~causal~models}
\author{En-Yu Lai$^1$, Jih-Chang Yu$^2$, and Yen-Tsung Huang$^1$\\\textit{\normalsize $^1$Institute of Statistical Science, Academia Sinica, Taipei, Taiwan}\\[-3mm]\textit{\normalsize $^2$Department of Statistics, National Taipei University, Taipei, Taiwan}}
\date{}
\begin{document}

\maketitle

\begin{abstract}%
Multi-mediator causal models provide a general framework for decomposing causal effects along specific mechanisms, but effect identification in the presence of treatment-induced confounding remains conceptually challenging. Path-specific effects (PSEs) provide one important formulation of such causal decomposition. In particular, existing approaches differ substantially in their counterfactual assumptions and in the interventions used to define effects through specific pathways. We use Single-World Intervention Graphs (SWIGs) as a unifying analytic device to formalize and compare three approaches to PSE identification: classical path-specific effects based on cross-world independence assumptions, interventional effects based on randomized mediator interventions, and separable effects based on decomposing the exposure into distinct causal components. Through this unified analysis, we show that these seemingly different approaches can, under appropriate conditions, lead to the same identification formulas despite relying on different causal assumptions and intervention semantics. The SWIG representation further reveals how each approach resolves the counterfactual conflicts created by treatment-induced confounding. Our results highlight a fundamental distinction between statistical identification and causal interpretation: the same observed-data functional can represent different causal questions depending on how the underlying counterfactual worlds are constructed. This unified perspective clarifies relationships among major approaches in multimediator settings and guides alignment of the estimand, identification assumptions, and scientific mechanism of interest.

\end{abstract}
\bigskip
\begin{flushleft}
\end{flushleft}

\section{Introduction}

Causal inference has a long history in statistics, with modern formulations largely developing along complementary counterfactual and graphical traditions.
The potential-outcomes framework, originating with Neyman's formulation of randomized experiments and subsequently developed into a general framework for statistical causal inference by Rubin and others~\cite{splawa1990application,rubin1974estimating,holland1986statistics}, defines causal effects through comparisons of outcomes that would be observed under alternative interventions.
Subsequent developments extended counterfactual reasoning to more complex longitudinal settings and observational studies~\cite{robins1986new}.
In parallel, structural causal models and causal directed acyclic graphs (DAGs) provided a graphical language for encoding causal assumptions and studying identification from observational data~\cite{pearl1995causal,greenland1999causal,pearl2009causality}.
These counterfactual and graphical perspectives now provide complementary languages for defining causal estimands and expressing the assumptions under which they can be identified from observed data.

\begin{figure}[t]
    \includegraphics[width=\textwidth]{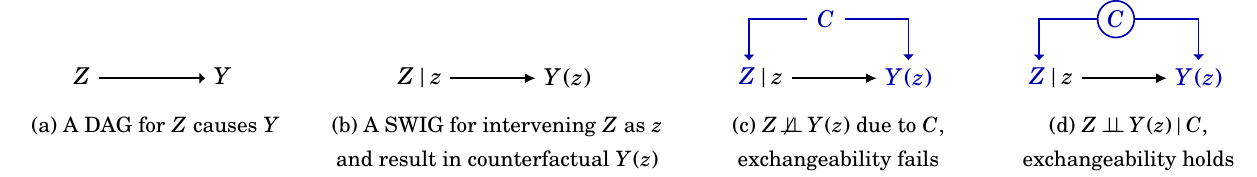}
    \caption{\textbf{A causal relationship.} 
    (a) A directed acyclic graph (DAG) illustrates the causal relationship between $Z$ and $Y$, where the edge direc tion indicates $Z$ as the exposure and $Y$ as the outcome. 
    (b) A single world intervention graph (SWIG) can show the independence of counterfactual outcome $Y(z)$ and the actual exposure $Z$. (c) If a confounder $C$ exists, then the blue path $Z\leftarrow C\rightarrow Y(z)$ illustrates the failure of exchangeability. (d) If we control $C$, the association path is blocked and exchangeability holds. }
    \label{fig:simple}
\end{figure}

A standard set of sufficient assumptions for identifying causal effects from observed data consists of \textit{exchangeability}, \textit{consistency}, and \textit{positivity}. 
Exchangeability~\cite{pearl2009causality} requires that the counterfactual outcome be independent of the actual exposure, consistency~\cite{cole2009consistency,vanderweele2009concerning,pearl2010consistency} ensures that the counterfactual outcome under the observed exposure equals the actual outcome, and positivity~\cite{westreich2010invited,petersen2012diagnosing} requires that each relevant exposure level is observed.
Let $Z$ denote the exposure and $Y$ the outcome, as shown in Figure~\ref{fig:simple}(a), then the causal effect of $Z$ on $Y$ can be identified through
\[\underbrace{P(Y(z)=y)}_{\substack{\text{counterfactual}\\[.5mm]\text{marginal probability}}}\hspace{-3mm}\eqtt{exchangeability}\;
\underbrace{P(Y(z)=y\gvn Z=z)}_{\substack{\text{counterfactual}\\[.5mm]\text{conditional probability}}}\hspace{1mm}\eqtt{consistency}
\underbrace{P(Y=y\gvn Z=z)}_{\substack{\text{actual}\\[.5mm]\text{conditional probability}}},\text{ and }P(Z=z)\overset{\text{positivity}}{>0}, \]
where $Y(z)$ denotes the counterfactual outcome of $Y$ under the intervention $Z=z$, as shown in Figure~\ref{fig:simple}(b). 
However, \textit{unmeasured confounding} between $Z$ and $Y$ can violate exchangeability.
As shown in Figure~\ref{fig:simple}(c), a confounder $C$ opens the path
$Z\leftarrow C\rightarrow Y(z)$, such that $Z\not\indp Y(z)$.
Conditioning on $C$ blocks this path and restores conditional exchangeability,
$Z\indp Y(z)\gvn C$, ensuring that differences in $Y(z)$ and $Y(z')$ can be attributed solely to the intervention on $Z$, as shown in Figure~\ref{fig:simple}(d).

Mediation analysis has its roots in path analysis, introduced by Sewall Wright to decompose associations among variables along specified paths~\cite{wright1921correlation,wright1934method}.
This idea was later developed into regression-based approaches to mediation, with the influential framework of Baron and Kenny~\cite{baron1986moderator} providing an operational procedure for studying the role of an intermediate variable through a sequence of regression relationships.
A counterfactual formulation of causal mediation was subsequently developed by Robins and Greenland~\cite{robins1992identifiability} and Pearl~\cite{pearl2001direct}, providing formal definitions of causal mediation effects and conditions for their identification.
Later developments further clarified the role of interventions, identification assumptions, estimation, and sensitivity analysis in causal mediation~\cite{vanderweele2009conceptual,imai2010identification}.
Within this framework, introducing an intermediate variable $M$ enables the total causal effect of $Z$ on $Y$ to be decomposed into direct and indirect effects.

\begin{figure}
\centering
    \includegraphics[width=\textwidth]{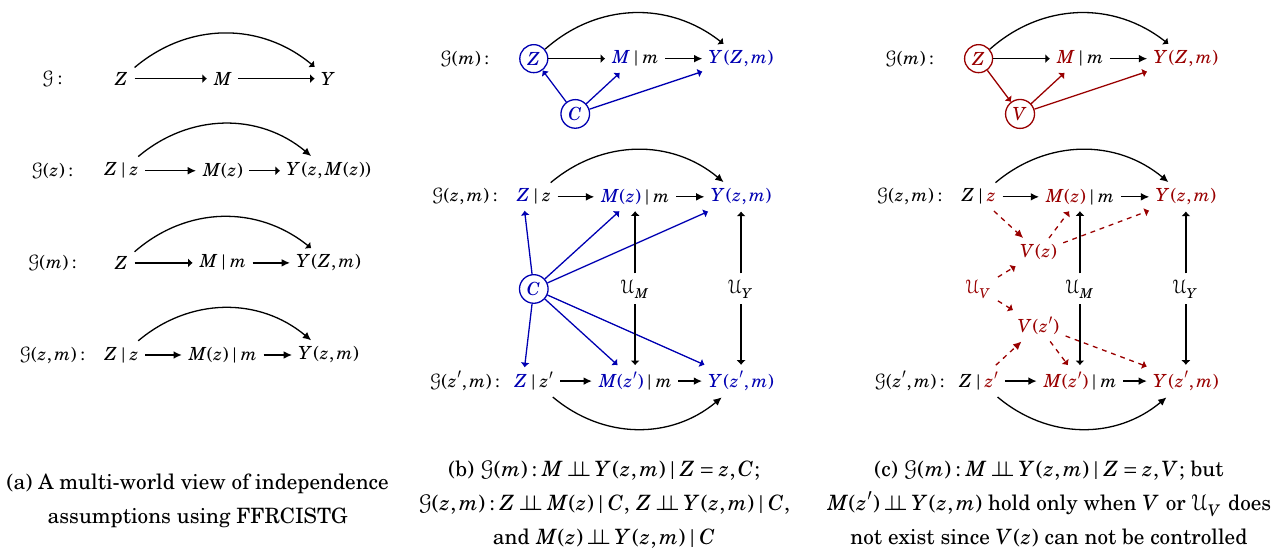}
    \caption{\textbf{Causal relationship with one mediator.} 
    (a) A DAG $\mathcal{G}$ illustrates the causal relationship between $Z$ and $Y$ with mediator $M$, and three SWIGs---$\mathcal{G}(z)$, $\mathcal{G}(m)$, and $\mathcal{G}(z,m)$---represent the \textit{Finest Fully Randomized Causally Interpretable Structured Tree Graphs} (FFRCISTG) independence~\cite{robins1986new}.
    (b) A confounder $C$ between any pair of $Z$, $M$, and $Y$ disrupts the independence conditions required for identifying natural direct and indirect effects. Although observed confounders can be controlled, the \textit{no unmeasured confounding} assumption remains essential to uphold these independencies.
    (c) The \textit{cross-world independence} assumption is violated when an unmeasured mediator $V$ induces \textit{cross-world confounding} between $M(z')$ and $Y(z,m)$. Even if $V$ were observed and controlled, the cross-world confounding would persist because $V(z)$ is not controlled. }
    \label{fig:mediation}
\end{figure}

Specifically, the \textit{natural indirect effect} quantifies the effect associated with the pathway $Z\rightarrow M\rightarrow Y$, contrasting $Y(z,M(z))$ and $Y(z,M(z'))$, while the \textit{natural direct effect} captures the effect associated with the remaining pathway, contrasting $Y(z,M(z))$ and $Y(z',M(z))$~\cite{pearl2001direct}.
The composition of $Y(z,M(z'))$ seems counterintuitive since $Z$ is intervened as $z$ directly but $z'$ indirectly, a joint intervention that cannot be realized within a single randomized experiment.
While consistency, exchangeability, and positivity suffice to identify the distribution of $Y(z,M(z))$ (Figure~\ref{fig:mediation}(a)), nonparametric identification of $Y(z,M(z'))$ can be established under the additional \textit{cross-world independence} assumption $Y(z,m)\indp M(z')$.
To see this, we construct an augmented DAG representing the joint structural model underlying multiple SWIGs, with shared exogenous variables linking counterfactual copies across worlds.
The joint distribution of the counterfactual variables factorizes according to the augmented DAG, which is proved in the Online Supplement, and therefore d-separation implies the corresponding conditional independence in the augmented DAG.

Exchangeability, although not empirically testable in general, can be supported by study design---most notably through randomization---or invoked conditionally by adjusting for measured confounders in observational studies (Figure~\ref{fig:mediation}(b)).
Cross-world independence poses a different challenge: because it relates counterfactual variables under mutually incompatible interventions, it cannot be established through an analogous experimental design~\cite{andrews2021insights}.
A key distinction lies in the source of confounding: the \textit{actual-world confounding} in Figure~\ref{fig:mediation}(b) involves a confounder $C$ unaffected by $Z$, thus identical across all counterfactual worlds (i.e., $\mathcal{G}(z)$, $\mathcal{G}(m)$, and $\mathcal{G}(z,m)$ in Figure~\ref{fig:mediation}(a)) as well as in the actual world (i.e., $\mathcal{G}$ in Figure~\ref{fig:mediation}(a)). 
Controlling $C$ in the actual world therefore controls $C$ in all counterfactual worlds. 
In contrast, Figure~\ref{fig:mediation}(c) illustrates \textit{cross-world confounding} between $M(z')$ and $Y(z,m)$ through a mediator $V$ affected by $Z$. 
Because $V(z)$ changes with $Z$, controlling $V$ in the actual world cannot remove the cross-world confounding it induces, and $V(z)$ itself cannot be directly observed.

We consider two strategies for identifying natural direct and indirect effects. 
The first, based on the \textit{nonparametric structural equation model with independent errors} (NPSEM-IE)~\cite{pearl2009causality}, assumes $Y(z,m)\indp M(z')$ and ignores $V$---the only viable choice when $V$ is unobserved, though potentially biased if $V(z)$ acts as a \textit{recanting witness}~\cite{vanderweele2014effect,vanderweele2014mediation,de2015mediation,vansteelandt2017interventional}. 
The second approach incorporates $V$ into the causal mechanism and treats $V$ as another mediator.
When multiple mediators are considered, natural direct and indirect effects generalize to \textit{path-specific effects} (PSEs)~\cite{avin2005identifiability,albert2011generalized,imai2013identification,daniel2015causal,steen2017flexible}, whose nonparametric identification can rely on stronger cross-world independence assumptions, as detailed in later sections.

Our main contribution is to use SWIGs as a unifying analytic device to show that three seemingly distinct approaches to path-specific effects can yield the same identification formula while relying on fundamentally different causal semantics. Cross-world assumptions required in the classical framework may instead be avoided through randomized interventions or reformulated as structural restrictions under alternative causal semantics. The remainder of this paper is organized as follows.
Section~\ref{sec:PSE} introduces notation for multi-mediator models and two strategies for decomposing PSEs. Sections~\ref{sec:classical}--\ref{sec:SE} develop the classical assumptions~\cite{avin2005identifiability,daniel2015causal}, interventional approach~\cite{vanderweele2014effect,lin2017interventional}, and separable effects~\cite{stensrud2021generalized,stensrud2022separable} using Single-World Intervention Graphs (SWIGs)~\cite{richardson2013single}, highlighting their respective reliance on cross-world independence, counterfactual mediator randomization, and component manipulability. The justification for applying \(d\)-separation across intervention-specific SWIGs linked by shared cross-world confounders is provided in Appendix~A. Section~\ref{sec:RT} reviews the recanting-twins approach for intermediate confounding~\cite{vo2026recanting,liu2024general} and relates it to the classical and separable-effects frameworks. Section~\ref{sec:example} illustrates these approaches using a cancer immunotherapy example. 
Section~\ref{sec:C} provides a unified comparison of the three frameworks, highlighting how similar identification formulas can arise under different causal assumptions and how cross-world independence is reframed across them. Section~\ref{sec:D} concludes with a broader discussion.

\section{Path-specific effects}\label{sec:PSE}
\begin{figure}
    \centering\includegraphics[width=.7\textwidth]{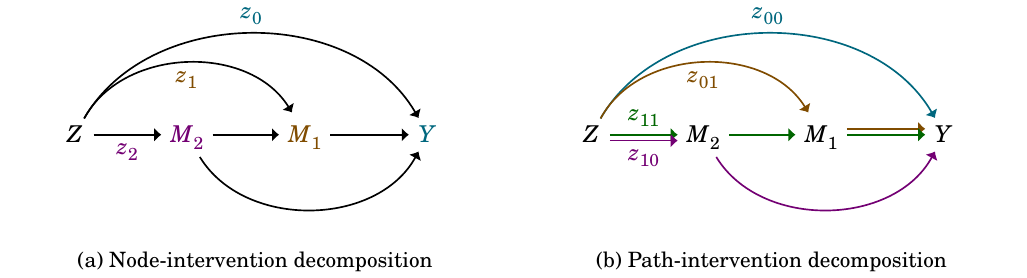}
    \caption{\textbf{Multimediator model with sequentially ordered mediators.}
    (a) node-intervention approach provides $p+1$ intervention targets. Here $\zi=\big\{z_0,z_1,z_2\big\}$ denote interventions corresponding to the direct effect (not mediated by $M_2$ or $M_1$), the effect only mediated by $M_1$, and the effect mediated by $M_2$, respectively. 
    (b) path-intervention approach provides $2^p$ intervention targets. Here $\zi=\big\{z_{00},z_{01},z_{10},z_{11}\big\}$ denote interventions corresponding to the direct effect, the effect only mediated by $M_1$, the effect only mediated by $M_2$, and the joint effect mediated by both $M_2$ and $M_1$, respectively.}
    \label{fig:PSE}
\end{figure}

We now consider a multi-mediator setting with $p$ sequentially ordered mediators $\big\{M_j\big\}_{j\in\mathbb{J}}$, where $\mathbb{J}=\{1,\ldots,p\}$.
For notational simplicity, mediators are indexed by their order upstream from $Y$, with $Y$ assigned index zero.
We present two approaches for PSE decomposition~\cite{vanderweele2014effect,vansteelandt2017interventional,daniel2015causal,loh2022nonlinear,tai2023complete}. 
The first, the \textit{node-intervention approach} (also called \textit{partial decomposition}), treats downstream variables as intervention targets. 
Because all mediators and the outcome can be intervened upon by $Z$ set to $z$, there are $p+1$ intervention settings, $\zi=\big\{z_i\big\}_{i\in\mathbb{I}}$ with $\mathbb{I}=\{0,\ldots,p\}$, as illustrated in Figure~\ref{fig:PSE}(a). 
Each intervention notion corresponds to the effect mediated by the intervened mediator. 
For example, $z_2$ identifies the effect mediated by $M_2$, corresponding to the subgraph of $Z\rightarrow M_2 \rightarrow (M_1)\rightarrow Y$. 
This approach does not distinguish between paths involving $M_1$ and those bypassing it; $z_2$ captures the total effect mediated by $M_2(z_2)$.

The second, the \textit{path-intervention approach} (or \textit{finest/complete decomposition}), assigns an intervention notion to every possible path. 
Consequently, there are $2^p$ intervention settings, $\zi=\big\{z_\bn{i}\big\}_{i\in\mathbb{I}}$, where $\mathbb{I}=\big\{0,\ldots,2^p-1\big\}$ and $\bn{\cdot}$ denotes the binary expression of $i$. 
For example, $z_\bn{2}$ corresponds to $z_{10}$ in binary form, representing the path $Z\rightarrow M_2\rightarrow Y$, where the leftmost ``1'' indicates that the path includes $M_2$, and the rightmost ``0'' indicates that it excludes $M_1$, as illusatrated in Figure~\ref{fig:PSE}(b).  
Similarly, $z_{00}$, $z_{01}$, and $z_{11}$ correspond to the paths $Z\rightarrow Y$, $Z\rightarrow M_1\rightarrow Y$, and $Z\rightarrow M_2\rightarrow M_1\rightarrow Y$, respectively.
Note that $M_2$ receives two intervention settings, $z_{10}$ and $z_{11}$; as discussed later, the independence between $M_2(z_{10})$ and $M_2(z_{11})$ is the cost of separately identifying the effects along $Z\rightarrow M_2\rightarrow Y$ and $Z\rightarrow M_2\rightarrow M_1\rightarrow Y$. 

\section{Classical assumptions}\label{sec:classical}
\begin{figure}
    \centering\includegraphics[width=.95\textwidth]{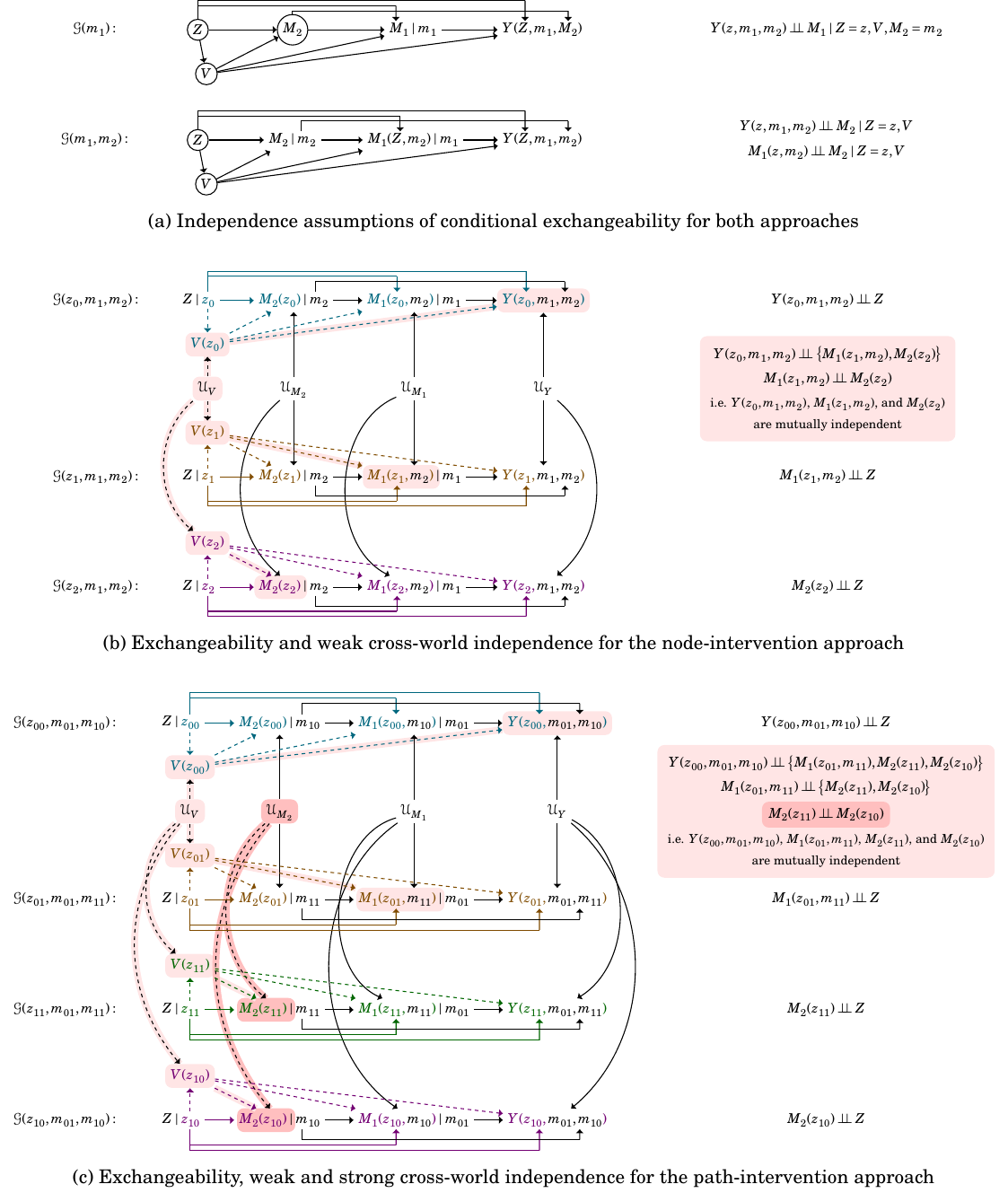}
\caption{\textbf{Classical assumptions.}
The independence assumptions in (a--c) without shading represent classical
exchangeability.
In (a), conditional exchangeability is presented in its general form
conditioning on $V$; when $V$ is absent, the corresponding conditioning
terms disappear.
For simplicity, the cross-world independence assumptions in (b--c) are
displayed for the case without $V$, while the $V(z)$ nodes and their dashed
cross-world connections are retained to illustrate the dependence structure
that would arise in the presence of a mediated confounder.
The red shading in (b--c) indicates weak cross-world independence, corresponding to the absence of $\UV$, while the dark red shading in (c) indicates strong cross-world independence, corresponding to the absence of $\mathcal{U}_{M_2}$.}
    \label{fig:CA}
\end{figure}

We refer to Pearl's NPSEM-IE as the \textit{classical assumptions}, as it explicitly outlines sufficient independence conditions for identification without further manipulation (i.e., \textit{randomizing the counterfactual mediator} in the interventional approach and \textit{decomposing the exposure} in the separable effects).
For illustration, Figure~\ref{fig:CA} presents three SWIGs of a two-mediator model that includes an exposure-induced confounder $V$.
Assuming no unmeasured confounding among the exposure, mediators, and outcome after conditioning on the measured confounders $C$, we omit $C$ from Figure~\ref{fig:CA}, Figure~\ref{fig:SE3s}, and Figure~\ref{fig:SE4} for visual clarity.
In Figure~\ref{fig:CA}(a), $\mathcal{G}(m_1)$ represents the SWIG in which $M_1$ is intervened as $m_1$. When we examine the relationship between $M_1$ and $Y$, the variables $Z$, $M_2$, and $V$ (along with the omitted $C$) act as confounders. By conditioning on these variables, we establish the exchangeability $Y(z,m_1,m_2)\indp M_1\gvn Z=z,V,M_2=m_2$, which reflects the independence between the actual variable $M_1$ and the counterfactual outcome $Y(m_1)$. Note that $Z$ and $M_2$ are conditioned upon, not intervened.
Similarly, $\mathcal{G}(m_1, m_2)$ in Figure~\ref{fig:CA}(a) depicts the SWIG with joint interventions $M_1=m_1$ and $M_2=m_2$. Conditioning on $Z$ and $V$ yields $Y(z,m_1,m_2)\indp M_2\gvn Z,V$ and $M_1(z,m_2)\indp M_2\gvn Z,V$. 
Finally, in Figure~\ref{fig:CA}(b-c), where $Z$ is intervened as $z$, the classical exchangeability assumptions hold: $Y(z,m_1,m_2)\indp Z$, $M_1(z,m_2)\indp Z$, and $M_2(z)\indp Z$. 

Each of the above assumptions can be represented within its corresponding SWIG; however, identifying PSEs requires additional assumptions concerning the independence of counterfactual variables across multiple SWIGs.
Under the node-intervention approach, the cross-world independence $Y(z,m)\indp M(z')$ must be generalized into three assumptions, corresponding to independence relationships across $\mathcal{G}(z_0,m_1,m_2)$, $\mathcal{G}(z_1,m_1,m_2)$, and $\mathcal{G}(z_2,m_1,m_2)$, as shown in Figure~\ref{fig:CA}(b). 
In this setting, the distribution of the counterfactual outcome $Y\big\{z_0,M_1\big[z_1,M_2(z_2)\big],M_2(z_2)\big\}$ 
is identifiable under the cross-world independence assumptions (see Appendix~B for details of all the identification formulas below):

\eqev{
        P\big(Y\big\{z_0,M_1\big[z_1,M_2(z_2)\big],M_2(z_2)\big\}=y\big)   
        =&\sum_{m_1}\sum_{m_2}P\big(Y\big(z_0,m_1,m_2\big)=y\big|M_1\big(z_1,m_2\big)=m_1,M_2(z_2)=m_2\big)\\[-4mm]
        &\quq P\big(M_1(z_1,m_2)=m_1\big|M_2(z_2)=m_2\big)P\big(M_2(z_2)=m_2\big)\\
        \eqbb{26mm}{\text{cross-world}}{Y(z_0,m_1,m_2)\indp \{M_1(\smash{z_1,m_2}),M_2(\smash{z_2})\}; \;M_1(\smash{z_1,m_2})\indp M_2(\smash{z_2})} &\sum_{m_1}\sum_{m_2}P\big(Y\big(z_0,m_1,m_2\big)=y\big)P\big(M_1(z_1,m_2)=m_1\big)P\big(M_2(z_2)=m_2\big)\\
        \eqtt{exchangeability \& consistency}&\sum_{m_1}\sum_{m_2}P\big(Y=y\big|{z_0,M_1=m_1,M_2=m_2}\big)\\[-4mm]
        &\quq P\big(M_1=m_1\big|{z_1,M_2=m_2}\big)P\big(M_2=m_2\big|{z_2}\big).
\label{eq:WC}}  
    
Analogous to Figure~\ref{fig:mediation}(c), Figure~\ref{fig:CA}(b) illustrates cross-world confounding among $M_2(z_2)$, $M_1(z_1,m_2)$, and $Y(z_0,m_1,m_2)$ (shaded in red). Traditionally, this is addressed by assuming that $V$ does not exist, ensuring the mutual independence between $Y(z_0,m_1,m_2)$, $M_1(z_1,m_2)$, and $M_2(z_2)$, as stated in Equation~\eqref{eq:WC}. Alternatively, one may assume that $V(z)\indp V(z')$, or the absence of $\UV$, to satisfy cross-world independence; as an illustrative example can be found in~\cite{robins2010alternative}. 
We use $\UV$ to denote the shared latent factor that induces dependence among intervention-specific versions of $V$.
If $V$ exists and is treated as a \textit{mediator}, similar to the ``approach 2'' suggested in~\cite{vanderweele2014effect}, then it becomes a three-mediator problem.
If $V$ is instead treated as a \textit{mediated confounder}---our term for an exposure-induced confounder that is not treated as an intervention target, emphasizing its dual role as both a downstream variable and a confounder---then assuming $V(z)\indp V(z')$ allows identification of the distribution of $Y\!\big\{z_0,M_1[z_1,M_2(z_2)],M_2(z_2)\big\}$:

\eqev{
        P\big(Y\big\{z_0,M_1\big[z_1,M_2(z_2)\big],M_2(z_2)\big\}=y\big)  
        =&\sum_{m_1}\sum_{m_2}\sum_{v_0}\sum_{v_1}\sum_{v_2}P\big(Y=y\big|z_0,M_1=m_1,M_2=m_2,V=v_0\big) P\big(V=v_0\big|z_0\big)\\[-1mm]
        &\quq P\big(M_1=m_1\big|z_1,M_2=m_2,V=v_1\big) P\big(V=v_1\big|z_1\big)\\
        &\quq P\big(M_2=m_2\big|z_2,V=v_2\big) P\big(V=v_2\big|z_2\big).
\label{eq:WCv}}

As shown in Equation~\eqref{eq:WCv}, assuming the absence of $\UV$ allows identification of the distribution of $Y\big\{z_0,M_1\big[z_1,M_2(z_2)\big],M_2(z_2)\big\}$ using a formula analogous to Equation~\eqref{eq:WC}, but weighted by $P(V=v\gvn Z=z)$. Here, $V$ acts as a mediated confounder rather than a mediator of scientific interest. Consequently, no separate PSE is defined for $V$; instead, its distribution is integrated out through the weighting terms $P(V=v\gvn Z=z)$. A similar estimator was proposed by Robins and Richardson in the single-mediator setting~\cite{robins2010alternative}. 
When $V$ is present, we refer to the absence of $\UV$ as \textit{weak cross-world independence}, in contrast to the stronger condition introduced later for the path-intervention approach.
Identifying the distribution of $Y\big\{z_{00},M_1\big[z_{01},M_2(z_{11})\big],M_2(z_{10})\big\}$ for all possible paths therefore demands stronger cross-world independence assumptions:

\eqev{
        P\big(Y\big\{z_{00},M_1\big[z_{01},M_2(z_{11})\big],M_2(z_{10})\big\}=y\big)
        =&\sum_{m_{01}}\sum_{m_{10}}P\big(Y\big(z_{00},M_1\big[z_{01},M_2(z_{11})\big],m_{10}\big)=y,M_1\big[z_{01},M_2(z_{11})\big]=m_{01}\big|M_2(z_{10})=m_{10}\big)\\[-4mm]
        &\qqq P\big(M_2(z_{10})=m_{10}\big)\\
        =&\sum_{m_{01}}\sum_{m_{10}}\tred{\sum_{m_{11}}}P\big(Y(z_{00},m_{01},m_{10})=y\big|\tred{M_1(z_{01},m_{11})=m_{01},M_2(z_{11})=m_{11}},M_2(z_{10})=m_{10}\big)\\[-4mm]
        &\qqq\tred{P\big(M_1(z_{01},m_{11})=m_{01}\big|M_2(z_{10})=m_{10},M_2(z_{11})=m_{11}\big)}\\[-2mm]
        &\qqq P\big(M_2(z_{11})=m_{11}\big|M_2(z_{10})=m_{10}\big) P\big(M_2(z_{10})=m_{10}\big)\\
        \eqbb{0mm}{\text{\textbf{strong} cross-world}}{\tred{M_2(\smash{z_{11}})\indp M_2(\smash{z_{10}})}} &\sum_{m_{01}}\sum_{m_{10}}\sum_{m_{11}}P\big(Y(z_{00},m_{01},m_{10})=y\big|M_1(z_{01},m_{11})=m_{01},M_2(z_{11})=m_{11},M_2(z_{10})=m_{10}\big)\\[-4mm]
        &\qqq P\big(M_1(z_{01},m_{11})=m_{01}\big|M_2(z_{10})=m_{10},M_2(z_{11})=m_{11}\big)\\[-2mm]
        &\qqq \tred{P\big(M_2(z_{11})=m_{11}\big)} P\big(M_2(z_{10})=m_{10}\big)\\
        \eqbb{22mm}{\text{cross-world}}{\substack{\tred{Y(z_{00},m_{01},m_{10})\indp \{M_1(z_{01},m_{11}),M_2(z_{10}),M_2(z_{11})\}};\\[.5mm]\tred{M_1(z_{01},m_{11})\indp \{M_2(z_{10}),M_2(z_{11})\}}}}
        &\sum_{m_{01}}\sum_{m_{10}}\sum_{m_{11}}P\big(Y(z_{00},m_{01},m_{10})=y\big)\tred{P\big(M_1(z_{01},m_{11})=m_{01}\big)}\\[-7mm]
        &\qqq P\big(M_2(z_{11})=m_{11}\big) P\big(M_2(z_{10})=m_{10}\big)\\ 
        \eqtt{exchangeability \& consistency}&\sum_{m_{01}}\sum_{m_{10}}\sum_{m_{11}}P\big(Y=y\big|{z_{00},M_1=m_{01},M_2=m_{10}}\big)  P\big(M_1=m_{01}\big|{z_{01},M_2=m_{11}}\big)\\[-4mm] 
        &\qqq P\big(M_2=m_{11}\big|z_{11}\big) P\big(M_2=m_{10}\big|z_{10}\big).
\label{eq:SC}}

Using the path-intervention approach, the effects along $Z\rightarrow M_2\rightarrow Y$ and $Z\rightarrow M_2\rightarrow M_1\rightarrow Y$ can be separately identified.
This distinction introduces two intervention-specific counterfactuals, $M_2(z_{10})$ and $M_2(z_{11})$, corresponding to the two paths through $M_2$.
Their independence, $M_2(z_{10})\indp M_2(z_{11})$, requires the absence of $\mathcal{U}_{M_2}$.
Together with weak cross-world independence, which corresponds to the absence of $\UV$ when $V$ is present, we refer to this additional restriction as \textit{strong cross-world independence}.
As illustrated by the dark-red cross-world confounding in Figure~\ref{fig:CA}(c), a shared $\mathcal{U}_{M_2}$ induces dependence between $M_2(z_{10})$ and $M_2(z_{11})$ across the two SWIGs.
The remaining cross-world independence conditions in Equation~\eqref{eq:SC} require mutual independence among three units: $Y(z_{00},m_{01},m_{10})$, $M_1(z_{01},m_{11})$, and the pair $\big\{M_2(z_{10}),M_2(z_{11})\big\}$, as indicated by the red shading.
These conditions are identical to the weak cross-world independence introduced under the node-intervention approach and require the absence of $\UV$ when $V$ is present. 
Thus, compared with the node-intervention approach, separately identifying the two paths through $M_2$ requires the additional cross-world restriction represented by the absence of $\mathcal{U}_{M_2}$.

The absence of $\mathcal{U}_{M_2}$ is a stronger condition than $V(z)\indp V(z')$ since $V$ may not exist whereas the mediators are fully observed. 
For a general case with $p$ mediators, identifying all possible paths requires the absence of $\big\{\mathcal{U}_{M_p},\mathcal{U}_{M_{p-1}},\ldots,\mathcal{U}_{M_2}\big\}$. We therefore refer to the absence of $\big\{\mathcal{U}_{M_j}\big\}_{j\in\mathbb{J},j\neq1}$ as the condition of \textit{strong cross-world independence}. 
Albert and Nelson also discussed this assumption in a two-mediator setting~\cite{albert2011generalized}. 
In brief, the weak cross-world independence is sufficient to identify $p+1$ PSEs under the node-intervention approach, whereas both weak and strong cross-world independence are required to identify $2^p$ PSEs using the path-intervention approach. 

\section{Interventional approach}\label{sec:interventional}

\begin{figure}
    \centering\includegraphics[width=.95\textwidth]{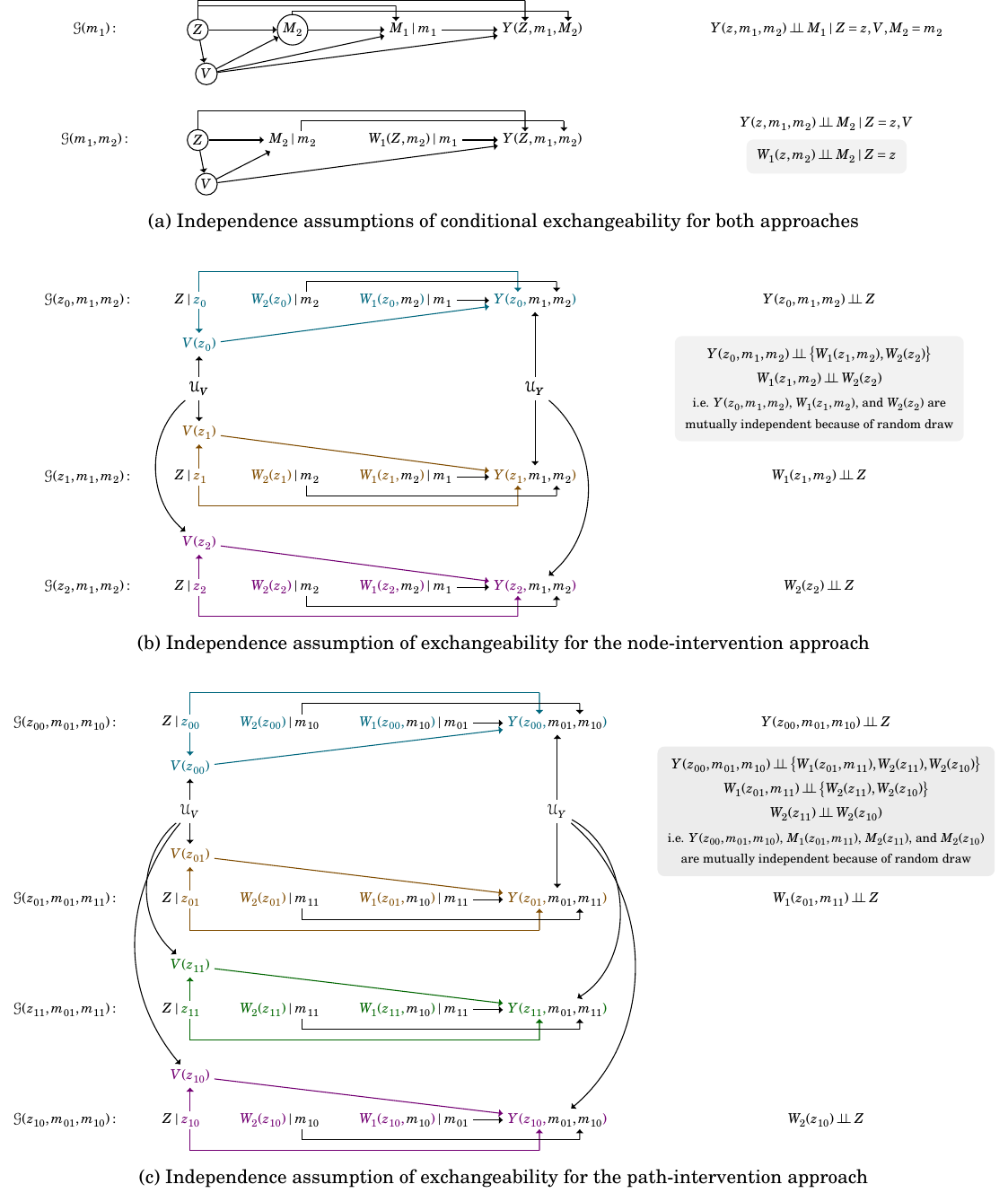}
\caption{\textbf{Interventional approach.}
The independence assumptions in (a--c) without shading represent classical
exchangeability.
As in Figure~\ref{fig:CA}, the independence conditions in (b--c) are
displayed for the case without $V$, while the $V(z)$ nodes are retained
to illustrate the corresponding cross-world dependence structure.
The gray shading in (b--c) indicates counterfactual independence induced
by replacing the counterfactual mediators $M(\cdot)$ with random draws
$W(\cdot)$.}
    \label{fig:IA}
\end{figure}

The \textit{interventional approach} was proposed by VanderWeele, Vansteelandt, and Robins~\cite{vanderweele2014effect}, and later extended by Lin and VanderWeele to identify PSEs~\cite{lin2017interventional}. 
From a graphical perspective, this approach removes the need to assume the weak cross-world independence induced by $\UV$.
The key idea is to replace the counterfactual mediator $M(\cdot)$ with a random draw $W(\cdot)$ that is independent of other counterfactual variables (i.e., no incoming edges into $W(\cdot)$) and follows the corresponding mediator distribution, i.e., \(W(\mathbf{a})\mid \mathbf{A}=\mathbf{a}\;\overset{d}{=}\;M(\mathbf{a})\mid \mathbf{A}=\mathbf{a},\)
where $\mathbf{A}=\mathbf{a}$ denotes the observed assignment of the variables corresponding to the intervention arguments $\mathbf{a}$.
This construction builds on earlier formulations that consider interventions drawing the mediator from specified distributions~\cite{didelez2006direct,geneletti2007identifying}.
Consequently, cross-world independence assumptions involving counterfactual mediators are no longer required. 
As illustrated in Figure~\ref{fig:IA}, the independence conditions involving counterfactual mediators are guaranteed by the definition of $W(\cdot)$ (shaded in gray).
Thus, apart from the classical identification assumptions (exchangeability, consistency, and positivity), no additional cross-world independence assumptions are required to identify the distribution of $Y\big\{z_0,W_1\big[z_1,W_2(z_2)\big],W_2(z_2)\big\}$ using the node-intervention approach:

\eqev{
   P\big(Y\big\{z_0,W_1\big[z_1,W_2(z_2)\big],W_2(z_2)\big\}=y\big)   
        =&\sum_{m_1}\sum_{m_2}P\big(Y\big(z_0,m_1,m_2\big)=y\big|W_1\big(z_1,m_2\big)=m_1,W_2(z_2)=m_2\big)\\[-4mm]
        &\quq P\big(W_1(z_1,m_2)=m_1\big|W_2(z_2)=m_2\big)P\big(W_2(z_2)=m_2\big)\\
        \eqbb{24mm}{\text{random draw}}{\tred{Y(z_0,m_1,m_2)\indp \{W_1(z_1,m_2),W_2(z_2)\};\;W_1(z_1,m_2)\indp W_2(z_2)}} &\sum_{m_1}\sum_{m_2}P\big(Y\big(z_0,m_1,m_2\big)=y\big)P\big(W_1(z_1,m_2)=m_1\big)P\big(W_2(z_2)=m_2\big)\\
        \eqtt{$W(\mathbf{a})\mid \mathbf{A}=\mathbf{a}\;\overset{d}{=}\;M(\mathbf{a})\mid \mathbf{A}=\mathbf{a}$ \& exchangeability}&\sum_{m_1}\sum_{m_2}P\big(Y\big(z_0,m_1,m_2\big)=y\big|{z_0,M_1=m_1,M_2=m_2}\big)\\[-4mm]
        &\quq P\big(\tred{M_1(z_1,m_2)=m_1}\big|z_1,M_2=m_2\big)P\big(\tred{M_2(z_2)=m_2}\big|z_2\big)\\
        \eqtt{consistency}&\sum_{m_1}\sum_{m_2}P\big(Y=y\big|{z_0,M_1=m_1,M_2=m_2}\big) P\big(M_1=m_1\big|{z_1,M_2=m_2}\big)P\big(M_2=m_2\big|{z_2}\big),
\label{eq:WCi}} 

The key advantage of the interventional approach is its ability to identify the counterfactual outcome even in the presence of $V$, without requiring additional assumptions---whereas the classical assumptions rely on the absence of $\UV$. When $V$ is present and treated as a mediated confounder, the counterfactual outcome $Y\big\{z_0,W_1\big[z_1,W_2(z_2)\big],W_2(z_2)\big\}$ can be identified as follows:

\eqev{
        P\big(Y\big\{z_0,W_1\big[z_1,W_2(z_2)\big],W_2(z_2)\big\}=y\big)  
        =&\sum_{m_1}\sum_{m_2}\sum_{v_0} P\big(Y=y\big|z_0,M_1=m_1,M_2=m_2,V=v_0\big) P\big(V=v_0\big|z_0\big)\\[-4mm]
        &\qqq P\big(M_1=m_1\big|z_1,M_2=m_2\big) P\big(M_2=m_2\big|z_2\big).
\label{eq:WCiv}}

Unlike Equation~\eqref{eq:WCv} under the classical assumptions, Equation~\eqref{eq:WCiv} does not weight mediator probabilities by $V$, as the counterfactual outcome uses randomly drawn mediators $W$, removing any association between $V$ and the original mediators $M$.  
Another advantage of the interventional approach is that it identifies the PSE associated with every path under the path-intervention approach without requiring additional cross-world independence assumptions.
As shown in Equation~\eqref{eq:SC}, the classical assumptions require the absence of both $\UV$ and $\mathcal{U}_{M_2}$ to identify the distribution of counterfactual outcomes, whereas the interventional approach only needs exchangeability to yield the same formula (see Appendix~B, Equation~\eqref{eq:SCi}). 

While the interventional approach eliminates the need for cross-world independence assumptions, it introduces two important limitations. 
First, the resulting estimands do not, in general, coincide with the corresponding classical counterfactual quantities (e.g., $Y(z)\neq Y\{z,W(z)\}$ when a mediated confounder is present), so the total effect may not be recovered. 
Second, the modified dependence structure may alter the sign of the causal effects. 
Miles demonstrated a case where the interventional effect reverses sign, even when all individual-level effects point in the same direction~\cite{miles2023causal}.

\section{Separable effects}\label{sec:SE}
\begin{figure}
    \centering\includegraphics[width=.7\textwidth]{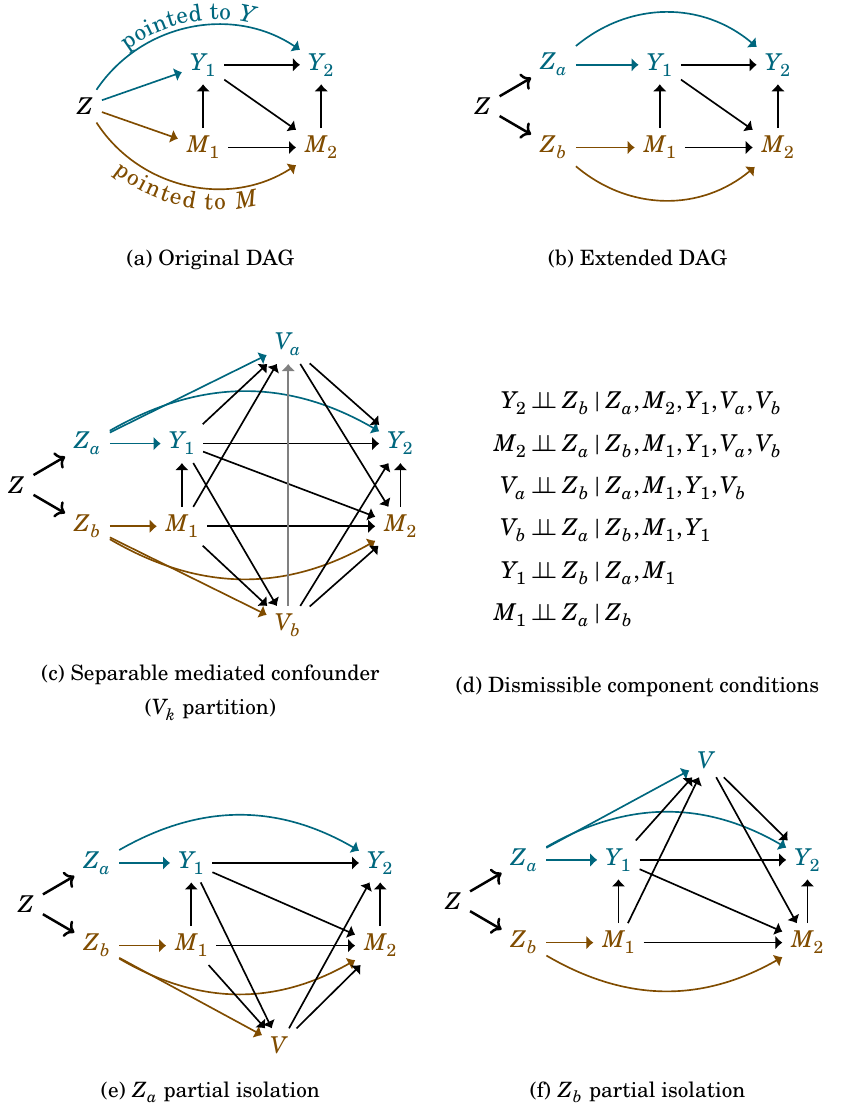}
    \caption{\textbf{Competing risk scenario without counterfactual notions.}
    (a) DAG from Stensrud et al.~\cite{stensrud2022separable, stensrud2021generalized}, where exposure $Z$ may cause two competing events $M$ and $Y$. 
    (b) extended DAG that conceptually detaches the observed $Z$ from the original DAG and introduces two manipulable components, $Z_a$ and $Z_b$, attributing effects to $Y$ and $M$, respectively. 
    (c) DAG with separable mediated confounders, i.e., no confounder has directed paths from both $Z_a$ and $Z_b$. (original Figure~5(b) in~\cite{stensrud2021generalized})
    (d) DCCs proposed in~\cite{stensrud2021generalized}. 
    The independence statements can be read from panel (c) via $d$-separation rules. 
    The condition implies that $V_b\rightarrow V_a$ is allowed even if it is not depicted in the original Figure~5(b) in~\cite{stensrud2021generalized}. 
    (e-f) Partial isolation scenarios are special cases of (c) that only one component of the exposure is followed by a mediated confounder $V$.}
    \label{fig:stLIFE}
\end{figure}

While the interventional approach achieves cross-world independence by randomizing the counterfactual mediators, the \textit{separable effects} framework takes a different route by introducing \textit{manipulable components} to replace the actual exposure. 
The idea originates from the extended DAG introduced by Robins and Richardson~\cite{robins2010alternative}; and was later developed in the context of competing events by Stensrud et al.~\cite{stensrud2021generalized,stensrud2022separable}.
We summarize the competing risk scenario in Figure~\ref{fig:stLIFE}, where the observed exposure $Z$ is replaced by two manipulable components, $Z_a$ and $Z_b$, attributing effects to $Y$ and $M$, respectively. 

Figure~\ref{fig:stLIFE}(c) illustrates what we refer to as a \textit{separable mediated confounder}, describing a graphical configuration in which no confounder has directed paths from both $Z_a$ and $Z_b$.
Under this structure, the causal effects remain identifiable if the \textit{dismissible component conditions} (DCCs)~\cite{stensrud2021generalized} listed in Figure~\ref{fig:stLIFE}(d) hold.
One may also consider partial isolation scenarios, as illustrated in Figure~\ref{fig:stLIFE}(e-f) (original Figure~4 in~\cite{stensrud2021generalized}), where the mediated confounder is affected by only one manipulable component. In the case of $Z_a$ partial isolation, the direct effect cannot be attributed solely to $Z_a$, since the pathway $Z_b \rightarrow V \rightarrow Y_2$ also contributes. Similarly, under $Z_b$ partial isolation, the indirect effect cannot be attributed solely to $Z_b$, due to the path $Z_a \rightarrow V \rightarrow M_2 \rightarrow Y_2$. 
Therefore, the effects attributed to $Z_a$ and $Z_b$ in Figure~\ref{fig:stLIFE}(e–f) are not direct and indirect effects, respectively.

\begin{figure}
    \centering\includegraphics[width=.7\textwidth]{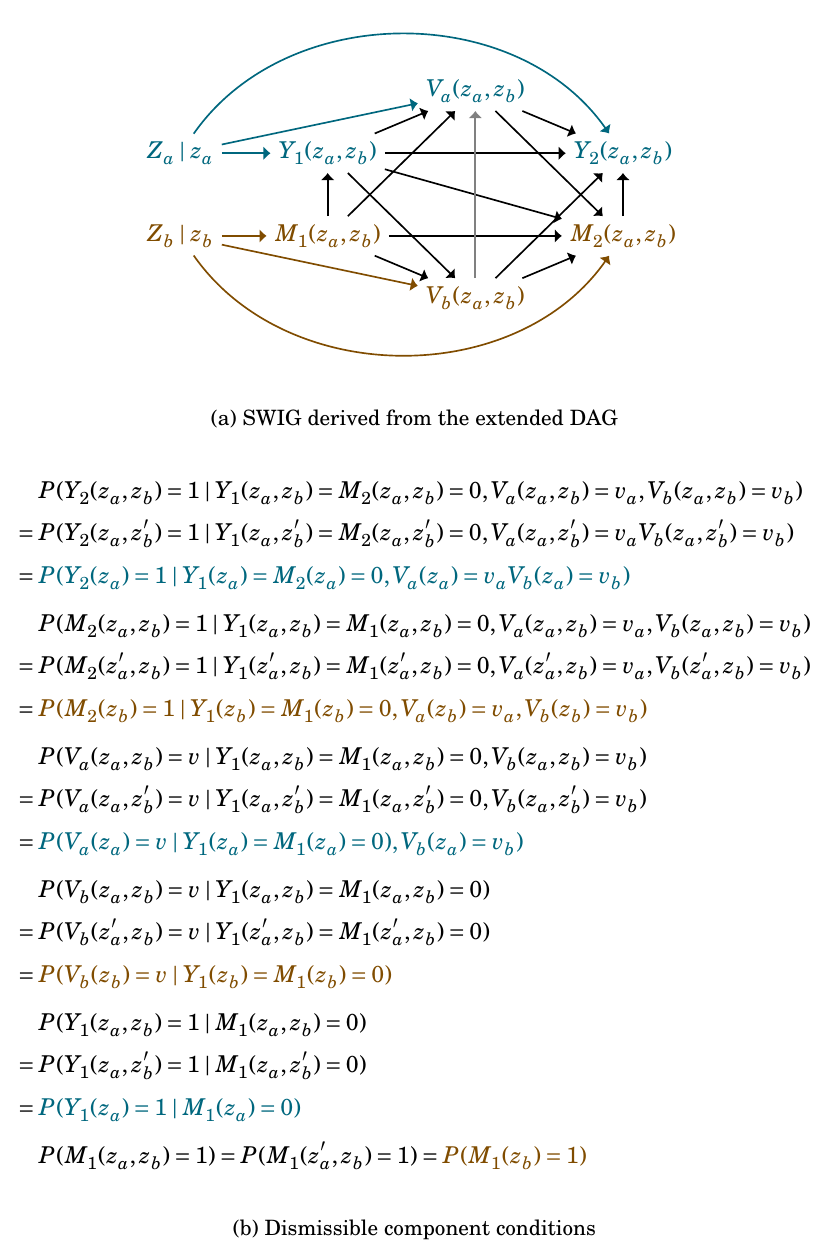}
    \caption{\textbf{Competing risk scenario with counterfactual notions.}  
    (a) Corresponding SWIG of Figure~\ref{fig:stLIFE}(c). There is no difference in graphical structure except that $Z_a$ and $Z_b$ are intervened as $z_a$ and $z_b$, respectively. (b) DCCs using counterfactual notation. One can see the intervention $z_a$ contributes to the effects toward $Y_1$, $V_a$, and $Y_2$; $z_b$ to the effects toward $M_1$, $V_b$, and $M_2$. }
    \label{fig:stJASA}
\end{figure}

\begin{figure}
    \centering\includegraphics[width=\textwidth]{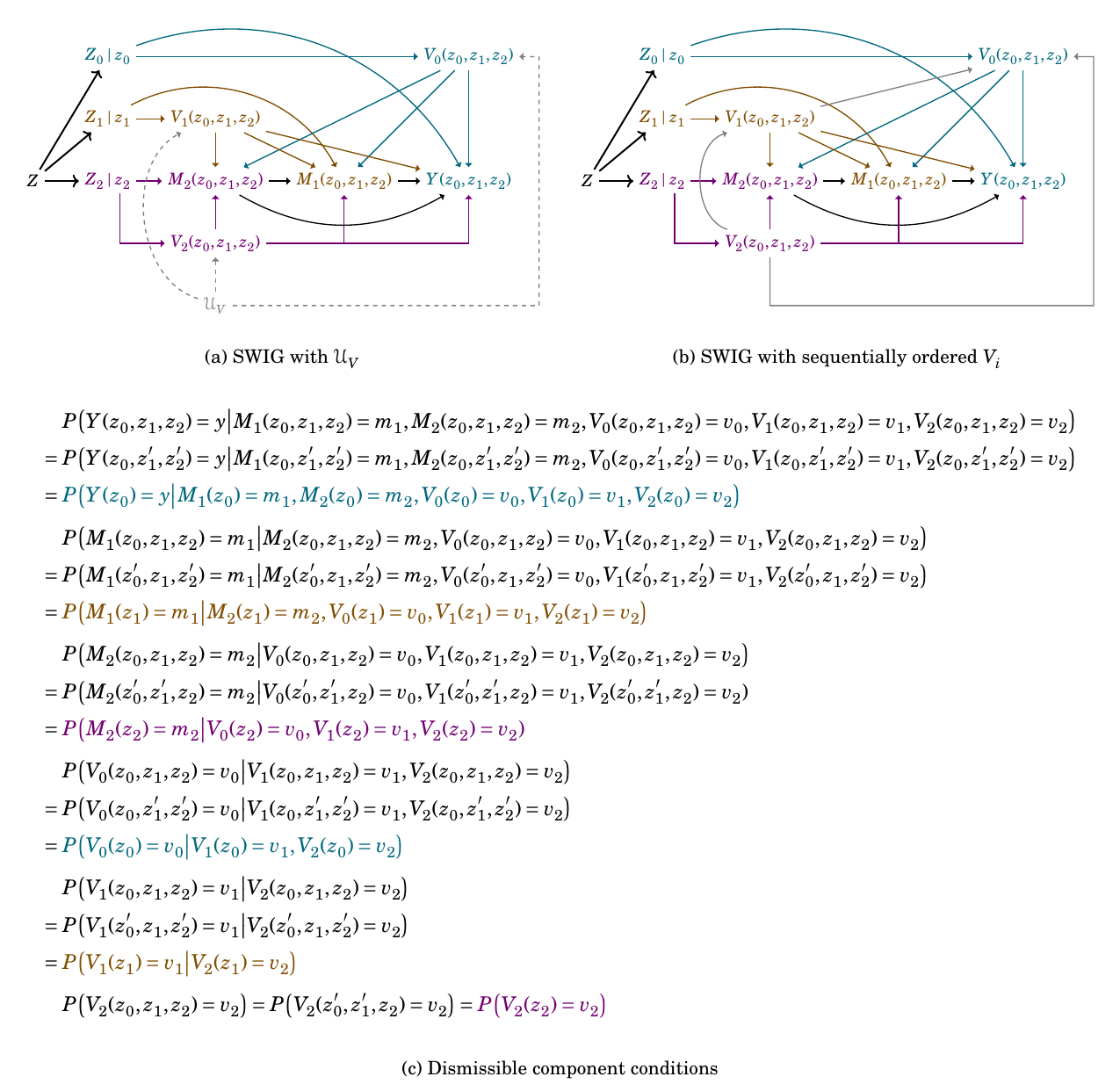}
    \caption{\textbf{Separable effects for the node-intervention approach.} 
    (a) SWIG of the extended DAG for the two-mediator model. The actual exposure $Z$ is detached from the DAG and replaced by three manipulable components $Z_0$, $Z_1$, and $Z_2$, each attributing one of the three PSEs. Three mediated confounders, $V_0$, $V_1$, and $V_2$, are also introduced; $\UV$ denotes an unmeasured confounder among them, whose presence violates DCCs.
    (b) Although $\UV$ is not allowed under DCCs, identification of PSEs remains possible if the joint distribution of $V_0\zpd$, $V_1\zpd$, and $V_2\zpd$ can be factorized.
    (c) DCCs for identifying PSEs in panel (b).}
    \label{fig:SE3}
\end{figure}

Unlike the classical or interventional approaches, Figure~\ref{fig:stLIFE}(c) involves no counterfactual notion. 
The DCCs in Figure~\ref{fig:stLIFE}(d) impose that, conditional on the relevant causal history, each variable is invariant to interventions on exposure components outside its causal ancestry.
Figure~\ref{fig:stJASA} illustrates this dependency structure using SWIG, and Figure~\ref{fig:SE3} applies the same principle to identify PSEs: three manipulable components, $Z_1$, $Z_2$, and $Z_3$, are introduced to attribute the PSEs in the node-intervention approach. We first consider the case without mediated confounders, the scenario also suggested by Chen et al~\cite{chen2025definition}.
Under DCCs in Figure~\ref{fig:SE3}(c), the distribution of $Y\zpd$ can be identified as follows:

\eqev{
        P\big(Y\zpd=y\big)
        =&\sum_{m_1}\sum_{m_2}P\big(Y\zpd=y\big|M_1\zpd=m_1,M_2\zpd=m_2\big)\\[-4mm]
        &\quq P\big(M_1\zpd=m_1\big|M_2\zpd=m_2\big) P\big(M_2\zpd=m_2\big)\\            
        \eqbb{3mm}{\text{DCC}}{\text{Figure~\ref{fig:SE3}(c)}} &\sum_{m_1}\sum_{m_2}P\big(Y(\tred{z_0})=y\big|M_1(\tred{z_0})=m_1,M_2(\tred{z_0})=m_2\big)\\[-4mm] &\quq P\big(M_1(\tred{z_1})=m_1\big|M_2(\tred{z_1})=m_2\big) P\big(M_2(\tred{z_2})=m_2\big)\\
        \eqtt{exchangeability \& consistency}&\sum_{m_1}\sum_{m_2}P\big(Y=y\big|{Z=z_0},M_1=m_1,M_2=m_2\big)\\[-4mm] 
        &\quq P\big(M_1=m_1\big|{Z=z_1},M_2=m_2\big) P\big(M_2=m_2\big|{Z=z_2}\big).
\label{eq:WCs}}

The separable effects framework differs from the classical assumptions in two key ways. 
First, the classical assumptions represent interventions through multiple SWIGs for $Z=z_0$, $Z=z_1$, and $Z=z_2$ in Figure~\ref{fig:CA}(b), whereas the separable effects framework uses a single SWIG with three manipulable components, $(Z_0,Z_1,Z_2)$ jointly assigned to $(z_0,z_1,z_2)$ in Figure~\ref{fig:SE3}(a). 
Besides DCCs, we use two properties to characterize the separable-effects representation: \textit{component manipulability} and \textit{component consistency}.
Component manipulability means that $Z_0$, $Z_1$, and $Z_2$ can, in principle, be manipulated independently under hypothetical interventions.
Component consistency refers to the deterministic equality $Z_0=Z_1=Z_2=Z$. Since the hypothetical components $(Z_0,Z_1,Z_2)$ are not directly observed, this equality plays a role analogous to consistency by linking the component-wise intervention representation back to the observed treatment.

Second, under the classical assumptions, variables associated with different interventions, such as $V(z_0)$, $V(z_1)$, and $V(z_2)$, are defined in different SWIGs and may depend on a common exogenous variable $\UV$. 
By contrast, the separable effects framework represents all variables within a single SWIG, jointly intervened by $(Z_0,Z_1,Z_2)$.
In Figure~\ref{fig:SE3}(a), $V_0\zpd$, $V_1\zpd$, and $V_2\zpd$ are each affected only by the corresponding components $Z_0$, $Z_1$, and $Z_2$.
If these variables are generated by distinct exogenous variables, and their joint distribution factorizes according to the graph in Figure~\ref{fig:SE3}(b), then DCCs hold.
However, if $V_0\zpd$, $V_1\zpd$, and $V_2\zpd$ are interpreted as component-specific analogues of $V(z_0)$, $V(z_1)$, and $V(z_2)$, it is natural to posit $\UV$, as shown in Figure~\ref{fig:SE3}(a), and its presence violates DCCs.
For example, $z_2\rightarrow\fbox{$V_2\zpd$}\leftarrow\UV\rightarrow V_1\zpd$ implies 
\[P\big(V_1\zpd=v_1\big|V_2\zpd=v_2\big)\neq P\big(V_1(z_0',z_1,z_2')=v_1\big|V_2(z_0',z_1,z_2')=v_2\big).\]
We want to emphasize that, even under the setting of separable effects, $\UV$ can not exist, just as under the classical assumption. 
This does not, however, require $V_0(z_0,z_1,z_2)$, $V_1(z_0,z_1,z_2)$, and $V_2(z_0,z_1,z_2)$ to be mutually independent. As long as their joint distribution remains compatible with the factorization implied by the graph, a sequential dependence structure among them is permitted, as illustrated in Figure~\ref{fig:SE3}(b). Nevertheless, requiring such a factorization can be highly restrictive and somewhat artificial, particularly when $V_0$, $V_1$, and $V_2$ are viewed as component-specific analogues of the same underlying variable $V$.
Under these conditions, the distribution of $Y\zpd$ can be identified as

\eqev{
        P\big(Y\zpd=y\big)  
        =&\sum_{m_1}\sum_{m_2}\sum_{v_0}\sum_{v_1}\sum_{v_2} P\big(Y=y\big|z_0,M_1=m_1,M_2=m_2,V_0=v_0,V_1=v_1,V_2=v_2\big)\\[-1mm]
        &\quq P\big(M_1=m_1\big|z_1,M_2=m_2,V_0=v_0,V_1=v_1,V_2=v_2\big)\\
        &\quq P\big(M_2=m_2\big|z_2,V_0=v_0,V_1=v_1,V_2=v_2\big)\\
        &\quq P\big(V_0=v_0\big|z_0,V_1=v_1,V_2=v_2\big) P\big(V_1=v_1\big|z_1,V_2=v_2\big) P\big(V_2=v_2\big|z_2\big).
\label{eq:WCsv}}

Here again, if we interpret $V_0\zpd$, $V_1\zpd$, and $V_2\zpd$ as component-specific analogues of $V(z_0)$, $V(z_1)$, and $V(z_2)$, and impose \textit{interventional component manipulability} (ICM), i.e., $V_0\zpd$, $V_1\zpd$, $V_2\zpd$ are mutaully independent, along with \textit{interventional component consistency} (ICC), i.e., $V_0(z)=V_1(z)=V_2(z)=V(z)$, then DCCs in Figure~\ref{fig:SE3s}(a) hold, yielding the same identification formula as in Equation~\eqref{eq:WCv} (see Appendix~B, Equation~\eqref{eq:WCsvv}). 
We give the formal definition of ICM and ICC in Appendix~C.
The intuition of ICC is illustrated in Figure~\ref{fig:SE3s}. In panel~(a), the first dismissible component condition states that $Y\zpd$ conditional on its ancestors depends only on $Z_0$. Under $(Z_0,Z_1,Z_2)=(z_0,z_0,z_0)$, panel~(b) shows that only $Z_0=z_0$ influences $Y(z_0)$ and panel~(c) suggests that $(Z_0,Z_1,Z_2)=(z_0,z_0,z_0)$ equals to $Z=z_0$ because of component consistency, therefore ICC implies $V_0(z_0)=V_1(z_0)=V_2(z_0)=V(z_0)$.

ICC not only resolves issues arising from mediated confounders but also underlies the path-intervention approach. Here, we apply DCCs together with ICC in Figure~\ref{fig:SE4}(a) to identify the distribution of $Y\zcd$:

\eqev{
        P\big(Y\zcd=y\big)
        \eqtt{ICM}&\sum_{m_{01}}\sum_{m_{10}}\sum_{m_{11}} P\big(Y\zcd=y\big|M_1\zcd=m_{01},\\[-4mm]
        &\qqq\qqq M_2^{10}\zcd=m_{10},M_2^{11}\zcd=m_{11}\big)\\[-2mm]
        &\qqq P\big(M_1\zcd=m_{01}\big|M_2^{10}\zcd=m_{10},\\[-2mm]
        &\qqq\qqq M_2^{11}\zcd=m_{11}\big)\\[-2mm]
        &\qqq \tred{P\big(M_2^{10}\zcd=m_{10}\big)P\big(M_2^{11}\zcd=m_{11}\big)}\\    
        \eqbb{4mm}{\text{DCC}}{\text{Figure~\ref{fig:SE4}(a)}} &\sum_{m_{01}}\sum_{m_{10}}\sum_{m_{11}} P\big(Y(\tred{z_{00}})=y\big|M_1(\tred{z_{00}})=m_{01},M_2^{10}(\tred{z_{00}})=m_{10},M_2^{11}(\tred{z_{00}})=m_{11}\big)\\[-4mm]
        &\qqq P\big(M_1(\tred{z_{01}})=m_{01}\big|M_2^{10}(\tred{z_{01}})=m_{10},M_2^{11}(\tred{z_{01}})=m_{11}\big)\\[-2mm]
        &\qqq P\big(M_2^{10}(\tred{z_{10}})=m_{10}\big) P\big(M_2^{11}(\tred{z_{11}})=m_{11}\big)\\
        \eqbb{10mm}{\text{ICC}}{M_2(z)=M_2^{10}(z)=M_2^{11}(z)} &\sum_{m_{01}}\sum_{m_{10}}\sum_{m_{11}} P\big(Y(z_{00})=y\big|M_1(z_{00})=m_{01},\tred{M_2(z_{00})=m_{10}}\big)\\[-4mm]
        &\qqq P\big(M_1(z_{01})=m_{01}\big|\tred{M_2(z_{01})=m_{11}}\big) P\big(\tred{M_2(z_{10})}=m_{10}\big) P\big(\tred{M_2(z_{11})}=m_{11}\big)\\           
        \eqtt{exchangeability \& consistency}&\sum_{m_{01}}\sum_{m_{10}}\sum_{m_{11}}P\big(Y=y\big|z_{00},M_1=m_{01},M_2=m_{10}\big) P\big(M_1=m_{01}\big|z_{01},M_2=m_{11}\big)\\[-4mm]
        &\qqq P\big(M_2=m_{11}\big|z_{11}\big) P\big(M_2=m_{10}\big|z_{10}\big).
\label{eq:SCs}}

To distinguish between the paths $Z \rightarrow M_2 \rightarrow M_1 \rightarrow Y$ and $Z \rightarrow M_2 \rightarrow Y$, we introduce two manipulable components, $M_2^{10}$ and $M_2^{11}$, to replace $M_2$ in Figure~\ref{fig:PSE}(b), and the SWIG of the extended DAG is shown in Figure~\ref{fig:SE4}(b). 
As shown in Figure~\ref{fig:SE4}(c), the first dismissible component condition ensures that $Y\zcd$ conditioned on its ancestors, depends only on $Z_{00}$. Finally, ICC in Figure~\ref{fig:SE4}(d) equates the manipulable components back to the original mediator $M_2$ within DCCs, i.e. $M_2^{10}(z_{00})=M_2^{11}(z_{00})=M_2(z_{00})$. 
The required independence between $M_2^{10}(z_{00}, z_{01}, z_{10}, z_{11})$ and $M_2^{11}(z_{00}, z_{01}, z_{10}, z_{11})$ is essentially as restrictive as requiring the absence of a shared exogenous $\mathcal{U}_{M_2}$ in the classical framework; likewise, mutual independence among $V_0(z_0, z_1, z_2)$, $V_1(z_0, z_1, z_2)$, and $V_2(z_0, z_1, z_2)$ is as restrictive as requiring the absence of a shared exogenous $\mathcal{U}_V$.

\begin{figure}
    \centering\includegraphics[width=\textwidth]{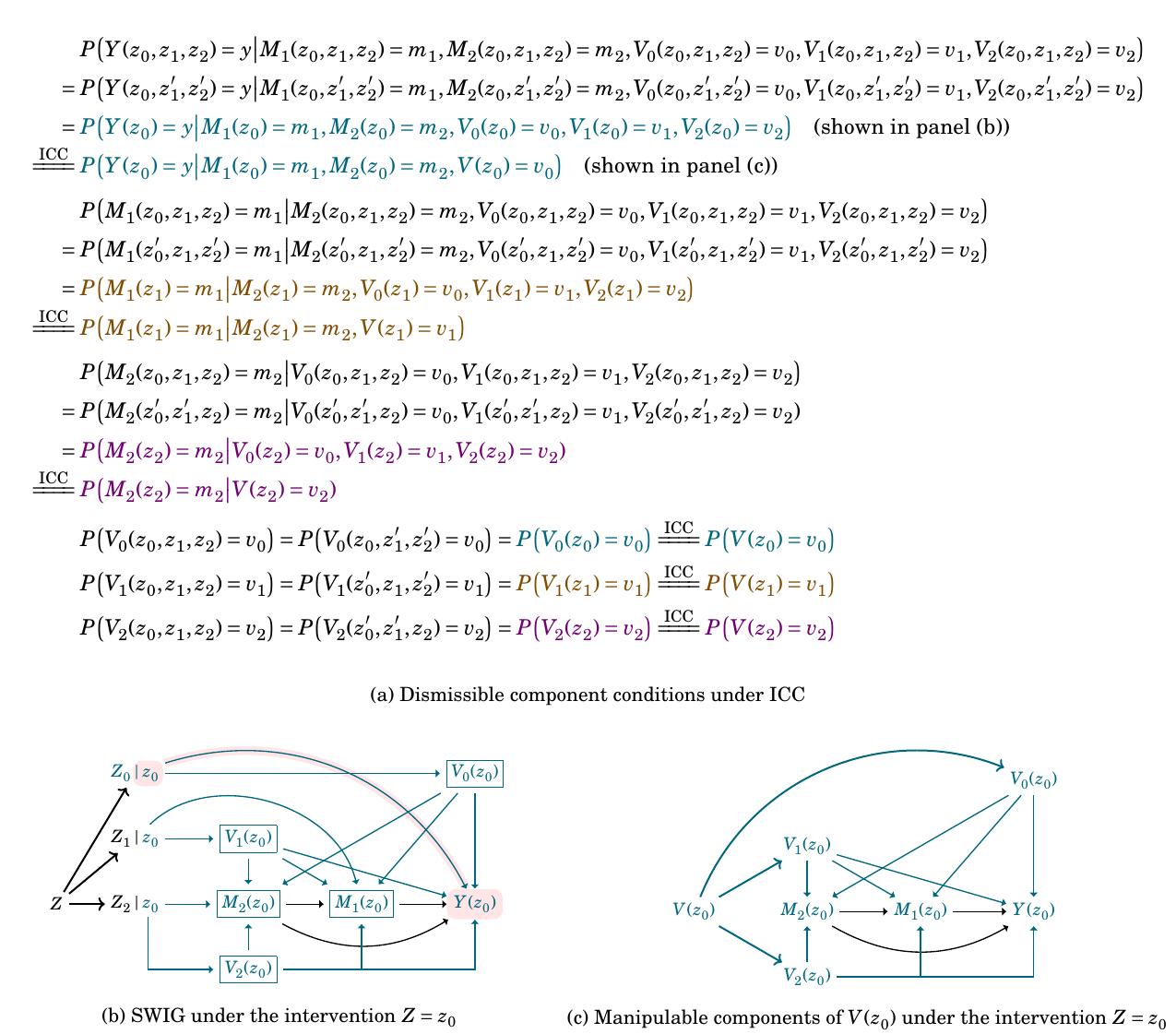}
    \caption{\textbf{Separable effects for the node-intervention approach with ICC.} 
    (a) Under DCCs, suppose $V_0$, $V_1$, and $V_2$ are manipulable components of $V$. 
    Then $V_0\zpd\indp V_1\zpd\indp V_2\zpd$, and ICC implies $V(z)=V_0(z)=V_1(z)=V_2(z)$.
    (b) SWIG illustrating the conditional dependencies underlying $P\big(Y(z_0)=y\big|M_1(z_0)=m_1,M_2(z_0)=m_2,V_0(z_0)=v_0,V_1(z_0)=v_1,V_2(z_0)=v_2\big)$ under the intervention $Z=z_0$. The red shaded path indicates that only $Z_0$ influences $Y(z_0)$.
    (c) Under the intervention $Z=z_0$, $V_0(z_0)$, $V_1(z_0)$, and $V_2(z_0)$ are the manipulable components of $V(z_0)$; by ICC, they coincide with $V(z_0)$, i.e., $V(z_0)=V_0(z_0)=V_1(z_0)=V_2(z_0)$.}
    \label{fig:SE3s}
\end{figure}

\begin{figure}
    \centering\includegraphics[width=\textwidth]{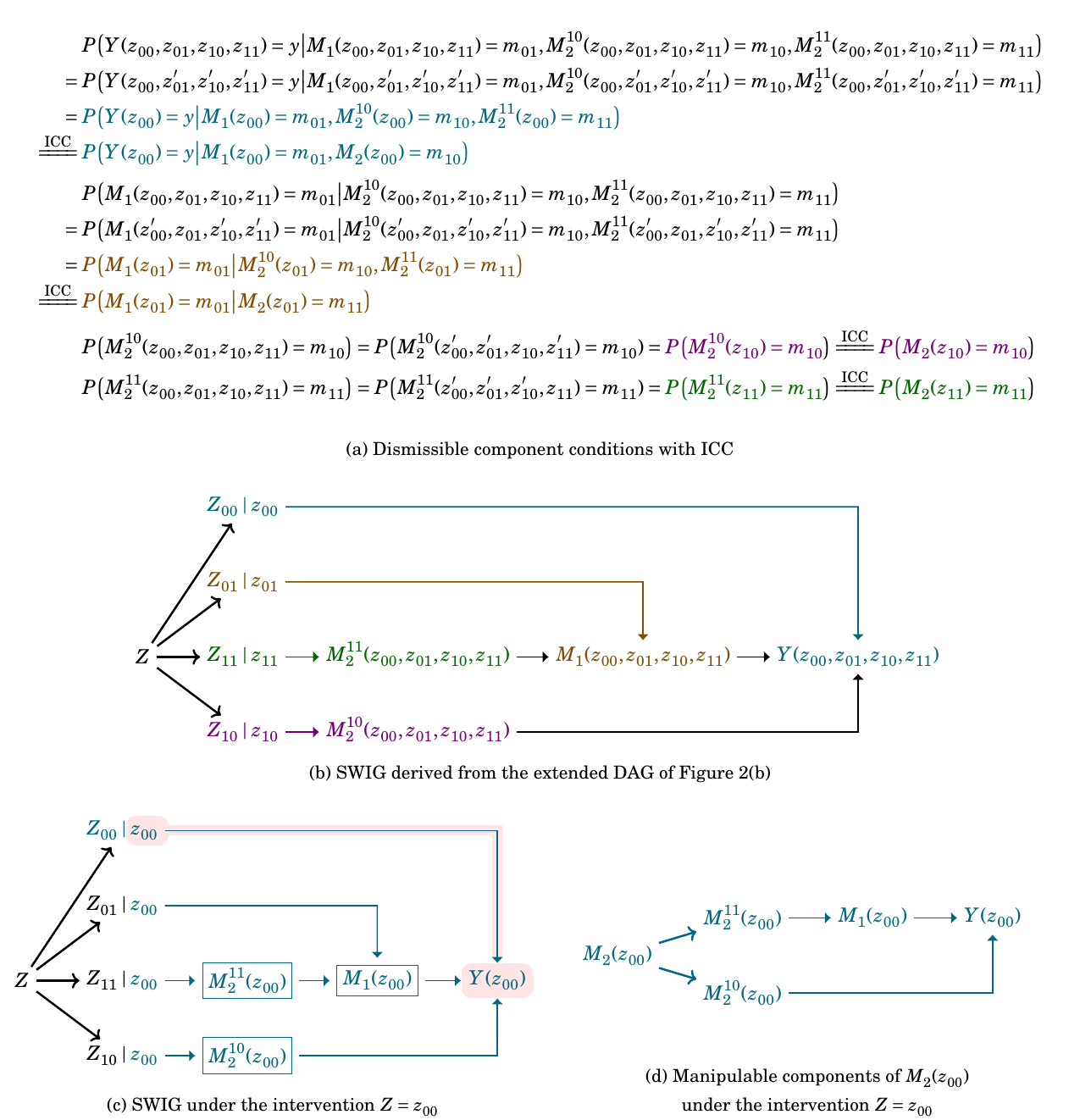}
    \caption{\textbf{Separable effects for the path-intervention approach with ICC.} 
    (a) Under DCCs, $M_2^{10}(\cdot)$ and $M_2^{11}(\cdot)$ are manipulable components of $M_2(\cdot)$. 
    Then $M_2^{10}\zcd\indp M_2^{11}\zcd$, and ICC implies $M_2(z_i)=M_2^{10}(z_i)=M_2^{11}(z_i)$.
    (b) SWIG illustrating the extended DAG with two manipulable components of $M_2$. 
    (c) SWIG illustrating the conditional dependencies underlying $P\big(Y(z_{00})=y\big|M_1(z_{00})=m_{01},M_2^{10}(z_{00})=m_{10},M_2^{11}(z_{00})=m_{11}\big)$ under the intervention $Z=z_{00}$. The red shaded path indicates only $Z_{00}$ influences $Y(z_{00})$.
    (d) Under the intervention $Z=z_{00}$, $M_2^{10}(z_{00})$ and $M_2^{11}(z_{00})$ are the manipulable components of $M_2(z_{00})$, so $M_2(z_{00})=M_2^{10}(z_{00})=M_2^{11}(z_{00})$. }
    \label{fig:SE4}
\end{figure}

\section{A hybrid approach: Recanting twins effects}\label{sec:RT}

\begin{figure}
    \centering\includegraphics[width=\textwidth]{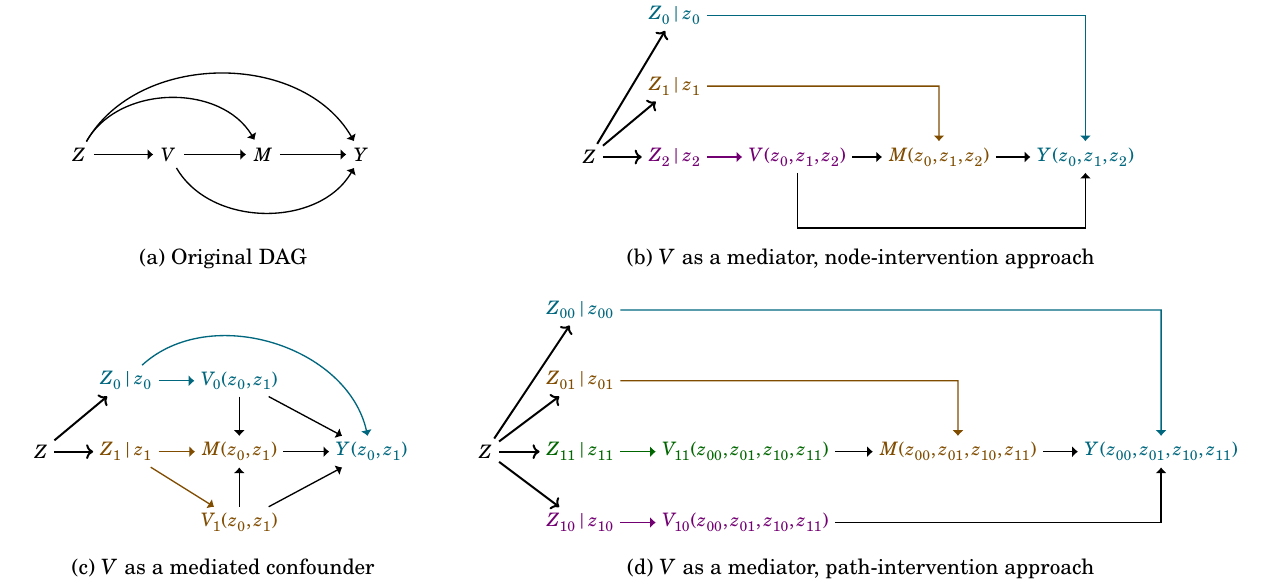}
    \caption{\textbf{Generalized Separable effects.} 
        (a) The causal mechanism of $Z$, $V$, $M$, and $Y$. The extended DAG depends on the role of $V$.
        (b) If $V$ is treated as a mediator and the node-intervention approach is used to identify the PSE, then only $Z$ is replaced by the three components.
        (c) If $V$ is treated as a mediated confounder, then $V$ is replaced by two components, aligned with the components of $Z$.
        (d) If $V$ is treated as a mediator and the path-intervention approach is used to identify the PSE, then $V$ is replaced by two components, aligned with the paths through which $V$ operates.}
    \label{fig:SEg}
\end{figure}

The treatment of $V$ depends on whether pathways through $V$ are themselves of interest. 
If $V$ is treated as a mediator, paths through $V$ contribute to the PSE decomposition; if $V$ is treated as a mediated confounder, we focus on PSEs involving the mediators $M$ and do not separately target paths through $V$. 
Figure~\ref{fig:SEg} illustrates these two representations and their corresponding joint interventions. In particular, when $V$ is treated as a mediated confounder, no intervention is defined specifically for paths through $V$; for example, $Z\rightarrow V\rightarrow Y$ is not separately identified in Figure~\ref{fig:SEg}(c).
The \textit{recanting twins} effect, proposed by Vo et al.~\cite{vo2026recanting} and further discussed in~\cite{liu2024general}, addresses this situation by combining elements of the classical and interventional approaches. 
It treats $V$ structurally as a mediator but replaces its counterfactuals with random draws, as in the interventional approach.
This hybrid construction preserves the classical definition of PSEs involving $M$ while avoiding cross-world assumptions for~$V$.

\begin{figure}
    \centering
    \includegraphics[width=.92\textwidth]{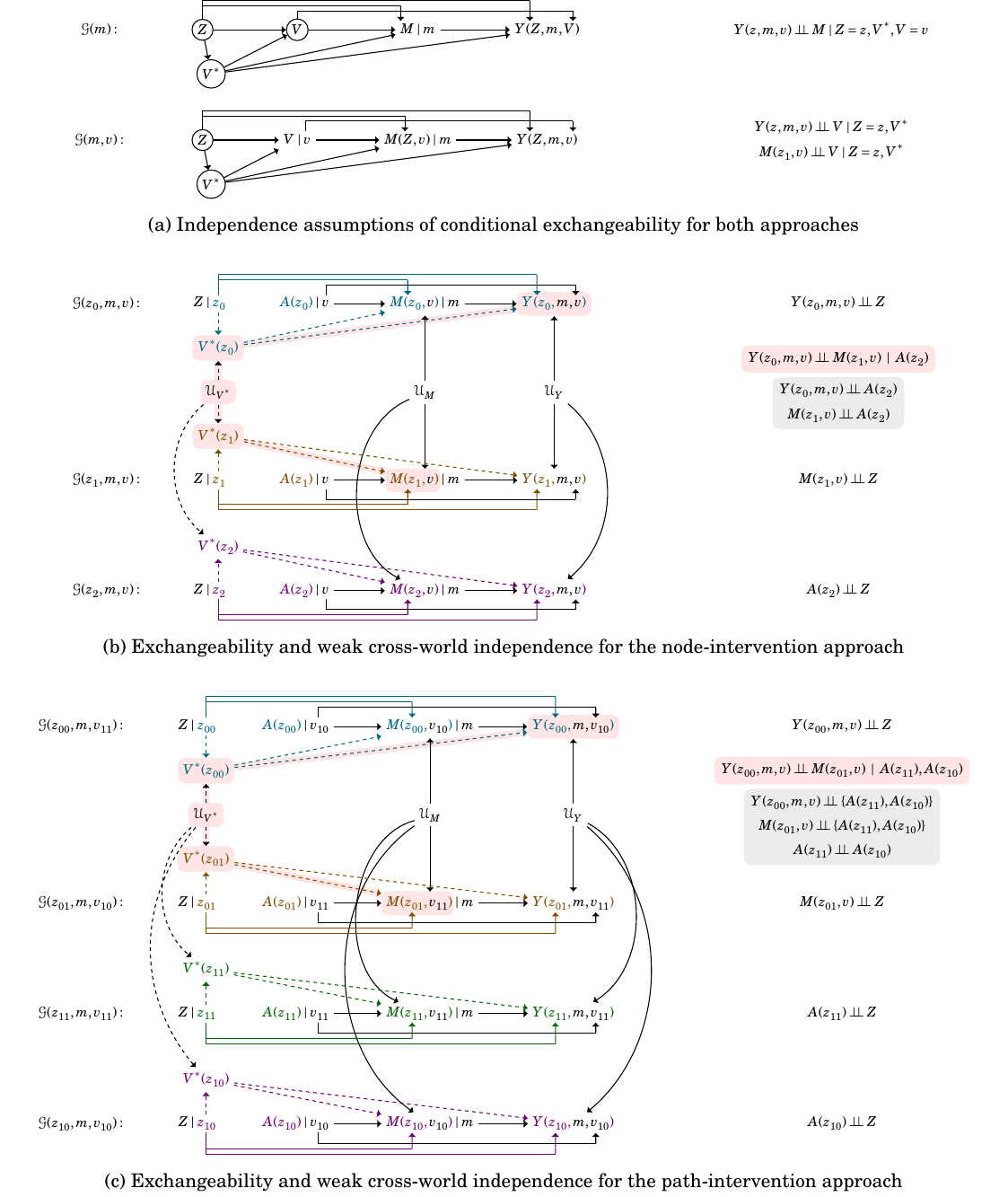}
    \caption{\textbf{Recanting twin effects.}
    The independence assumptions in (a--c) without shading represent classical exchangeability.
    The gray shading in (b--c) indicates counterfactual independence that follows from the definition of the randomized draws $A(\cdot)$ of $V(\cdot)$.
    The red shading in (b--c) indicates weak cross-world independence, corresponding to the absence of $\mathcal{U}_{V^*}$.}
    \label{fig:RTE}
\end{figure}

In accordance with their assumptions, we replace $M_2$ with $V$ in Figure~\ref{fig:RTE}, yielding a model with one mediator $M$ and one mediated confounder $V$ (along with an additional confounder $V^*$). 
The counterfactual $V(\cdot)$ is replaced by a randomized draw $A(\cdot)$ that follows the corresponding distribution, \(A(z)\mid Z=z \;\overset{d}{=}\; V(z)\mid Z=z,\) and is independent of the other counterfactual variables.
Consequently, cross-world independence assumptions involving $V(\cdot)$
are no longer required, leaving only those involving $M(\cdot)$.
In other words, the required independence assumptions are identical to those in a one-mediator model, and we can identify the distribution of $Y\big\{z_0,M\big[z_1,A(z_2)\big],A(z_2)\big\}$ using the node-intervention approach:

\eqev{
 P\big(Y\big\{z_0,M\big[z_1,A(z_2)\big],A(z_2)\big\}=y\big)  
  =&\sum_{m}\sum_{v} P\big(Y\big(z_0,m,v\big)=y\big|M\big(z_1,v\big)=m,A(z_2)=v\big) P\big(M(z_1,v)=m\big|A(z_2)=v\big) P\big(A(z_2)=v\big)\\
  \eqbb{8mm}{\text{cross-world}}{Y(z_0,m,v)\indp M(\smash{z_1,v})\gvn A(z_2)} &\sum_{m}\sum_{v} P\big(Y\big(z_0,m,v\big)=y\big|A(z_2)=v\big) P\big(M(z_1,v)=m\big|A(z_2)=v\big) P\big(A(z_2)=v\big)\\
  \eqbb{11mm}{\text{random draw}}{Y(z_0,m,v)\indp A(\smash{z_2});\;M(\smash{z_1,v})\indp A(\smash{z_2})} &\sum_{m}\sum_{v} P\big(Y\big(z_0,m,v\big)=y\big) P\big(M(z_1,v)=m\big) P\big(A(z_2)=v\big)\\
  \eqtt{$A(z)\mid Z=z \;\overset{d}{=}\; V(z)\mid Z=z$ \& exchangeability}
  &\sum_{m}\sum_{v} P\big(Y\big(z_0,m,v\big)=y\big|z_0,M=m,V=v\big) P\big(M(z_1,v)=m\big|z_1,V=v\big) P\big(\tred{V(z_2)}=v\big|z_2\big)\\  
  \eqtt{consistency}&\sum_{m}\sum_{v}P\big(Y=y\big|z_0,M=m,V=v\big) P\big(M=m\big|z_1,V=v\big) P\big(V=v\big|z_2\big).
\label{eq:RTn}}

Similarly, the distribution of $Y\big\{z_{00},M\big[z_{01},A(z_{11})\big],A(z_{10})\big\}$ can be identified under the path-intervention approach:

\eqev{
   P\big(Y\big\{z_{00},M\big[z_{01},A(z_{11})\big],A(z_{10})\big\}=y\big)
  =&\sum_{m}\sum_{v_{10}}P\big(Y\big(z_{00},M\big[z_{01},A(z_{11})\big],v_{10}\big)=y,M\big[z_{01},A(z_{11})\big]=m\big|A(z_{10})=v_{10}\big)P\big(A(z_{10})=v_{10}\big)\\
  =&\sum_{m}\sum_{v_{10}}\sum_{v_{11}}P\big(Y(z_{00},m,v_{10})=y\big|\tred{M(z_{01},v_{11})=m},A(z_{11})=v_{11},A(z_{10})=v_{10}\big)\\[-4mm]
  &\qqq\tred{P\big(M(z_{01},v_{11})=m\big|A(z_{10})=v_{10},A(z_{11})=v_{11}\big)}\\[-2mm]
  &\qqq P\big(A(z_{11})=v_{11}\big|A(z_{10})=v_{10}\big) P\big(A(z_{10})=v_{10}\big)\\  
  \eqbb{16mm}{\text{cross-world}}{\tred{Y(z_{00},m,v_{10})\indp M(\smash{z_{01},v_{11}})\gvn A(z_{11}),A(z_{10})}} &\sum_{m}\sum_{v_{10}}\sum_{v_{11}}\tred{P\big(Y(z_{00},m,v_{10})=y\big|A(z_{11})=v_{11},A(z_{10})=v_{10}\big)}\\[-4mm]
  &\qqq P\big(M(z_{01},v_{11})=m\big|A(z_{10})=v_{10},A(z_{11})=v_{11}\big)\\[-2mm]
  &\qqq P\big(A(z_{11})=v_{11}\big|A(z_{10})=v_{10}\big) P\big(A(z_{10})=v_{10}\big)\\   
  \eqbb{16mm}{\text{random draw}}{\substack{\tred{Y(z_{00},m,v_{10})\indp \{A(\smash{z_{11}}),A(\smash{z_{10}})\} };\\[.5mm]\tred{M(\smash{z_{01},v_{11}})\indp \{A(\smash{z_{11}}),A(\smash{z_{10}})\}};\;A(\smash{z_{11}})\indp A(\smash{z_{10}})}} &\sum_{m}\sum_{v_{10}}\sum_{v_{11}}\tred{P\big(Y(z_{00},m,v_{10})=y\big)}\tred{P\big(M(z_{01},v_{11})=m\big)}\tred{P\big(A(z_{11})=v_{11}\big)} P\big(A(z_{10})=v_{10}\big)\\[1mm]
  \eqtt{$A(z)\mid Z=z \;\overset{d}{=}\; V(z)\mid Z=z$ \& exchangeability} &\sum_{m}\sum_{v_{10}}\sum_{v_{11}}P\big(Y(z_{00},m,v_{10})=y\big|z_{00},M=m,V=v_{10}\big) P\big(M(z_{01},v_{11})=m\big|z_{01},V=v_{11}\big)\\[-4mm]
  &\qqq P\big(\tred{V(z_{11})=v_{11}}\big|z_{11}\big)P\big(\tred{V(z_{10})=v_{10}}|z_{10}\big)\\
  \eqtt{consistency}&\sum_{m}\sum_{v_{10}}\sum_{v_{11}}P\big(Y=y\big|{z_{00},M=m,V=v_{10}}\big) P\big(M=m\big|{z_{01},V=v_{11}}\big) P\big(V=v_{11}\big|z_{11}\big) P\big(V=v_{10}\big|z_{10}\big).
\label{eq:RTp}}

The recanting twins effects in Vo et al.~\cite{vo2026recanting} are defined by $Y\big\{z_0,M\big[z_1,V(z_2)\big],V(z_2)\big\}$, $Y\big\{z_{00},M\big[z_{01},\allowbreak V(z_{11})\big],A(z_{10})\big\}$, and $Y\big\{z_{00},M\big[z_{01},A(z_{11})\big],V(z_{10})\big\}$. 
The first can be identified by Equation~\eqref{eq:WC}, while the latter two can be identified using a similar strategy in Equation~\eqref{eq:RTp}. 
By replacing $V(\cdot)$ with $A(\cdot)$, this approach removes the need for strong cross-world independence assumptions involving $V(\cdot)$, while preserving the classical definition of the natural indirect effect through $M(\cdot)$. 
Notice that, in multimediator settings, only $M_1$ can retain its original counterfactual definition. 
From the perspective of separable effects, replacing $V(\cdot)$ with the random draw $A(\cdot)$ is essentially equivalent to rendering the $V$-specific component interventionally manipulable; the random draw breaks the dependence carried by the original counterfactual $V(\cdot)$ (i.e., \textit{interventional component manipulability} for $V(\cdot)$), thereby removing the corresponding cross-world restriction, as illustrated in Figure~\ref{fig:SEg}(d). 
Therefore, recanting twins effects can be naturally accommodated within the separable effects framework, as derived in Equation~\eqref{eq:RTs} using DCCs in Figure~\ref{fig:SE4}(a) (replace $M_2$ with $V$, similar to Equation~\eqref{eq:SCs}).

\section{Cancer immunotherapy example}~\label{sec:example}

\begin{figure}
\centering\includegraphics[width=\textwidth]{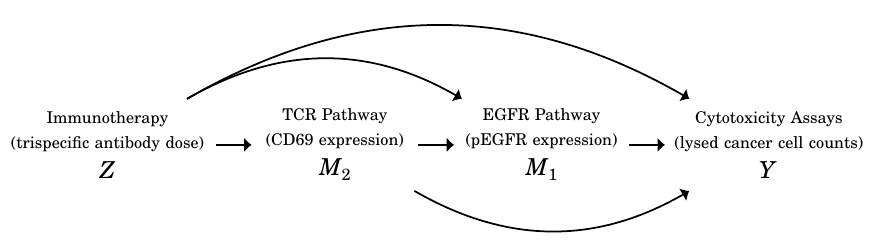}
    \caption{\textbf{Cancer immunotherapy example.}
    Illustrative causal structure relating the dose of a trispecific antibody ($Z$) to TCR pathway activation ($M_2$), EGFR pathway activation ($M_1$), and cancer cell lysis ($Y$). The example is used to compare the assumptions and interpretations of PSEs under the classical, interventional, and separable-effects frameworks.}
    \label{fig:demo}
\end{figure}

We use the cancer immunotherapy example~\cite{tapia2022trispecific} in Figure~\ref{fig:demo} to illustrate the causal assumptions under the three causal frameworks. 
Here, the exposure $Z$ denotes the dose of a trispecific antibody, $M_2$ the activation of the TCR pathway, $M_1$ the activation of the EGFR pathway, and $Y$ the level of cancer cell lysis.
Under the classical framework, weak cross-world independence assumes the absence of an exposure-induced response, such as NK cell activation, that simultaneously affects the EGFR pathway and cytotoxicity. 
Strong cross-world independence further assumes no residual source of dependence between $M_2(z)$ and $M_2(z')$; in this example, TCR pathway activation is therefore fully determined by the treatment antibody and baseline factors without shared residual variation across intervention levels.

The interventional approach replaces counterfactual mediators with randomized draws, thereby changing the interpretation from the natural biological behavior of the pathways to stochastic mediator assignments.
For example, rather than preserving an individual's TCR pathway activation under a particular antibody dose, the interventional approach considers a TCR activation level randomly drawn from the distribution induced by that dose and assigns it independently of the individual's natural TCR response.
The resulting PSE therefore describes the effect transmitted through a stochastic intervention on TCR pathway activation, rather than through the individual's naturally occurring TCR response under that antibody dose.

Under the separable-effects framework, the trispecific antibody can be decomposed into three manipulable components---anti-CD3, anti-EGFR, and anti-EpCAM---each targeting distinct biological mechanisms involving the TCR pathway, EGFR pathway, and cytotoxicity, respectively. 
Component consistency ($Z_0=Z_1=Z_2=Z$) represents administration of the trispecific antibody as the joint intervention of all three components, whereas component manipulability ($Z_0$, $Z_1$, $Z_2$ are mutually independent) allows hypothetical interventions assigning these components separately, such as administering anti-CD3 while withholding anti-EGFR and anti-EpCAM, or any other combination of component assignments.
Thus, separable effects provide a mechanistic interpretation when the exposure admits a meaningful component decomposition, but rely on prior knowledge that supports the proposed component-specific causal structure.

\section{{Comparisons of the three causal semantics}}\label{sec:C}

{In this paper, we compared three causal semantics for identifying PSEs, whose identification conditions are summarized in Table~\ref{tbl:ac}; positivity conditions are discussed in Appendix~D. When no mediated confounder is present, the classical, interventional, and separable-effects frameworks lead to the same identification formulas, despite defining the corresponding effects through different intervention semantics. This equivalence is illustrated by Equations~\eqref{eq:WC}, \eqref{eq:WCi}, and~\eqref{eq:WCs} for the node-intervention approach, and by Equations~\eqref{eq:SC}, \eqref{eq:SCi}, and~\eqref{eq:SCs} for the path-intervention approach. Thus, equality of identification formulas does not imply equality of the underlying causal assumptions or interpretations.
When a mediated confounder is present, however, the identification conditions diverge. 
If the mediated confounder is unmeasured, its contribution to the pathway-specific decomposition, as reflected by the difference between Equations~\eqref{eq:WC} and~\eqref{eq:WCv}, cannot be directly assessed. Consequently, the originally defined PSEs under the classical and separable-effects frameworks may be contaminated by pathways involving the mediated confounder, whereas the interventional PSEs remain unbiased through randomized mediator interventions.}

{Observing the mediated confounder does not by itself resolve this problem. Under the node-intervention approach, the classical framework requires the absence of the cross-world dependence induced by \(\UV\), as represented by the red-shaded relations in Figure~\ref{fig:CA}(b); under this condition, Equation~\eqref{eq:WCv} identifies the originally defined PSE by weighting over the mediated confounder. Under separable effects, the same \(\UV\) violates the DCC in Figure~\ref{fig:SE3}(c). Further decomposing the mediated confounder into manipulable components, as represented in Figure~\ref{fig:SE3}(a), provides an alternative structural restriction and yields the same identification formula in Equation~\eqref{eq:WCsvv}. Alternatively, when the mediated confounder admits the factorization represented in Figure~\ref{fig:SE3}(b), Equation~\eqref{eq:WCsv} identifies a different separable-effects estimand, for which there is no direct counterpart under the classical framework. 
Under the path-intervention approach, the finer decomposition additionally requires the absence of the cross-world dependence induced by \(U_{M_2}\) in the classical framework, corresponding to the strong cross-world independence represented by the dark-red-shaded relations in Figure~\ref{fig:CA}(c). Similarly, under separable effects, the presence of \(\mathcal{U}_{M_2}\) would violate the DCC represented in Figure~\ref{fig:SE4}(a). Further decomposing \(M_2\) into manipulable components, as shown in Figure~\ref{fig:SE4}(b), provides an alternative structural restriction.
Thus, the cross-world dependence represented explicitly by shared counterfactual causes in the classical framework reappears as violations of the corresponding DCCs under separable effects.}

More broadly, the three frameworks can be viewed as handling cross-world dependence in different ways. The classical framework retains natural counterfactuals and therefore states cross-world independence explicitly, as illustrated in Figure~\ref{fig:CA}. The interventional framework removes the corresponding requirement by replacing counterfactual mediators with independent random draws \(W(\cdot)\), as illustrated in Figure~\ref{fig:IA}, at the cost of changing the causal estimand. 
Separable effects instead reformulate cross-world dependence through structural restrictions involving DCCs, component manipulability, and component consistency, as illustrated in Figures~\ref{fig:SE3}, \ref{fig:SE3s}, and~\ref{fig:SE4}. 
The recanting-twins approach lies between these strategies: it retains the natural mediator counterfactuals but replaces the mediated-confounder counterfactuals with randomized draws \(A(\cdot)\), as illustrated in Figure~\ref{fig:RTE}, thereby removing the corresponding cross-world restrictions while preserving the remaining natural counterfactual structure. Consequently, cross-world independence is not simply present in one framework and absent in another; rather, it is assumed directly, avoided through randomization, or reformulated as structural restrictions, depending on the causal semantics and the level of pathway decomposition.

These different treatments of cross-world dependence have consequences not only for identification, but also for the causal interpretation of the resulting PSEs.
Suppose our interest lies in the pathway $Z \rightarrow M_1 \rightarrow Y$ in Figure~\ref{fig:demo}, representing the effect of immunotherapy through inhibition of the EGFR pathway and subsequent cytolysis. Under the classical assumptions, the corresponding PSE compares $Y\big\{z,M_2(z),M_1[z',M_2(z)]\big\}$ with $Y\big\{z,M_2(z),M_1[z,M_2(z)]\big\}$ and is interpreted as the change in cytolysis attributable to setting the EGFR pathway to the level it would attain under high-dose antibody, while holding the remaining pathways at their low-dose levels. This interpretation relies on cross-world independence assumptions, such as $M_2(z) \indp M_1(z',m_2)$. Under the interventional approach, these cross-world assumptions are avoided, but the estimand instead reflects stochastic mediator assignments, e.g., $Y\big\{z,M_2^*(z),M_1^*[z',M_2^*(z)]\big\}$, and therefore departs from the natural joint behavior of the underlying biological mechanism. 
Under separable effects, the trispecific antibody is decomposed into manipulable components---anti-CD3, anti-EGFR, and anti-EpCAM---so the same pathway-specific contrast is interpreted as the effect induced by the anti-EGFR component. Thus, although the frameworks may target the same pathway, they differ in how the intervention is conceptualized and what mechanistic interpretation the resulting effect can support.

\section{Discussion}\label{sec:D}

In this paper, we compared three causal frameworks for identifying path-specific effects in multimediator settings. Although these frameworks may lead to the same identification formula, they rely on different causal assumptions and intervention semantics. The classical framework imposes cross-world independence assumptions directly, whereas the interventional and separable-effects frameworks replace them with randomized interventions or component-specific structural restrictions. Thus, cross-world assumptions required in the classical framework may correspond to equally strong structural restrictions in the interventional or separable-effects frameworks. This comparison also highlights that the same observed-data functional can represent different causal questions.

These three approaches also echo broader themes in the philosophy of causation. Classical effects align naturally with counterfactual accounts of causation, in which causal relations are understood through contrasts between alternative counterfactual states~\cite{lewis1973causation,pearl2009causality}.
The interventional approach resembles a transitive view of causal transmission, combining the effect of exposure on the mediator distribution with the effect of intervening on the mediator to characterize an effect operating through that mediator; importantly, causation is not transitive in general~\cite{hesslow1981transitivity,pearl2009causality}.
Separable effects instead resemble the principle of modularity or autonomy of causal mechanisms, under which components of a causal system can be intervened upon separately while other mechanisms remain invariant~\cite{woodward2005making}.
These connections are not formal equivalences, but illustrate how the three frameworks reflect different conceptions of causal transmission through a pathway.

Different scientific questions motivate different choices of causal assumptions, leading to different definitions and estimands of PSEs and, consequently, different mechanistic interpretations of the resulting estimates~\cite{lee2021guideline}. 
This is reflected in applications across domains: in algorithmic fairness, PSEs characterize discrimination as effects transmitted through impermissible pathways and guide models that remove these effects while preserving admissible ones~\cite{nabi2018fair,chiappa2019path}. 
In biomedical settings, they characterize clinically meaningful mechanisms, including adherence-mediated treatment effects~\cite{miles2020semiparametric}, hormone-related live birth pathway~\cite{lai2020mediation}, and personalized treatment policies~\cite{shpitser2017personalizing}. 
These applications highlight that mechanistic interpretation fundamentally depends on identification: without identifying the relevant PSE under the assumed model, conclusions about fairness, biological mechanisms, or personalized interventions are not empirically justified.


In summary, we show that the same identification formula can arise under different causal assumptions and intervention semantics. Thus, the same formula does not necessarily imply the same causal interpretation. In addition, cross-world assumptions required in the classical framework may be replaced by equally strong restrictions in the interventional or separable-effects frameworks, rather than being completely removed. A finer decomposition of causal pathways also generally requires stronger causal assumptions or structural restrictions. Overall, these comparisons show that path-specific effects are not determined by identification formulas alone. Their interpretation depends on how the counterfactual variables or interventions are defined, and therefore the estimand, identification assumptions, and scientific mechanism of interest should be considered together.


\begin{landscape}
\begin{table}
    \caption{\textbf{Comparison of identification conditions for PSEs.}
    We summarize the number of identifiable PSEs under different settings of $V$. 
    For the classical assumptions, conditions (C1)–(C3) correspond to whether cross-world independence holds for $V$ or whether $V$ is treated as a mediator. 
    For the separable effects framework, conditions (S1)–(S3) indicate whether $V$ can be factorized or decomposed into manipulable components. 
    We also highlight key advantages and limitations of each framework.}
    \label{tbl:ac}
    \centering\includegraphics[width=1.3\textheight]{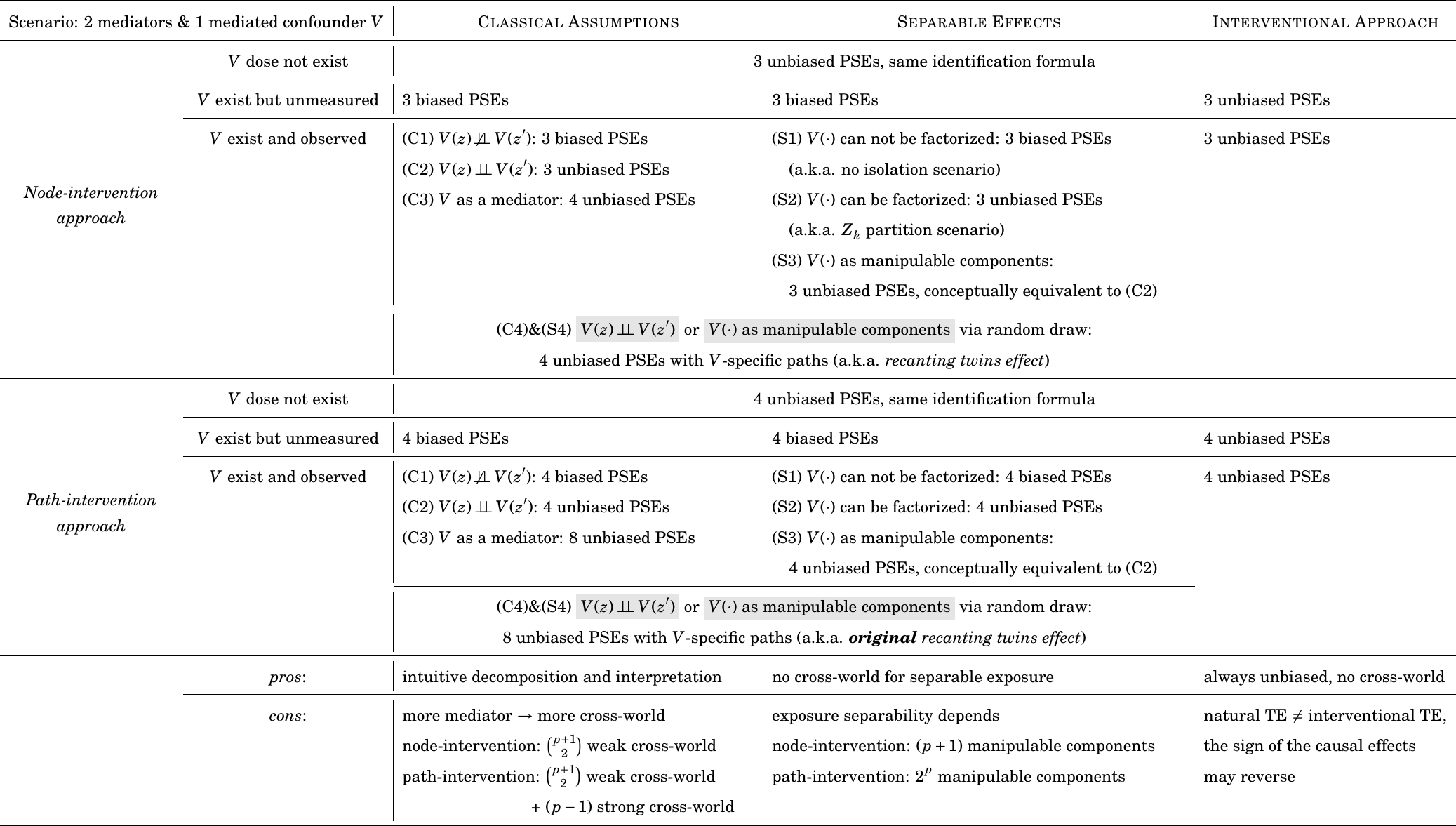}
\end{table}    
\end{landscape}

\appendix
\clearpage

\renewcommand{\thetable}{S\arabic{table}}  
\renewcommand{\thefigure}{S\arabic{figure}}
\renewcommand{\theequation}{S\arabic{equation}}
\setcounter{figure}{0}
\setcounter{table}{0}
\setcounter{equation}{0}
\setcounter{page}{1}
\renewcommand{\theHfigure}{S\arabic{figure}}
\renewcommand{\theHtable}{S\arabic{table}}
\renewcommand{\theHequation}{S\arabic{equation}}



\section{Justification of $d$-Separation in the Augmented DAG}
\begin{figure}[h]
    \centering
    \includegraphics[width=.85\linewidth]{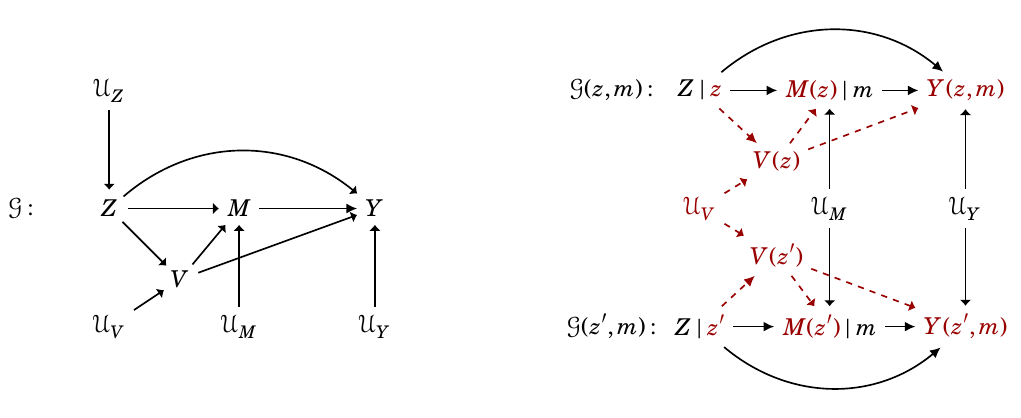}
    \caption{\textbf{Augmented DAG with Cross-World Confounders}. 
    The original DAG $\mathcal{G}$ (left) induces two SWIGs under interventions $(z,m)$ and $(z',m)$ (right). Shared exogenous variables generate cross-world confounding between the counterfactual variables. By explicitly representing the exogenous variables, the augmented graph forms a DAG whose induced joint distribution factorizes according to the graph, allowing conditional independences to be characterized by $d$-separation.}
    \label{fig:placeholder}
\end{figure}

In this study, we illustrate the dependency structure of a DAG constructed from SWIGs with explicit exogenous variables representing cross-world confounding. 
Each intervention-specific SWIG inherits the topological ordering of the original structural causal models (SCM). The shared exogenous variables are root nodes and only have directed edges into their corresponding counterfactual copies across intervention settings. Since no directed edges are introduced from one intervention-specific world to another, combining the SWIGs through shared exogenous variables cannot create a directed cycle. The resulting augmented graph is therefore a DAG.

Here we show that the joint distribution induced by the SCM factorizes according to the augmented DAG, thereby justifying the use of \(d\)-separation in this graph. 
Consider the single mediator model with a mediated confounder $V$ illustrated in the above figure, we have the following structural equations:
\begin{align*}
Z &= f_Z(U_Z), & V &= f_V(Z, U_V), & M &= f_M(Z, V, U_M), & Y &= f_Y(Z, V, M, U_Y),
\end{align*}
where $\mathbb{U}=\{U_Z, U_V, U_M, U_Y\}$ are mutually independent exogenous variables, so this causal model is \textit{Markovian}~\cite{pearl2009causality}. Under the \textit{NPSEM counterfactual existence assumption}~\cite{richardson2013single}, we have the structural equations for $\mathcal{G}(z,m)$ and $\mathcal{G}(z',m)$:
\begin{align*}
V(z) &= f_V(z, U_V), & V(z') &= f_V(z', U_V), \\
M(z) &= f_M(z, V(z), U_M), & M(z') &= f_M(z', V(z'), U_M), \\
Y(z,m) &= f_Y(z, V(z), m, U_Y), & Y(z',m) &= f_Y(z', V(z'), m, U_Y),
\end{align*}
where $U_V, U_M, U_Y$ naturally become cross-world confounders.
The joint distribution of the original DAG $\mathcal{G}$ factorizes naturally after marginalizing $\mathbb{U}$:
\begin{align*}
P(Z,V,M,Y)&=P(Z) P(V\mid Z) P(M\mid Z,V) P(Y\mid Z,V,M),
\end{align*}
and because of \textit{modularity condition}, the factorization associated with $\mathcal{G}(z,m)$ also holds: 
\begin{align*}
P(Z,V(z),M(z),Y(z,m))&=P(Z) P(V(z)\mid Z) P(M(z)\mid Z,V(z)) P(Y(z,m)\mid Z,M(z),V(z)),\\
    &=P(Z) P(V(z)) P(M(z)\mid V(z)) P(Y(z,m)\mid V(z)).
\end{align*}
To consider the roles of $\mathbb{U}$ between $\mathcal{G}(z,m)$ and  $\mathcal{G}(z',m)$, the joint distribution becomes
\begin{align*}
&P(U_Z,U_V,U_M,U_Y,Z,V(z),M(z),Y(z,m),V(z'),M(z'),Y(z',m)) \\
&= P(U_Z,U_V,U_M,U_Y) \times P(Z\gvn U_Z,U_V,U_M,U_Y) \\
&\quad \times P(V(z)\gvn U_Z,U_V,U_M,U_Y,Z) \\
&\quad \times P(M(z)\gvn U_Z,U_V,U_M,U_Y,Z,V(z)) \\
&\quad \times P(Y(z,m)\gvn U_Z,U_V,U_M,U_Y,Z,V(z),M(z)) \\
&\quad \times P(V(z')\gvn U_Z,U_V,U_M,U_Y,Z,V(z),M(z),Y(z,m)) \\
&\quad \times P(M(z')\gvn U_Z,U_V,U_M,U_Y,Z,V(z),M(z),Y(z,m),V(z')) \\
&\quad \times P(Y(z',m)\gvn U_Z,U_V,U_M,U_Y,Z,V(z),M(z),Y(z,m),V(z'),M(z')) \\
&= P(U_Z)\times P(U_V)\times P(U_M)\times P(U_Y)\times P(Z\gvn U_Z) \\
&\quad \times P(V(z)\gvn U_V) \times P(M(z)\gvn V(z),U_M) \times P(Y(z,m)\gvn V(z),U_Y) \\
&\quad \times P(V(z')\gvn U_V) \times P(M(z')\gvn V(z'),U_M) \times P(Y(z',m)\gvn V(z'),U_Y).
\end{align*}
Because each counterfactual variable is a deterministic function of the arguments appearing in its structural equation, the joint distribution therefore admits the factorization displayed above, which corresponds to the augmented DAG with explicit exogenous variables. Consequently, by the global Markov property, conditional independences implied by the augmented DAG can be read using $d$-separation.

The same argument applies to the two-mediator setting considered in the main text. 
Suppose that $M_2$ precedes $M_1$, with structural equations
\begin{align*}
Z &= f_Z(U_Z), &
V &= f_V(Z,U_V), \\
M_2 &= f_{M_2}(Z,V,U_{M_2}), &
M_1 &= f_{M_1}(Z,V,M_2,U_{M_1}), &
Y &= f_Y(Z,V,M_1,M_2,U_Y),
\end{align*}
where
$\mathbb{U}=\{U_Z,U_V,U_{M_2},U_{M_1},U_Y\}$
are mutually independent exogenous variables.
Under the node-intervention approach, there are $p+1$ intervention settings indexed by $i\in\mathbb{I}=\{0,\ldots,p\}$, with the corresponding intervention written as $z_i$. 
For the two-mediator setting considered here, the three intervention-specific SWIGs in Figure~\ref{fig:CA}(b) are
\[\mathcal{G}(z_i,m_1,m_2),\qquad i\in\mathbb{I}=\{0,1,2\}.\]
Each SWIG contains the intervention-specific random variables \(V(z_i),M_2(z_i),M_1(z_i,m_2),Y(z_i,m_1,m_2).\)
Under the NPSEM counterfactual existence assumption, their structural equations are
\begin{align*}
V(z_i) &= f_V(z_i,U_V),\\
M_2(z_i) &= f_{M_2}\big(z_i,V(z_i),U_{M_2}\big),\\
M_1(z_i,m_2) &= f_{M_1}\big(z_i,V(z_i),m_2,U_{M_1}\big),\\
Y(z_i,m_1,m_2) &= f_Y\big(z_i,V(z_i),m_1,m_2,U_Y\big).
\end{align*}
By modularity, the distribution associated with each
$\mathcal{G}(z_i,m_1,m_2)$ factorizes as
\begin{align*}
&P\big(Z,V(z_i),M_2(z_i),M_1(z_i,m_2),Y(z_i,m_1,m_2)\big)\\
&\qquad =
P(Z)P\big(V(z_i)\big)
P\big(M_2(z_i)\mid V(z_i)\big)
P\big(M_1(z_i,m_2)\mid V(z_i)\big)
P\big(Y(z_i,m_1,m_2)\mid V(z_i)\big).
\end{align*}
When the three SWIGs are jointly represented with their shared exogenous variables, the joint distribution factorizes as
\begin{align*}
&P(U_Z)P(U_V)P(U_{M_2})P(U_{M_1})P(U_Y)P(Z\gvn U_Z)\\
&\quad\times
\prod_{i=0}^{2}
\Big[
P\big(V(z_i)\mid U_V\big)
P\big(M_2(z_i)\mid V(z_i),U_{M_2}\big)
P\big(M_1(z_i,m_2)\mid V(z_i),U_{M_1}\big)
P\big(Y(z_i,m_1,m_2)\mid V(z_i),U_Y\big)
\Big].
\end{align*}
Thus, the augmented graph for the node-intervention approach again admits a DAG factorization, with $U_V$, $U_{M_2}$, $U_{M_1}$, and $U_Y$ explicitly representing dependencies across intervention-specific SWIGs.

Under the path-intervention approach, there are $2^p$ intervention settings indexed by $i\in\mathbb{I}=\{0,\ldots,2^p-1\}$, with the corresponding intervention written as $z_{\bn{i}}$, where $\bn{i}$ denotes the binary expression of $i$.
For the two-mediator setting, $\mathbb{I}=\{0,1,2,3\}$, with \(z_{\bn{0}}=z_{00},z_{\bn{1}}=z_{01},z_{\bn{2}}=z_{10},z_{\bn{3}}=z_{11},\)
and the four intervention-specific SWIGs in Figure~\ref{fig:CA}(c) are
\[
\mathcal{G}(z_{00},m_{01},m_{10}),\quad
\mathcal{G}(z_{01},m_{01},m_{11}),\quad
\mathcal{G}(z_{10},m_{01},m_{10}),\quad
\mathcal{G}(z_{11},m_{01},m_{11}).
\]
Under the NPSEM counterfactual existence assumption, their counterfactual
variables follow the same structural equations as the original SCM.
For all $i\in\mathbb{I}$,
\begin{align*}
V(z_{\bn{i}})
&=f_V(z_{\bn{i}},U_V),\\
M_2(z_{\bn{i}})
&=f_{M_2}\big(z_{\bn{i}},V(z_{\bn{i}}),U_{M_2}\big).
\end{align*}
For the paths that do not pass through $M_1$, corresponding to
$i\in\{0,2\}$, $M_2$ is intervened as $m_{10}$, and
\begin{align*}
M_1(z_{\bn{i}},m_{10})
&=f_{M_1}\big(z_{\bn{i}},V(z_{\bn{i}}),m_{10},U_{M_1}\big),\\
Y(z_{\bn{i}},m_{01},m_{10})
&=f_Y\big(z_{\bn{i}},V(z_{\bn{i}}),m_{01},m_{10},U_Y\big).
\end{align*}
For the paths that pass through $M_1$, corresponding to
$i\in\{1,3\}$, $M_2$ is intervened as $m_{11}$, and
\begin{align*}
M_1(z_{\bn{i}},m_{11})
&=f_{M_1}\big(z_{\bn{i}},V(z_{\bn{i}}),m_{11},U_{M_1}\big),\\
Y(z_{\bn{i}},m_{01},m_{11})
&=f_Y\big(z_{\bn{i}},V(z_{\bn{i}}),m_{01},m_{11},U_Y\big).
\end{align*}
By the modularity condition, the factorization associated with each
intervention-specific SWIG also holds. For example, in $\mathcal{G}(z_{01},m_{01},m_{11})$, 
\begin{align*}
&P\big\{Z,V(z_{01}),M_2(z_{01}),M_1(z_{01},m_{11}),
Y(z_{01},m_{01},m_{11})\big\}\\
&\quad =
P(Z)P\big\{V(z_{01})\big\}
P\big\{M_2(z_{01})\mid V(z_{01})\big\}
P\big\{M_1(z_{01},m_{11})\mid V(z_{01})\big\}
P\big\{Y(z_{01},m_{01},m_{11})\mid V(z_{01})\big\}.
\end{align*}
The same factorization applies to
$\mathcal{G}(z_{00},m_{01},m_{10})$,
$\mathcal{G}(z_{10},m_{01},m_{10})$, and
$\mathcal{G}(z_{11},m_{01},m_{11})$ under their respective intervention
assignments.

To consider the roles of $\mathbb{U}$ across the four intervention-specific
SWIGs, consider the joint distribution
\begin{align*}
&P\Big(
U_Z,U_V,U_{M_2},U_{M_1},U_Y,Z,\\
&\qquad
V(z_{00}),M_2(z_{00}),M_1(z_{00},m_{10}),
Y(z_{00},m_{01},m_{10}),\\
&\qquad
V(z_{01}),M_2(z_{01}),M_1(z_{01},m_{11}),
Y(z_{01},m_{01},m_{11}),\\
&\qquad
V(z_{10}),M_2(z_{10}),M_1(z_{10},m_{10}),
Y(z_{10},m_{01},m_{10}),\\
&\qquad
V(z_{11}),M_2(z_{11}),M_1(z_{11},m_{11}),
Y(z_{11},m_{01},m_{11})
\Big).
\end{align*}
Following the same chain-rule argument as in the single-mediator case,
the structural equations reduce this joint distribution to
\begin{align*}
& P(U_Z)\times P(U_V)\times P(U_{M_2})\times
P(U_{M_1})\times P(U_Y)\times P(Z\mid U_Z)\\
&\quad\times
P\big\{V(z_{00})\mid U_V\big\}
P\big\{M_2(z_{00})\mid V(z_{00}),U_{M_2}\big\}\\
&\quad\times
P\big\{M_1(z_{00},m_{10})\mid V(z_{00}),U_{M_1}\big\}
P\big\{Y(z_{00},m_{01},m_{10})\mid V(z_{00}),U_Y\big\}\\
&\quad\times
P\big\{V(z_{01})\mid U_V\big\}
P\big\{M_2(z_{01})\mid V(z_{01}),U_{M_2}\big\}\\
&\quad\times
P\big\{M_1(z_{01},m_{11})\mid V(z_{01}),U_{M_1}\big\}
P\big\{Y(z_{01},m_{01},m_{11})\mid V(z_{01}),U_Y\big\}\\
&\quad\times
P\big\{V(z_{10})\mid U_V\big\}
P\big\{M_2(z_{10})\mid V(z_{10}),U_{M_2}\big\}\\
&\quad\times
P\big\{M_1(z_{10},m_{10})\mid V(z_{10}),U_{M_1}\big\}
P\big\{Y(z_{10},m_{01},m_{10})\mid V(z_{10}),U_Y\big\}\\
&\quad\times
P\big\{V(z_{11})\mid U_V\big\}
P\big\{M_2(z_{11})\mid V(z_{11}),U_{M_2}\big\}\\
&\quad\times
P\big\{M_1(z_{11},m_{11})\mid V(z_{11}),U_{M_1}\big\}
P\big\{Y(z_{11},m_{01},m_{11})\mid V(z_{11}),U_Y\big\}.
\end{align*}
Because each counterfactual variable is a deterministic function of the arguments appearing in its structural equation, conditioning on those arguments renders it independent of the non-descendant variables.
Therefore, the joint distribution factorizes according to the augmented DAG in Figure~\ref{fig:CA}(c) and is Markov relative to this DAG.
Consequently, conditional independences implied by the augmented DAG can again be read using $d$-separation.

To generalize the above argument, for a counterfactual variable $X$, we partition its endogenous parents in an intervention-specific SWIG into two sets. Let $\mathbf{pa}_{do}(X)$ denote the parents whose values are fixed by the intervention, and let $\mathbf{pa}_{rv}(X)$ denote the parents that remain random variables in the intervention-specific SWIG. The corresponding counterfactual variable is written as
$X\{\mathbf{pa}_{do}(X)\}$ and satisfies
\[
X\{\mathbf{pa}_{do}(X)\}
=
f_X\big(
\mathbf{pa}_{do}(X),
\mathbf{pa}_{rv}(X),
U_X
\big).
\]
Because $\mathbf{pa}_{do}(X)$ is fixed under the intervention, the
corresponding factor in the joint distribution is
\[
P\left[
X\{\mathbf{pa}_{do}(X)\}
\,\middle|\,
\mathbf{pa}_{rv}(X),U_X
\right].
\]
Thus, $\mathbf{pa}_{do}(X)$ specifies the intervention under which the counterfactual variable is defined, whereas $\mathbf{pa}_{rv}(X)$, together with $U_X$, specifies its stochastic parent set in the augmented DAG.

The node-intervention construction extends to an arbitrary number $p$
of sequential mediators by induction.
Under the node-intervention approach with $p$ mediators, there are
$p+1$ intervention settings $\{z_i\}_{i\in\mathbb{I}}$, where
$\mathbb{I}=\{0,\ldots,p\}$.
The joint distribution of the exogenous variables and all
intervention-specific counterfactual variables factorizes as
\begin{align*}
& P(U_Z)P(U_V)P(U_Y)
  \prod_{j=1}^{p}P(U_{M_j})P(Z\gvn U_Z)\\
&\quad\times
\prod_{i\in\mathbb{I}}
\Bigg[
P\{V(z_i)\gvn U_V\}\times
\prod_{j=1}^{p}
P\left[
M_j\{\mathbf{pa}_{do,i}(M_j)\}
\,\middle|\,
\mathbf{pa}_{rv,i}(M_j),U_{M_j}
\right]\times
P\left[
Y\{\mathbf{pa}_{do,i}(Y)\}
\,\middle|\,
\mathbf{pa}_{rv,i}(Y),U_Y
\right]
\Bigg].
\end{align*}

Here, $\mathbf{pa}_{do,i}(X)$ and $\mathbf{pa}_{rv,i}(X)$ denote,
respectively, the parent values fixed by intervention and the parents
remaining random variables for $X$ in the $i$th intervention-specific
SWIG. Thus, these parent sets are indexed by the intervention-specific SWIG and are defined with respect to the augmented graph under consideration.
The one-mediator construction established above provides the base case. Suppose that this factorization holds for $p$ mediators. 
After adding a sequential mediator $M_{p+1}$, the resulting augmented graph contains $p+2$ intervention-specific SWIGs, indexed by $\mathbb{I}=\{0,\ldots,p+1\}$. 
The additional mediator introduces the independent exogenous variable $U_{M_{p+1}}$. In this new augmented graph, each intervention-specific counterfactual variable has the same parent-indexed local factorization form, with
$\mathbf{pa}_{do,i}(X)$ and $\mathbf{pa}_{rv,i}(X)$ defined according
to its corresponding intervention-specific SWIG. 
Consequently, the joint distribution for $p+1$ mediators factorizes as
\begin{align*}
& P(U_Z)P(U_V)P(U_Y)
  \prod_{j=1}^{p+1}P(U_{M_j})P(Z\gvn U_Z)\\
&\quad\times
\prod_{i\in\mathbb{I}}
\Bigg[
P\{V(z_i)\gvn U_V\}\times
\prod_{j=1}^{p+1}
P\left[
M_j\{\mathbf{pa}_{do,i}(M_j)\}
\,\middle|\,
\mathbf{pa}_{rv,i}(M_j),U_{M_j}
\right]\times
P\left[
Y\{\mathbf{pa}_{do,i}(Y)\}
\,\middle|\,
\mathbf{pa}_{rv,i}(Y),U_Y
\right]
\Bigg].
\end{align*}
Thus, the factorization is preserved from $p$ to $p+1$.
Since it holds for the base case, it follows by induction for any finite number of sequential mediators.
The resulting joint distribution is therefore Markov relative to the
corresponding augmented DAG, and its implied conditional independences can be read using $d$-separation.

The path-intervention construction can be extended to an arbitrary
number $p$ of sequential mediators in a similar manner.
Under the path-intervention approach with $p$ mediators, there are
$2^p$ intervention settings $\big\{z_{\bn{i}}\big\}_{i\in\mathbb{I}}$, where
$\mathbb{I}=\{0,\ldots,2^p-1\}$ and $\bn{i}$ denotes the binary
expression of $i$.
The joint distribution of the exogenous variables and all
intervention-specific counterfactual variables factorizes as
\begin{align*}
& P(U_Z)P(U_V)P(U_Y)
  \prod_{j=1}^{p}P(U_{M_j})P(Z\gvn U_Z)\\
&\quad\times
\prod_{i\in\mathbb{I}}
\Bigg[
P\{V(z_{\bn{i}})\gvn U_V\}\times
\prod_{j=1}^{p}
P\left[
M_j\{\mathbf{pa}_{do,i}(M_j)\}
\,\middle|\,
\mathbf{pa}_{rv,i}(M_j),U_{M_j}
\right]\times
P\left[
Y\{\mathbf{pa}_{do,i}(Y)\}
\,\middle|\,
\mathbf{pa}_{rv,i}(Y),U_Y
\right]
\Bigg].
\end{align*}
The one-mediator construction established above provides the base case.
Suppose that this factorization holds for $p$ mediators.
After adding a sequential mediator $M_{p+1}$, the resulting augmented graph contains $2^{p+1}$ intervention-specific SWIGs, indexed by $\mathbb{I}=\{0,\ldots,2^{p+1}-1\}$. 
The additional mediator introduces the independent exogenous variable $U_{M_{p+1}}$. In this new augmented graph, each intervention-specific counterfactual variable has the same parent-indexed local factorization form, with
$\mathbf{pa}_{do,i}(X)$ and $\mathbf{pa}_{rv,i}(X)$ defined according
to its corresponding intervention-specific SWIG. 
Consequently, the joint distribution for $p+1$ mediators factorizes as
\begin{align*}
& P(U_Z)P(U_V)P(U_Y)
  \prod_{j=1}^{p+1}P(U_{M_j})P(Z\gvn U_Z)\\
&\quad\times
\prod_{i\in\mathbb{I}}
\Bigg[
P\{V(z_{\bn{i}})\gvn U_V\}\times
\prod_{j=1}^{p+1}
P\left[
M_j\{\mathbf{pa}_{do,i}(M_j)\}
\,\middle|\,
\mathbf{pa}_{rv,i}(M_j),U_{M_j}
\right]\times
P\left[
Y\{\mathbf{pa}_{do,i}(Y)\}
\,\middle|\,
\mathbf{pa}_{rv,i}(Y),U_Y
\right]
\Bigg].
\end{align*}
Thus, the factorization is preserved from $p$ to $p+1$.
Since it holds for the base case, it follows by induction for any finite number of sequential mediators.
The resulting joint distribution is therefore Markov relative to the corresponding augmented DAG, and its implied conditional independences can be read using $d$-separation.

Finally, the above argument assumes that each intervention-specific SWIG is correctly constructed according to its intervention assignment and that variables representing the same counterfactual quantity across different intervention settings are consistently aligned in the augmented DAG. Under this construction, the augmented graph admits the stated factorization and the corresponding $d$-separation statements follow from the global Markov property. Readers should therefore avoid applying the graphical criterion to arbitrary collections of intervention-specific graphs without first establishing their joint structural representation. A related caution for parallel-world graph representations was discussed by Shpitser and Pearl~\cite{shpitser2008complete}, where duplicated nodes corresponding to the same random variable require explicit graph transformation before graphical separation can be interpreted.

\section{Identification Formula}

\subsection{Equation~\eqref{eq:WC}: classical assumptions (Figure~\ref{fig:CA}), node-intervention approach, and $V$ is absent}
\eqevsn{
        P\big(Y\big\{z_0,M_1\big[z_1,M_2(z_2)\big],M_2(z_2)\big\}=y\big)   
        =&\sum_{m_1}\sum_{m_2}P\big(Y\big(z_0,m_1,m_2\big)=y\big|M_1\big(z_1,m_2\big)=m_1,M_2(z_2)=m_2\big)\\[-4mm]
        &\quq P\big(M_1(z_1,m_2)=m_1\big|M_2(z_2)=m_2\big)P\big(M_2(z_2)=m_2\big)\\
        \eqbb{13mm}{\text{cross-world}}{\tred{Y(z_0,m_1,m_2)\indp \{M_1(\smash{z_1,m_2}),M_2(\smash{z_2})\}}} &\sum_{m_1}\sum_{m_2}P\big(Y\big(z_0,m_1,m_2\big)=y\big)\\[-4mm]
        &\quq P\big(M_1(z_1,m_2)=m_1\big|M_2(z_2)=m_2\big)P\big(M_2(z_2)=m_2\big)\\
        \eqbb{5mm}{\text{cross-world}}{M_1(\smash{z_1,m_2})\indp M_2(\smash{z_2})} &\sum_{m_1}\sum_{m_2}P\big(Y\big(z_0,m_1,m_2\big)=y\big)P\big(M_1(z_1,m_2)=m_1\big)P\big(M_2(z_2)=m_2\big)\\
        \eqbb{2mm}{\mathcal{G}(z_0,m_1,m_2);\;\mathcal{G}(z_1,m_1,m_2);\;\mathcal{G}(z_2,m_1,m_2)}{Y(z_0,m_1,m_2)\indp Z;\;M_1(z_1,m_2)\indp Z;\;M_2(z_2)\indp Z}
        &\sum_{m_1}\sum_{m_2}P\big(Y\big(z_0,m_1,m_2\big)=y\big|\tred{z_0}\big)\\[-4mm]
        &\quq P\big(M_1(z_1,m_2)=m_1\big|\tred{z_1}\big)P\big(M_2(z_2)=m_2\big|\tred{z_2}\big)\\
        \eqbb{15.5mm}{\mathcal{G}(m_1,m_2)}{Y(z,m_1,m_2)\indp M_2\gvn Z;\;M_1(z,m_2)\indp M_2\gvn Z}
        &\sum_{m_1}\sum_{m_2}P\big(Y\big(z_0,m_1,m_2\big)=y\big|z_0,\tred{M_2=m_2}\big)\\[-4mm]
        &\quq P\big(M_1(z_1,m_2)=m_1\big|z_1,\tred{M_2=m_2}\big)P\big(M_2(z_2)=m_2\big|z_2\big)\\
        \eqbb{9.5mm}{\mathcal{G}(m_1)}{Y(z,m_1,m_2)\indp M_1\gvn Z,M_2}
        &\sum_{m_1}\sum_{m_2}P\big(Y\big(z_0,m_1,m_2\big)=y\big|z_0,\tred{M_1=m_1},M_2=m_2\big)\\[-4mm]
        &\quq P\big(M_1(z_1,m_2)=m_1\big|z_1,M_2=m_2\big)P\big(M_2(z_2)=m_2\big|z_2\big)\\
        \eqtt{consistency}&\sum_{m_1}\sum_{m_2}P\big(Y=y\big|{z_0,M_1=m_1,M_2=m_2}\big)\\[-4mm]
        &\quq P\big(M_1=m_1\big|{z_1,M_2=m_2}\big)P\big(M_2=m_2\big|{z_2}\big).
}
\vfill

\subsection{Equation~\eqref{eq:WCv}: classical assumptions (Figure~\ref{fig:CA}), node-intervention approach, and $V$ is present}
\eqevsn{
        P\big(Y\big\{z_0,M_1\big[z_1,M_2(z_2)\big],M_2(z_2)\big\}=y\big)  
        =&\sum_{m_1}\sum_{m_2}\sum_{v_0}\sum_{v_1}\sum_{v_2} P\big(Y\big(z_0,m_1,m_2\big)=y\big|M_1\big(z_1,m_2\big)=m_1,M_2(z_2)=m_2,\\[-4mm]
        &\qqq\quq V(z_0)=v_0,V(z_1)=v_1,V(z_2)=v_2\big)\\[-2mm]
        &\quq P\big(M_1(z_1,m_2)=m_1\big|M_2(z_2)=m_2,V(z_0)=v_0,V(z_1)=v_1,V(z_2)=v_2\big)\\[-2mm]
        &\quq P\big(M_2(z_2)=m_2\big|V(z_0)=v_0,V(z_1)=v_1,V(z_2)=v_2\big)\\[-2mm]
        &\quq P\big(V(z_0)=v_0,V(z_1)=v_1,V(z_2)=v_2\big)\\
        \eqbb{23mm}{\text{cross-world}}{\substack{\tred{Y(z_0,m_1,m_2)\indp \{V(z_1),V(z_2)\}\gvn M_1(z_1,m_2),M_2(z_2),V(z_0);}\\[.5mm]\tred{M_1(z,m_2)\indp \{V(z_0),V(z_2)\}\gvn M_2(z_2),V(z_1);}\\[.5mm] \tred{M_2(z)\indp \{V(z_0),V(z_1)\}\gvn V(z_2);}\\[.5mm]\tred{V(z_0), V(z_1), V(z_2)\text{ are mutually independent}}}} &\sum_{m_1}\sum_{m_2}\sum_{v_0}\sum_{v_1}\sum_{v_2} P\big(Y\big(z_0,m_1,m_2\big)=y\big|M_1\big(z_1,m_2\big)=m_1,M_2(z_2)=m_2,V(z_0)=v_0\big)\\[-10mm]
        &\quq P\big(M_1(z_1,m_2)=m_1\big|M_2(z_2)=m_2,V(z_1)=v_1\big)\\[-2mm]
        &\quq P\big(M_2(z_2)=m_2\big|V(z_2)=v_2\big) P\big(V(z_0)=v_0\big)P\big(V(z_1)=v_1\big)P\big(V(z_2)=v_2\big)\\
        \eqbb{16mm}{\text{cross-world}}{\tred{Y(z_0,m_1,m_2)\indp \{M_1(\smash{z_1,m_2}),M_2(\smash{z_2})\}\gvn V(z_0)}} &\sum_{m_1}\sum_{m_2}\sum_{v_0}\sum_{v_1}\sum_{v_2} P\big(Y\big(z_0,m_1,m_2\big)=y\big|V(z_0)=v_0\big) P\big(V(z_0)=v_0\big)\\[-2mm]
        &\quq P\big(M_1(z_1,m_2)=m_1\big|M_2(z_2)=m_2,V(z_1)=v_1\big) P\big(V(z_1)=v_1\big)\\[-2mm]
        &\quq P\big(M_2(z_2)=m_2\big|V(z_2)=v_2\big) P\big(V(z_2)=v_2\big)\\
        \eqbb{8mm}{\text{cross-world}}{\tred{M_1(\smash{z_1,m_2})\indp M_2(\smash{z_2})\gvn V(z_1)}} &\sum_{m_1}\sum_{m_2}\sum_{v_0}\sum_{v_1}\sum_{v_2} P\big(Y\big(z_0,m_1,m_2\big)=y\big|V(z_0)=v_0\big)P\big(V(z_0)=v_0\big)\\[-2mm]
        &\quq P\big(M_1(z_1,m_2)=m_1\big|V(z_1)=v_1\big)P\big(V(z_1)=v_1\big)\\[-2mm]
        &\quq P\big(M_2(z_2)=m_2\big|V(z_2)=v_2\big) P\big(V(z_2)=v_2\big)\\
        \eqbb{3mm}{\mathcal{G}(z_0,m_1,m_2);\;\mathcal{G}(z_1,m_1,m_2);\;\mathcal{G}(z_2,m_1,m_2)}{\substack{\tred{Y(z_0,m_1,m_2)\indp Z\gvn V(z_0);\;M_1(z_1,m_2)\indp Z\gvn V(z_1)};\\[.5mm] \tred{M_2(z_2)\indp Z\gvn V(z_2);}\;V(z)\indp Z}}
        &\sum_{m_1}\sum_{m_2}\sum_{v_0}\sum_{v_1}\sum_{v_2} P\big(Y\big(z_0,m_1,m_2\big)=y\big|\tred{z_0,V=v_0}\big) P\big(V(z_0)=v_0\big|\tred{z_0}\big)\\[-4mm]
        &\quq P\big(M_1(z_1,m_2)=m_1\big|\tred{z_1,V=v_1}\big) P\big(V(z_1)=v_1\big|\tred{z_1}\big)\\[-2mm]
        &\quq P\big(M_2(z_2)=m_2\big|\tred{z_2,V=v_2}\big) P\big(V(z_2)=v_2\big|\tred{z_2}\big)\\
        \eqbb{18mm}{\mathcal{G}(m_1,m_2)}{Y(z,m_1,m_2)\indp M_2\gvn Z,V;\;M_1(z,m_2)\indp M_2\gvn Z,V}
        &\sum_{m_1}\sum_{m_2}\sum_{v_0}\sum_{v_1}\sum_{v_2} P\big(Y\big(z_0,m_1,m_2\big)=y\big|z_0,\tred{M_2=m_2},V=v_0\big) P\big(V(z_0)=v_0\big|z_0\big)\\[-2mm]
        &\quq P\big(M_1(z_1,m_2)=m_1\big|z_1,\tred{M_2=m_2},V=v_1\big) P\big(V(z_1)=v_1\big|z_1\big)\\[-2mm]
        &\quq P\big(M_2(z_2)=m_2\big|z_2,V=v_2\big) P\big(V(z_2)=v_2\big|z_2\big)\\
        \eqbb{10.5mm}{\mathcal{G}(m_1)}{Y(z,m_1,m_2)\indp M_1\gvn Z,V,M_2}
        &\sum_{m_1}\sum_{m_2}\sum_{v_0}\sum_{v_1}\sum_{v_2} P\big(Y\big(z_0,m_1,m_2\big)=y\big|z_0,\tred{M_1=m_1},M_2=m_2,V=v_0\big) P\big(V(z_0)=v_0\big|z_0\big)\\[-2mm]
        &\quq P\big(M_1(z_1,m_2)=m_1\big|z_1,M_2=m_2,V=v_1\big) P\big(V(z_1)=v_1\big|z_1\big)\\[-2mm]
        &\quq P\big(M_2(z_2)=m_2\big|z_2,V=v_2\big) P\big(V(z_2)=v_2\big|z_2\big)\\
        \eqtt{consistency}&\sum_{m_1}\sum_{m_2}\sum_{v_0}\sum_{v_1}\sum_{v_2}P\big(Y=y\big|z_0,M_1=m_1,M_2=m_2,V=v_0\big) P\big(V=v_0\big|z_0\big)\\[-2mm]
        &\quq P\big(M_1=m_1\big|z_1,M_2=m_2,V=v_1\big) P\big(V=v_1\big|z_1\big)\\[-2mm]
        &\quq P\big(M_2=m_2\big|z_2,V=v_2\big) P\big(V=v_2\big|z_2\big).
}
\vfill

\subsection{Equation~\eqref{eq:SC}: classical assumptions (Figure~\ref{fig:CA}), path-intervention approach, and $V$ is absent}
\eqevsn{
        P\big(Y\big\{z_{00},M_1\big[z_{01},M_2(z_{11})\big],M_2(z_{10})\big\}=y\big)
        =&\sum_{m_{01}}\sum_{m_{10}}P\big(Y\big(z_{00},M_1\big[z_{01},M_2(z_{11})\big],m_{10}\big)=y,M_1\big[z_{01},M_2(z_{11})\big]=m_{01}\big|M_2(z_{10})=m_{10}\big)P\big(M_2(z_{10})=m_{10}\big)\\
        =&\sum_{m_{01}}\sum_{m_{10}}\tred{\sum_{m_{11}}}P\big(Y\big(z_{00},M_1\big[z_{01},m_{11}\big],m_{10}\big)=y,M_1\big[z_{01},m_{11}\big]=m_{01}\big|\tred{M_2(z_{11})=m_{11}},M_2(z_{10})=m_{10}\big)\\[-4mm]
        &\qqq \tred{P\big(M_2(z_{11})=m_{11}\big|M_2(z_{10})=m_{10}\big)} P\big(M_2(z_{10})=m_{10}\big)\\
        =&\sum_{m_{01}}\sum_{m_{10}}\sum_{m_{11}}P\big(Y(z_{00},m_{01},m_{10})=y\big|\tred{M_1(z_{01},m_{11})=m_{01}},M_2(z_{11})=m_{11},M_2(z_{10})=m_{10}\big)\\[-4mm]
        &\qqq\tred{P\big(M_1(z_{01},m_{11})=m_{01}\big|M_2(z_{10})=m_{10},M_2(z_{11})=m_{11}\big)}\\[-2mm]
        &\qqq P\big(M_2(z_{11})=m_{11}\big|M_2(z_{10})=m_{10}\big) P\big(M_2(z_{10})=m_{10}\big)\\
        \eqbb{0mm}{\text{\textbf{strong} cross-world}}{\tred{M_2(\smash{z_{11}})\indp M_2(\smash{z_{10}})}} &\sum_{m_{01}}\sum_{m_{10}}\sum_{m_{11}}P\big(Y(z_{00},m_{01},m_{10})=y\big|M_1(z_{01},m_{11})=m_{01},M_2(z_{11})=m_{11},M_2(z_{10})=m_{10}\big)\\[-4mm]
        &\qqq P\big(M_1(z_{01},m_{11})=m_{01}\big|M_2(z_{10})=m_{10},M_2(z_{11})=m_{11}\big)\\[-2mm]
        &\qqq \tred{P\big(M_2(z_{11})=m_{11}\big)} P\big(M_2(z_{10})=m_{10}\big)\\
        \eqbb{20mm}{\text{cross-world}}{\tred{Y(z_{00},m_{01},m_{10})\indp \{M_1(z_{01},m_{11}),M_2(z_{10}),M_2(z_{11})\}}} &\sum_{m_{01}}\sum_{m_{10}}\sum_{m_{11}}\tred{P\big(Y(z_{00},m_{01},m_{10})=y\big)} P\big(M_1(z_{01},m_{11})=m_{01}\big|M_2(z_{10})=m_{10},M_2(z_{11})=m_{11}\big)\\[-4mm]
        &\qqq P\big(M_2(z_{11})=m_{11}\big) P\big(M_2(z_{10})=m_{10}\big)\\ 
        \eqbb{11mm}{\text{cross-world}}{\tred{M_1(z_{01},m_{11})\indp \{M_2(z_{10}),M_2(z_{11})\}}}
        &\sum_{m_{01}}\sum_{m_{10}}\sum_{m_{11}}P\big(Y(z_{00},m_{01},m_{10})=y\big)\tred{P\big(M_1(z_{01},m_{11})=m_{01}\big)}\\[-4mm]
        &\qqq P\big(M_2(z_{11})=m_{11}\big) P\big(M_2(z_{10})=m_{10}\big)\\ 
        \eqbb{2mm}{\substack{\mathcal{G}(z_{00},m_{01},m_{10});\;\mathcal{G}(z_{01},m_{01},m_{11});\\\mathcal{G}(z_{11},m_{01},m_{11});\;\mathcal{G}(z_{10},m_{01},m_{10})}}{\substack{Y(z_{00},m_{01},m_{10})\indp Z;\;M_1(z_{01},m_{11})\indp Z;\\[.5mm]M_2(z_{11})\indp Z;\;M_2(z_{10})\indp Z}}
        &\sum_{m_{01}}\sum_{m_{10}}\sum_{m_{11}}P\big(Y(z_{00},m_{01},m_{10})=y\big|\tred{z_{00}}\big)P\big(M_1(z_{01},m_{11})=m_{01}\big|\tred{z_{01}}\big)\\[-7mm]
        &\qqq P\big(M_2(z_{11})=m_{11}\big|\tred{z_{11}}\big)P\big(M_2(z_{10})=m_{10}|\tred{z_{10}}\big)\\
        \eqbb{15mm}{\mathcal{G}(m_1,m_2)}{Y(z,m_1,m_2)\indp M_2\gvn Z;\;M_1(z,m_2)\indp M_2\gvn Z}
        &\sum_{m_{01}}\sum_{m_{10}}\sum_{m_{11}}P\big(Y(z_{00},m_{01},m_{10})=y\big|z_{00},\tred{M_2=m_{10}}\big) P\big(M_1(z_{01},m_{11})=m_{01}\big|z_{01},\tred{M_2=m_{11}}\big)\\[-4mm]
        &\qqq P\big(M_2(z_{11})=m_{11}\big|z_{11}\big)P\big(M_2(z_{10})=m_{10}|z_{10}\big)\\
        \eqbb{9mm}{\mathcal{G}(m_1)}{Y(z,m_1,m_2)\indp M_1\gvn Z,M_2}
        &\sum_{m_{01}}\sum_{m_{10}}\sum_{m_{11}}P\big(Y(z_{00},m_{01},m_{10})=y\big|z_{00},\tred{M_1=m_{01}},M_2=m_{10}\big) P\big(M_1(z_{01},m_{11})=m_{01}\big|z_{01},M_2=m_{11}\big)\\[-4mm]
        &\qqq P\big(M_2(z_{11})=m_{11}\big|z_{11}\big)P\big(M_2(z_{10})=m_{10}|z_{10}\big)\\
        \eqtt{consistency}&\sum_{m_{01}}\sum_{m_{10}}\sum_{m_{11}}P\big(Y=y\big|{z_{00},M_1=m_{01},M_2=m_{10}}\big)  P\big(M_1=m_{01}\big|{z_{01},M_2=m_{11}}\big)\\[-4mm] 
        &\qqq P\big(M_2=m_{11}\big|z_{11}\big) P\big(M_2=m_{10}\big|z_{10}\big).
}

\subsection{Equation~\eqref{eq:WCi}: interventional approach (Figure~\ref{fig:IA}), node-intervention approach, and $V$ is absent}
\eqevsn{
        P\big(Y\big\{z_0,W_1\big[z_1,W_2(z_2)\big],W_2(z_2)\big\}=y\big)   
        =&\sum_{m_1}\sum_{m_2}P\big(Y\big(z_0,m_1,m_2\big)=y\big|W_1\big(z_1,m_2\big)=m_1,W_2(z_2)=m_2\big)\\[-4mm]
        &\quq P\big(W_1(z_1,m_2)=m_1\big|W_2(z_2)=m_2\big)P\big(W_2(z_2)=m_2\big)\\
        \eqbb{11mm}{\text{random draw}}{\tred{Y(z_0,m_1,m_2)\indp \{W_1(z_1,m_2),W_2(z_2)\}}} &\sum_{m_1}\sum_{m_2}P\big(Y\big(z_0,m_1,m_2\big)=y\big) P\big(W_1(z_1,m_2)=m_1\big|W_2(z_2)=m_2\big)P\big(W_2(z_2)=m_2\big)\\
        \eqbb{3.5mm}{\text{random draw}}{W_1(\smash{z_1,m_2})\indp W_2(\smash{z_2})} &\sum_{m_1}\sum_{m_2}P\big(Y\big(z_0,m_1,m_2\big)=y\big)P\big(W_1(z_1,m_2)=m_1\big)P\big(W_2(z_2)=m_2\big)\\
        \eqbb{1.5mm}{\mathcal{G}(z_0,m_1,m_2),\;\mathcal{G}(z_1,m_1,m_2),\;\mathcal{G}(z_2,m_1,m_2)}{Y(z_0,m_1,m_2)\indp Z,\;W_1(z_1,m_2)\indp Z,\;W_2(z_2)\indp Z}
        &\sum_{m_1}\sum_{m_2}P\big(Y\big(z_0,m_1,m_2\big)=y\big|\tred{z_0}\big) P\big(W_1(z_1,m_2)=m_1\big|\tred{z_1}\big)P\big(W_2(z_2)=m_2\big|\tred{z_2}\big)\\
        \eqbb{15.5mm}{\mathcal{G}(m_1,m_2)}{Y(z,m_1,m_2)\indp M_2\gvn Z,\;W_1(z,m_2)\indp M_2\gvn Z}
        &\sum_{m_1}\sum_{m_2}P\big(Y\big(z_0,m_1,m_2\big)=y\big|z_0,\tred{M_2=m_2}\big)\\[-4mm] 
        &\quq P\big(W_1(z_1,m_2)=m_1\big|z_1,\tred{M_2=m_2}\big)P\big(W_2(z_2)=m_2\big|z_2\big)\\
        \eqbb{9mm}{\mathcal{G}(m_1)}{Y(z,m_1,m_2)\indp M_1\gvn Z,M_2}
        &\sum_{m_1}\sum_{m_2}P\big(Y\big(z_0,m_1,m_2\big)=y\big|z_0,\tred{M_1=m_1},M_2=m_2\big)\\[-4mm]
        &\quq P\big(W_1(z_1,m_2)=m_1\big|z_1,M_2=m_2\big)P\big(W_2(z_2)=m_2\big|z_2\big)\\
        \eqtt{$W(\mathbf{a})\mid \mathbf{A}=\mathbf{a}\;\overset{d}{=}\;M(\mathbf{a})\mid \mathbf{A}=\mathbf{a}$}&\sum_{m_1}\sum_{m_2}P\big(Y\big(z_0,m_1,m_2\big)=y\big|{z_0,M_1=m_1,M_2=m_2}\big)\\[-4mm]
        &\quq P\big(\tred{M_1(z_1,m_2)=m_1}\big|z_1,M_2=m_2\big)P\big(\tred{M_2(z_2)=m_2}\big|z_2\big)\\
        \eqtt{consistency}&\sum_{m_1}\sum_{m_2}P\big(Y=y\big|{z_0,M_1=m_1,M_2=m_2}\big) P\big(M_1=m_1\big|{z_1,M_2=m_2}\big)P\big(M_2=m_2\big|{z_2}\big),
} \small where $\mathbf{A}=\mathbf{a}$ denotes the observed assignment of the variables corresponding to the intervention arguments $\mathbf{a}$.

\subsection{Equation~\eqref{eq:WCiv}: interventional approach (Figure~\ref{fig:IA}), node-intervention approach, and $V$ is present}
\eqevsn{
        P\big(Y\big\{z_0,W_1\big[z_1,W_2(z_2)\big],W_2(z_2)\big\}=y\big)  
        =&\sum_{m_1}\sum_{m_2}\sum_{v_0}\sum_{v_1}\sum_{v_2} P\big(Y\big(z_0,m_1,m_2\big)=y\big|W_1\big(z_1,m_2\big)=m_1,W_2(z_2)=m_2,\\[-4mm]
        &\qqq\qqq V(z_0)=v_0,V(z_1)=v_1,V(z_2)=v_2\big)\\[-2mm]
        &\quq P\big(W_1(z_1,m_2)=m_1\big|W_2(z_2)=m_2,V(z_0)=v_0,V(z_1)=v_1,V(z_2)=v_2\big)\\[-2mm]
        &\quq P\big(W_2(z_2)=m_2\big|V(z_0)=v_0,V(z_1)=v_1,V(z_2)=v_2\big)\\[-2mm] 
        &\quq P\big(V(z_0)=v_0,V(z_1)=v_1,V(z_2)=v_2\big)\\
        \eqbb{21mm}{\text{random draw}}{\tred{Y(z_0,m_1,m_2)\indp \{W_1(z_1,m_2),W_2(z_2)\}\gvn V(z_0),V(z_1),V(z_2)}} &\sum_{m_1}\sum_{m_2}\sum_{v_0}\sum_{v_1}\sum_{v_2} P\big(Y\big(z_0,m_1,m_2\big)=y\big|V(z_0)=v_0,V(z_1)=v_1,V(z_2)=v_2\big)\\[-2mm]
        &\quq P\big(W_1(z_1,m_2)=m_1\big|W_2(z_2)=m_2,V(z_0)=v_0,V(z_1)=v_1,V(z_2)=v_2\big)\\[-2mm]
        &\quq P\big(W_2(z_2)=m_2\big|V(z_0)=v_0,V(z_1)=v_1,V(z_2)=v_2\big)\\[-2mm]
        &\quq P\big(V(z_0)=v_0,V(z_1)=v_1,V(z_2)=v_2\big)\\
        \eqbb{13.5mm}{\text{random draw}}{\substack{\tred{W_1(\smash{z_1,m_2})\indp W_2(\smash{z_2})\gvn V(z_0),V(z_1),V(z_2);}\\[.5mm]\tred{W_1(\smash{z_1,m_2})\indp \{V(z_0),V(z_1),V(z_2)\};}\\[.5mm] \tred{W_2(\smash{z_2})\indp \{V(z_0),V(z_1),V(z_2)\}}}} &\sum_{m_1}\sum_{m_2}\sum_{v_0}\sum_{v_1}\sum_{v_2} P\big(Y\big(z_0,m_1,m_2\big)=y\big|V(z_0)=v_0,V(z_1)=v_1,V(z_2)=v_2\big)\\[-7mm]
        &\quq P\big(W_1(z_1,m_2)=m_1\big) P\big(W_2(z_2)=m_2\big) P\big(V(z_0)=v_0,V(z_1)=v_1,V(z_2)=v_2\big)\\
        \eqtt{marginalization over $v_1,v_2$}&\sum_{m_1}\sum_{m_2}\sum_{v_0} P\big(Y\big(z_0,m_1,m_2\big)=y\big|V(z_0)=v_0\big) P(V(z_0)=v_0)\\[-4mm] 
        &\qqq P\big(W_1(z_1,m_2)=m_1\big) P\big(W_2(z_2)=m_2\big)\\ 
        \eqbb{1.5mm}{\mathcal{G}(z_0,m_1,m_2);\;\mathcal{G}(z_1,m_1,m_2);\;\mathcal{G}(z_2,m_1,m_2)}{Y(z_0,m_1,m_2)\indp Z;\;W_1(z_1,m_2)\indp Z;\;W_2(z_2)\indp Z}
        &\sum_{m_1}\sum_{m_2}\sum_{v_0} P\big(Y\big(z_0,m_1,m_2\big)=y\big|\tred{z_0,V=v_0}\big) P(V(z_0)=v_0\big|\tred{z_0})\\[-4mm]
        &\qqq P\big(W_1(z_1,m_2)=m_1\big|\tred{z_1}\big)  P\big(W_2(z_2)=m_2\big|\tred{z_2}\big)\\
        \eqbb{16mm}{\mathcal{G}(m_1,m_2)}{Y(z,m_1,m_2)\indp M_2\gvn Z,V;\; W(z,m_2)\indp M_2\gvn Z}
        &\sum_{m_1}\sum_{m_2}\sum_{v_0} P\big(Y\big(z_0,m_1,m_2\big)=y\big|z_0,\tred{M_2=m_2},V=v_0\big) P\big(V(z_0)=v_0\big|z_0\big)\\[-4mm]
        &\qqq P\big(W_1(z_1,m_2)=m_1\big|z_1,\tred{M_2=m_2}\big) P\big(W_2(z_2)=m_2\big|z_2\big)\\
        \eqbb{10mm}{\mathcal{G}(m_1)}{Y(z,m_1,m_2)\indp M_1\gvn Z,V,M_2}
        &\sum_{m_1}\sum_{m_2}\sum_{v_0} P\big(Y\big(z_0,m_1,m_2\big)=y\big|z_0,\tred{M_1=m_1},M_2=m_2,V=v_0\big) P\big(V(z_0)=v_0\big|z_0\big)\\[-4mm]
        &\qqq P\big(W_1(z_1,m_2)=m_1\big|z_1,M_2=m_2\big) P\big(W_2(z_2)=m_2\big|z_2\big)\\
        \eqtt{$W(\mathbf{a})\mid \mathbf{A}=\mathbf{a}\;\overset{d}{=}\;M(\mathbf{a})\mid \mathbf{A}=\mathbf{a}$}&\sum_{m_1}\sum_{m_2}\sum_{v_0} P\big(Y\big(z_0,m_1,m_2\big)=y\big|z_0,M_1=m_1,M_2=m_2,V=v_0\big) P\big(V(z_0)=v_0\big|z_0\big)\\[-4mm]
        &\qqq P\big(\tred{M_1(z_1,m_2)=m_1}\big|z_1,M_2=m_2\big) P\big(\tred{M_2(z_2)=m_2}\big|z_2\big)\\
        \eqtt{consistency}&\sum_{m_1}\sum_{m_2}\sum_{v_0} P\big(Y=y\big|z_0,M_1=m_1,M_2=m_2,V=v_0\big) P\big(V=v_0\big|z_0\big)\\[-4mm]
        &\qqq P\big(M_1=m_1\big|z_1,M_2=m_2\big) P\big(M_2=m_2\big|z_2\big).
}
\vfill

\subsection{Equation~\eqref{eq:SCi}: interventional approach (Figure~\ref{fig:IA}), path-intervention approach, and $V$ is absent}
\eqevs{
        P\big(Y\big\{z_{00},W_1\big[z_{01},W_2(z_{11})\big],W_2(z_{10})\big\}=y\big)
        =&\sum_{m_{01}}\sum_{m_{10}}P\big(Y\big(z_{00},W_1\big[z_{01},W_2(z_{11})\big],m_{10}\big)=y,W_1\big[z_{01},W_2(z_{11})\big]=m_{01}\big|W_2(z_{10})=m_{10}\big)P\big(W_2(z_{10})=m_{10}\big)\\
        =&\sum_{m_{01}}\sum_{m_{10}}\tred{\sum_{m_{11}}}P\big(Y\big(z_{00},W_1\big[z_{01},m_{11}\big],m_{10}\big)=y,W_1\big[z_{01},m_{11}\big]=m_{01}\big|\tred{W_2(z_{11})=m_{11}},W_2(z_{10})=m_{10}\big)\\[-4mm]
        &\qqq \tred{P\big(W_2(z_{11})=m_{11}\big|W_2(z_{10})=m_{10}\big)} P\big(W_2(z_{10})=m_{10}\big)\\
        =&\sum_{m_{01}}\sum_{m_{10}}\sum_{m_{11}}P\big(Y(z_{00},m_{01},m_{10})=y\big|\tred{W_1(z_{01},m_{11})=m_{01}},W_2(z_{11})=m_{11},W_2(z_{10})=m_{10}\big)\\[-4mm]
        &\qqq\tred{P\big(W_1(z_{01},m_{11})=m_{01}\big|W_2(z_{10})=m_{10},W_2(z_{11})=m_{11}\big)}\\[-2mm]
        &\qqq P\big(W_2(z_{11})=m_{11}\big|W_2(z_{10})=m_{10}\big) P\big(W_2(z_{10})=m_{10}\big)\\  
        \eqbb{19mm}{\text{random draw}}{\tred{Y(z_{00},m_{01},m_{10})\indp \{W_1(\smash{z_{01},m_{11}}),W_2(z_{11}),W_2(z_{10})\}}} &\sum_{m_{01}}\sum_{m_{10}}\sum_{m_{11}}\tred{P\big(Y(z_{00},m_{01},m_{10})=y\big)} P\big(W_1(z_{01},m_{11})=m_{01}\big|W_2(z_{10})=m_{10},W_2(z_{11})=m_{11}\big)\\[-4mm]
        &\qqq P\big(W_2(z_{11})=m_{11}\big|W_2(z_{10})=m_{10}\big) P\big(W_2(z_{10})=m_{10}\big)\\   
        \eqbb{10mm}{\text{random draw}}{\tred{W_1(\smash{z_{01},m_{11}})\indp \{W_2(\smash{z_{11}}),W_2(\smash{z_{10}})\}}} &\sum_{m_{01}}\sum_{m_{10}}\sum_{m_{11}}P\big(Y(z_{00},m_{01},m_{10})=y\big)\tred{P\big(W_1(z_{01},m_{11})=m_{01}\big)}\\[-4mm]
        &\qqq P\big(W_2(z_{11})=m_{11}\big|W_2(z_{10})=m_{10}\big) P\big(W_2(z_{10})=m_{10}\big)\\  
        \eqbb{2mm}{\text{random draw}}{W_2(\smash{z_{11}})\indp W_2(\smash{z_{10}})} &\sum_{m_{01}}\sum_{m_{10}}\sum_{m_{11}} P\big(Y(z_{00},m_{01},m_{10})=y\big) P\big(W_1(z_{01},m_{11})=m_{01}\big) \tred{P\big(W_2(z_{11})=m_{11}\big)}P\big(W_2(z_{10})=m_{10}\big)\\
        \eqbb{2mm}{\substack{\mathcal{G}(z_{00},m_{01},m_{10});\;\mathcal{G}(z_{01},m_{01},m_{11});\\\mathcal{G}(z_{11},m_{01},m_{11});\;\mathcal{G}(z_{10},m_{01},m_{10})}}{\substack{Y(z_{00},m_{01},m_{10})\indp Z;\;W_1(z_{01},m_{11})\indp Z;\\[.5mm]W_2(z_{11})\indp Z;\;W_2(z_{10})\indp Z}}
        &\sum_{m_{01}}\sum_{m_{10}}\sum_{m_{11}}P\big(Y(z_{00},m_{01},m_{10})=y\big|\tred{z_{00}}\big)P\big(W_1(z_{01},m_{11})=m_{01}\big|\tred{z_{01}}\big)\\[-7mm]
        &\qqq P\big(W_2(z_{11})=m_{11}\big|\tred{z_{11}}\big)P\big(W_2(z_{10})=m_{10}|\tred{z_{10}}\big)\\
        \eqbb{15.5mm}{\mathcal{G}(m_1,m_2)}{Y(z,m_1,m_2)\indp M_2\gvn Z;\;W_1(z,m_2)\indp M_2\gvn Z}
        &\sum_{m_{01}}\sum_{m_{10}}\sum_{m_{11}}P\big(Y(z_{00},m_{01},m_{10})=y\big|z_{00},\tred{M_2=m_{10}}\big) P\big(W_1(z_{01},m_{11})=m_{01}\big|z_{01},\tred{M_2=m_{11}}\big)\\[-4mm]
        &\qqq P\big(W_2(z_{11})=m_{11}\big|z_{11}\big)P\big(W_2(z_{10})=m_{10}|z_{10}\big)\\
        \eqbb{9mm}{\mathcal{G}(m_1)}{Y(z,m_1,m_2)\indp M_1\gvn Z,M_2}
        &\sum_{m_{01}}\sum_{m_{10}}\sum_{m_{11}}P\big(Y(z_{00},m_{01},m_{10})=y\big|z_{00},\tred{M_1=m_{01}},M_2=m_{10}\big)\\[-4mm]
        &\qqq P\big(W_1(z_{01},m_{11})=m_{01}\big|z_{01},M_2=m_{11}\big)\\[-2mm]
        &\qqq P\big(W_2(z_{11})=m_{11}\big|z_{11}\big)P\big(W_2(z_{10})=m_{10}|z_{10}\big)\\
        \eqtt{$W(\mathbf{a})\mid \mathbf{A}=\mathbf{a}\;\overset{d}{=}\;M(\mathbf{a})\mid \mathbf{A}=\mathbf{a}$} &\sum_{m_{01}}\sum_{m_{10}}\sum_{m_{11}}P\big(Y(z_{00},m_{01},m_{10})=y\big|z_{00},M_1=m_{01},M_2=m_{10}\big)\\[-4mm]
        &\qqq P\big(\tred{M_1(z_{01},m_{11})=m_{01}}\big|z_{01},M_2=m_{11}\big)\\[-2mm]
        &\qqq P\big(\tred{M_2(z_{11})=m_{11}}\big|z_{11}\big)P\big(\tred{M_2(z_{10})=m_{10}}|z_{10}\big)\\
        \eqtt{consistency}&\sum_{m_{01}}\sum_{m_{10}}\sum_{m_{11}}P\big(Y=y\big|{z_{00},M_1=m_{01},M_2=m_{10}}\big) P\big(M_1=m_{01}\big|{z_{01},M_2=m_{11}}\big)\\[-4mm]
        &\qqq P\big(M_2=m_{11}\big|z_{11}\big) P\big(M_2=m_{10}\big|z_{10}\big).\label{eq:SCi}
}

\subsection{Equation~\eqref{eq:WCs}: separable effects (Figure~\ref{fig:SE3}), node-intervention approach, and $V$ is absent}
\eqevsn{
        P\big(Y\zpd=y\big)
        =&\sum_{m_1}\sum_{m_2}P\big(Y\zpd=y\big|M_1\zpd=m_1,M_2\zpd=m_2\big)\\[-4mm]
        &\quq P\big(M_1\zpd=m_1\big|M_2\zpd=m_2\big) P\big(M_2\zpd=m_2\big)\\            
        \eqbb{0mm}{\text{dismissible component condition}}{\text{Figure~\ref{fig:SE3}(c)}} &\sum_{m_1}\sum_{m_2}P\big(Y(\tred{z_0})=y\big|M_1(\tred{z_0})=m_1,M_2(\tred{z_0})=m_2\big)\\[-4mm] &\quq P\big(M_1(\tred{z_1})=m_1\big|M_2(\tred{z_1})=m_2\big) P\big(M_2(\tred{z_2})=m_2\big)\\
        \eqbb{11mm}{\text{exchangeability}}{Y(z_0)\indp Z_0;\;M_1(z_1)\indp Z_1;\;M_2(z_2)\indp Z_2}&\sum_{m_1}\sum_{m_2}P\big(Y({z_0})=y\big|\tred{Z_0=z_0},M_1({z_0})=m_1,M_2({z_0})=m_2\big)\\[-4mm]
        &\quq P\big(M_1({z_1})=m_1\big|\tred{Z_1=z_1},M_2({z_1})=m_2\big) P\big(M_2({z_2})=m_2\big|\tred{Z_2=z_2}\big)\\  
        \eqtt{consistency}&\sum_{m_1}\sum_{m_2}P\big(Y=y\big|{Z_0=z_0},M_1=m_1,M_2=m_2\big)\\[-4mm] 
        &\quq P\big(M_1=m_1\big|{Z_1=z_1},M_2=m_2\big) P\big(M_2=m_2\big|{Z_2=z_2}\big)\\
        \eqbb{0mm}{\text{component consistency}}{Z=Z_0=Z_1=Z_2}&\sum_{m_1}\sum_{m_2}P\big(Y=y\big|{Z=z_0},M_1=m_1,M_2=m_2\big)\\[-4mm] 
        &\quq P\big(M_1=m_1\big|{Z=z_1},M_2=m_2\big) P\big(M_2=m_2\big|{Z=z_2}\big).
}

\subsection{Equation~\eqref{eq:WCsv}: separable effects (Figure~\ref{fig:SE3}), node-intervention approach, and $V_i$ is sequential}
\eqevsn{
        P\big(Y\zpd=y\big)  
        =&\sum_{m_1}\sum_{m_2}\sum_{v_0}\sum_{v_1}\sum_{v_2} P\big(Y\zpd=y\big|M_1\zpd=m_1,M_2\zpd=m_2,\\[-4mm]
        &\qqq\quq V_0\zpd=v_0,V_1\zpd=v_1,V_2\zpd=v_2\big)\\[-2mm]
        &\quq P\big(M_1\zpd=m_1\big|M_2\zpd=m_2,\\[-2mm]
        &\qqq\quq V_0\zpd=v_0,V_1\zpd=v_1,V_2\zpd=v_2\big)\\[-2mm]
        &\quq P\big(M_2\zpd=m_2\big|V_0\zpd=v_0,V_1\zpd=v_1,V_2\zpd=v_2\big)\\[-2mm]
        &\quq P\big(V_0\zpd=v_0\big|V_1\zpd=v_1,V_2\zpd=v_2\big)\\[-2mm]
        &\quq P\big(V_1\zpd=v_1\big|V_2\zpd=v_2\big)\\[-2mm]
        &\quq P\big(V_2\zpd=v_2\big)\\
        \eqbb{0mm}{\text{dismissible component condition}}{\text{Figure~\ref{fig:SE3}(c)}} &\sum_{m_1}\sum_{m_2}\sum_{v_0}\sum_{v_1}\sum_{v_2} P\big(Y(\tred{z_0})=y\big|M_1(\tred{z_0})=m_1,M_2(\tred{z_0})=m_2,V_0(\tred{z_0})=v_0,V_1(\tred{z_0})=v_1,V_2(\tred{z_0})=v_2\big)\\[-2mm]
        &\quq P\big(M_1(\tred{z_1})=m_1\big|M_2(\tred{z_1})=m_2,V_0(\tred{z_1})=v_0,V_1(\tred{z_1})=v_1,V_2(\tred{z_1})=v_2\big)\\[-2mm]
        &\quq P\big(M_2(\tred{z_2})=m_2\big|V_0(\tred{z_2})=v_0,V_1(\tred{z_2})=v_1,V_2(\tred{z_2})=v_2\big)\\[-2mm]
        &\quq P\big(V_0(\tred{z_0})=v_0\big|V_1(\tred{z_0})=v_1,V_2(\tred{z_0})=v_2\big) P\big(V_1(\tred{z_1})=v_1\big|V_2(\tred{z_1})=v_2\big) P\big(V_2(\tred{z_2})=v_2\big)\\
        \eqbb{0mm}{\text{exchangeability \& consistency}}{\text{\& component consistency}} &\sum_{m_1}\sum_{m_2}\sum_{v_0}\sum_{v_1}\sum_{v_2} P\big(Y=y\big|z_0,M_1=m_1,M_2=m_2,V_0=v_0,V_1=v_1,V_2=v_2\big)\\[-2mm]
        &\quq P\big(M_1=m_1\big|z_1,M_2=m_2,V_0=v_0,V_1=v_1,V_2=v_2\big)\\[-2mm]
        &\quq P\big(M_2=m_2\big|z_2,V_0=v_0,V_1=v_1,V_2=v_2\big)\\[-2mm]
        &\quq P\big(V_0=v_0\big|z_0,V_1=v_1,V_2=v_2\big) P\big(V_1=v_1\big|z_1,V_2=v_2\big) P\big(V_2=v_2\big|z_2\big).
}

\subsection{Equation~\eqref{eq:WCsvv}: separable effects (Figure~\ref{fig:SE3s}), node-intervention approach, and $V$ is present}
\eqevs{
        P\big(Y\zpd=y\big)  
        \eqtt{ICM}&\sum_{m_1}\sum_{m_2}\sum_{v_0}\sum_{v_1}\sum_{v_2} P\big(Y\zpd=y\big|M_1\zpd=m_1,M_2\zpd=m_2,\\[-4mm]
        &\qqq\quq V_0\zpd=v_0,V_1\zpd=v_1,V_2\zpd=v_2\big)\\[-2mm]
        &\quq P\big(M_1\zpd=m_1\big|M_2\zpd=m_2,\\[-2mm]
        &\qqq\quq V_0\zpd=v_0,V_1\zpd=v_1,V_2\zpd=v_2\big)\\[-2mm]
        &\quq P\big(M_2\zpd=m_2\big|V_0\zpd=v_0,V_1\zpd=v_1,V_2\zpd=v_2\big)\\[-2mm]
        &\quq \tred{P\big(V_0\zpd=v_0\big)P\big(V_1\zpd=v_1\big)P\big(V_2\zpd=v_2\big)}\\
        \eqbb{0mm}{\text{dismissible component condition}}{\text{Figure~\ref{fig:SE3s}(a)}} &\sum_{m_1}\sum_{m_2}\sum_{v_0}\sum_{v_1}\sum_{v_2} P\big(Y(\tred{z_0})=y\big|M_1(\tred{z_0})=m_1,M_2(\tred{z_0})=m_2,\\[-4mm]
        &\qqq\qqq V_0(\tred{z_0})=v_0,V_1(\tred{z_0})=v_1,V_2(\tred{z_0})=v_2\big)\\[-2mm]
        &\quq P\big(M_1(\tred{z_1})=m_1\big|M_2(\tred{z_1})=m_2,V_0(\tred{z_1})=v_0,V_1(\tred{z_1})=v_1,V_2(\tred{z_1})=v_2\big)\\[-2mm]
        &\quq P\big(M_2(\tred{z_2})=m_2\big|V_0(\tred{z_2})=v_0,V_1(\tred{z_2})=v_1,V_2(\tred{z_2})=v_2\big)\\[-2mm]
        &\quq P\big(V_0(\tred{z_0})=v_0\big) P\big(V_1(\tred{z_1})=v_1\big) P\big(V_2(\tred{z_2})=v_2\big)\\
        \eqbb{12mm}{\text{ICC}}{V(z)=V_0(z)=V_1(z)=V_2(z)}&\sum_{m_1}\sum_{m_2}\sum_{v_0}\sum_{v_1}\sum_{v_2} P\big(Y(z_0)=y\big|M_1(z_0)=m_1,M_2(z_0)=m_2,\tred{V(z_0)=v_0}\big) P\big(\tred{V(z_0)}=v_0\big)\\[-2mm]
        &\quq P\big(M_1(z_1)=m_1\big|M_2(z_1)=m_2,\tred{V(z_1)=v_1}\big) P\big(\tred{V(z_1)}=v_1\big)\\[-2mm]
        &\quq P\big(M_2(z_2)=m_2\big|\tred{V(z_2)=v_2}\big) P\big(\tred{V(z_2)}=v_2\big)\\
        \eqbb{0mm}{\text{exchangeability \& consistency}}{\text{\& component consistency}} &\sum_{m_1}\sum_{m_2}\sum_{v_0}\sum_{v_1}\sum_{v_2} P\big(Y=y\big|z_0,M_1=m_1,M_2=m_2,V=v_0\big) P\big(V=v_0|z_0\big)\\[-2mm]
        &\quq P\big(M_1=m_1\big|z_1,M_2=m_2,V=v_1\big) P\big(V=v_1|z_1\big)\\[-2mm]
        &\quq P\big(M_2=m_2\big|z_2,V=v_2\big) P\big(V=v_2|z_2\big).\label{eq:WCsvv}
}

\subsection{Equation~\eqref{eq:RTn}: recanting twins effects (Figure~\ref{fig:RTE}), node-intervention approach, single mediator and $V$ is present}

\eqevsn{
  P\big(Y\big\{z_0,M\big[z_1,A(z_2)\big],A(z_2)\big\}=y\big)  
  =&\sum_{m}\sum_{v} P\big(Y\big(z_0,m,v\big)=y\big|M\big(z_1,v\big)=m,A(z_2)=v\big)\\[-4mm]
  &\quq P\big(M(z_1,v)=m\big|A(z_2)=v\big) P\big(A(z_2)=v\big)\\
  \eqbb{8mm}{\text{cross-world}}{Y(z_0,m,v)\indp M(\smash{z_1,v})\gvn A(z_2)} &\sum_{m}\sum_{v} P\big(Y\big(z_0,m,v\big)=y\big|A(z_2)=v\big) P\big(M(z_1,v)=m\big|A(z_2)=v\big) P\big(A(z_2)=v\big)\\
  \eqbb{11mm}{\text{random draw}}{Y(z_0,m,v)\indp A(\smash{z_2});\;M(\smash{z_1,v})\indp A(\smash{z_2})} &\sum_{m}\sum_{v} P\big(Y\big(z_0,m,v\big)=y\big) P\big(M(z_1,v)=m\big) P\big(A(z_2)=v\big)\\
  \eqbb{3mm}{\mathcal{G}(z_0,m,v);\;\mathcal{G}(z_1,m,v);\;\mathcal{G}(z_2,m,v)}{\substack{Y(z_0,m,v)\indp Z;\;M(z_1,v)\indp Z;\;A(z_2)\indp Z}}
  &\sum_{m}\sum_{v} P\big(Y\big(z_0,m,v\big)=y\big|\tred{z_0}\big) P\big(M(z_1,v)=m\big|\tred{z_1}\big) P\big(A(z_2)=v\big|\tred{z_2}\big)\\
  \eqbb{13mm}{\mathcal{G}(m,v)}{Y(z,m,v)\indp V\gvn Z;\;M(z,v)\indp V\gvn Z}
  &\sum_{m}\sum_{v} P\big(Y\big(z_0,m,v\big)=y\big|z_0,\tred{V=v}\big) P\big(M(z_1,v)=m\big|z_1,\tred{V=v}\big) P\big(A(z_2)=v\big|z_2\big)\\
  \eqbb{7mm}{\mathcal{G}(m)}{Y(z,m,v)\indp M\gvn Z,V}
  &\sum_{m}\sum_{v} P\big(Y\big(z_0,m,v\big)=y\big|z_0,\tred{M=m},V=v\big)\\[-4mm]
  &\quq P\big(M(z_1,v)=m\big|z_1,V=v\big) P\big(A(z_2)=v\big|z_2\big)\\
  \eqtt{$A(z)\mid Z=z \;\overset{d}{=}\; V(z)\mid Z=z$}
  &\sum_{m}\sum_{v} P\big(Y\big(z_0,m,v\big)=y\big|z_0,M=m,V=v\big)\\[-4mm]
  &\quq P\big(M(z_1,v)=m\big|z_1,V=v\big) P\big(\tred{V(z_2)}=v\big|z_2\big)\\  
  \eqtt{consistency}&\sum_{m}\sum_{v}P\big(Y=y\big|z_0,M=m,V=v\big) P\big(M=m\big|z_1,V=v\big) P\big(V=v\big|z_2\big).
}

\subsection{Equation~\eqref{eq:RTp}: recanting twins effects (Figure~\ref{fig:RTE}), path-intervention approach, single mediator and $V$ is present}

\eqevsn{
  P\big(Y\big\{z_{00},M\big[z_{01},A(z_{11})\big],A(z_{10})\big\}=y\big)
  =&\sum_{m}\sum_{v_{10}}P\big(Y\big(z_{00},M\big[z_{01},A(z_{11})\big],v_{10}\big)=y,M\big[z_{01},A(z_{11})\big]=m\big|A(z_{10})=v_{10}\big)P\big(A(z_{10})=v_{10}\big)\\
  =&\sum_{m}\sum_{v_{10}}\tred{\sum_{v_{11}}}P\big(Y\big(z_{00},M\big[z_{01},v_{11}\big],v_{10}\big)=y,M\big[z_{01},v_{11}\big]=m\big|\tred{A(z_{11})=v_{11}},A(z_{10})=v_{10}\big)\\[-4mm]
  &\qqq \tred{P\big(A(z_{11})=v_{11}\big|A(z_{10})=v_{10}\big)} P\big(A(z_{10})=v_{10}\big)\\
  =&\sum_{m}\sum_{v_{10}}\sum_{v_{11}}P\big(Y(z_{00},m,v_{10})=y\big|\tred{M(z_{01},v_{11})=m},A(z_{11})=v_{11},A(z_{10})=v_{10}\big)\\[-4mm]
  &\qqq\tred{P\big(M(z_{01},v_{11})=m\big|A(z_{10})=v_{10},A(z_{11})=v_{11}\big)}\\[-2mm]
  &\qqq P\big(A(z_{11})=v_{11}\big|A(z_{10})=v_{10}\big) P\big(A(z_{10})=v_{10}\big)\\  
  \eqbb{15mm}{\text{cross-world}}{\tred{Y(z_{00},m,v_{10})\indp M(\smash{z_{01},v_{11}})\gvn A(z_{11}),A(z_{10})}} &\sum_{m}\sum_{v_{10}}\sum_{v_{11}}\tred{P\big(Y(z_{00},m,v_{10})=y\big|A(z_{11})=v_{11},A(z_{10})=v_{10}\big)}\\[-4mm]
  &\qqq P\big(M(z_{01},v_{11})=m\big|A(z_{10})=v_{10},A(z_{11})=v_{11}\big)\\[-2mm]
  &\qqq P\big(A(z_{11})=v_{11}\big|A(z_{10})=v_{10}\big) P\big(A(z_{10})=v_{10}\big)\\   
  \eqbb{16mm}{\text{random draw}}{\substack{\tred{Y(z_{00},m,v_{10})\indp \{A(\smash{z_{11}}),A(\smash{z_{10}})\} };\\[.5mm]\tred{M(\smash{z_{01},v_{11}})\indp \{A(\smash{z_{11}}),A(\smash{z_{10}})\}};\;A(\smash{z_{11}})\indp A(\smash{z_{10}})}} &\sum_{m}\sum_{v_{10}}\sum_{v_{11}}\tred{P\big(Y(z_{00},m,v_{10})=y\big)}\tred{P\big(M(z_{01},v_{11})=m\big)}\tred{P\big(A(z_{11})=v_{11}\big)} P\big(A(z_{10})=v_{10}\big)\\[1mm]
  \eqbb{2mm}{\substack{\mathcal{G}(z_{00},m,v_{10});\;\mathcal{G}(z_{01},m,v_{11});\\\mathcal{G}(z_{11},m,v_{11});\;\mathcal{G}(z_{10},m,v_{10})}}{\substack{Y(z_{00},m,v_{10})\indp Z;\;M(z_{01},v_{11})\indp Z;\\[.5mm]A(z_{11})\indp Z;\;A(z_{10})\indp Z}}
  &\sum_{m}\sum_{v_{10}}\sum_{v_{11}}P\big(Y(z_{00},m,v_{10})=y\big|\tred{z_{00}}\big)P\big(M(z_{01},v_{11})=m\big|\tred{z_{01}}\big)\\[-7mm]
  &\qqq P\big(A(z_{11})=v_{11}\big|\tred{z_{11}}\big)P\big(A(z_{10})=v_{10}|\tred{z_{10}}\big)\\
  \eqbb{13mm}{\mathcal{G}(m,v)}{Y(z,m,v)\indp V\gvn Z;\;M(z,v)\indp V\gvn Z}
  &\sum_{m}\sum_{v_{10}}\sum_{v_{11}}P\big(Y(z_{00},m,v_{10})=y\big|z_{00},\tred{V=v_{10}}\big) P\big(M(z_{01},v_{11})=m\big|z_{01},\tred{V=v_{11}}\big)\\[-4mm]
  &\qqq P\big(A(z_{11})=v_{11}\big|z_{11}\big)P\big(A(z_{10})=v_{10}|z_{10}\big)\\
  \eqbb{9mm}{\mathcal{G}(m)}{Y(z,m,v)\indp M\gvn Z,V}
  &\sum_{m}\sum_{v_{10}}\sum_{v_{11}}P\big(Y(z_{00},m,v_{10})=y\big|z_{00},\tred{M=m},V=v_{10}\big)P\big(M(z_{01},v_{11})=m\big|z_{01},V=v_{11}\big)\\[-4mm]
  &\qqq P\big(A(z_{11})=v_{11}\big|z_{11}\big)P\big(A(z_{10})=v_{10}|z_{10}\big)\\
  \eqtt{$A(z)\mid Z=z \;\overset{d}{=}\; V(z)\mid Z=z$} &\sum_{m}\sum_{v_{10}}\sum_{v_{11}}P\big(Y(z_{00},m,v_{10})=y\big|z_{00},M=m,V=v_{10}\big) P\big(M(z_{01},v_{11})=m\big|z_{01},V=v_{11}\big)\\[-4mm]
  &\qqq P\big(\tred{V(z_{11})=v_{11}}\big|z_{11}\big)P\big(\tred{V(z_{10})=v_{10}}|z_{10}\big)\\
  \eqtt{consistency}&\sum_{m}\sum_{v_{10}}\sum_{v_{11}}P\big(Y=y\big|{z_{00},M=m,V=v_{10}}\big) P\big(M=m\big|{z_{01},V=v_{11}}\big) P\big(V=v_{11}\big|z_{11}\big) P\big(V=v_{10}\big|z_{10}\big).
}

\subsection{Equation~\eqref{eq:RTs}: recanting twins effects (Figure~\ref{fig:SE4}, path-intervention approach, single mediator and $V$ is present}
\eqevs{
        P\big(Y\zcd=y\big)
        \eqbb{3mm}{\text{ICM via}}{\text{random draw}}&\sum_{m_{01}}\sum_{m_{10}}\sum_{m_{11}} P\big(Y\zcd=y\big|M\zcd=m_{01},\\[-4mm]
        &\qqq\quq A_{10}\zcd=m_{10},A_{11}\zcd=m_{11}\big)\\[-2mm]
        &\qqq P\big(M\zcd=m_{01}\big|A_{10}\zcd=m_{10},A_{11}\zcd=m_{11}\big) \\[-2mm] 
        &\qqq\tred{P\big(A_{10}\zcd=m_{10}\big)P\big(A_{11}\zcd=m_{11}\big)}\\   
        \eqtt{random draw}&\sum_{m_{01}}\sum_{m_{10}}\sum_{m_{11}} P\big(Y\zcd=y\big|M\zcd=m_{01}\big) P\big(M\zcd=m_{01}\big)\\[-4mm] 
        &\qqq P\big(A_{10}\zcd=m_{10}\big)P\big(A_{11}\zcd=m_{11}\big)\\  
        \eqbb{25mm}{\text{cross-world}}{\substack{Y\zcd\indp V\zcd\gvn M\zcd;\\M\zcd\indp V\zcd}}&\sum_{m_{01}}\sum_{m_{10}}\sum_{m_{11}} P\big(Y\zcd=y\big|M\zcd=m_{01},\tred{V\zcd=m_{10}}\big)\\[-6mm]
        &\qqq P\big(M\zcd=m_{01}\big|\tred{V\zcd=m_{11}}\big) \\[-2mm] 
        &\qqq P\big(A_{10}\zcd=m_{10}\big)P\big(A_{11}\zcd=m_{11}\big)\\   
        \eqtt{$A(\cdot)\overset{d}{=}V(\cdot)$}&\sum_{m_{01}}\sum_{m_{10}}\sum_{m_{11}} P\big(Y\zcd=y\big|M\zcd=m_{01},V\zcd=m_{10}\big)\\[-4mm]
        &\qqq P\big(M\zcd=m_{01}\big|V\zcd=m_{11}\big) \\[-1mm] 
        &\qqq \tred{P\big(V_{10}\zcd=m_{10}\big)P\big(V_{11}\zcd=m_{11}\big)}\\   
        \eqbb{4mm}{\text{DCC}}{\text{Figure~\ref{fig:SE4}(a)}} &\sum_{m_{01}}\sum_{m_{10}}\sum_{m_{11}} P\big(Y(\tred{z_{00}})=y\big|M(\tred{z_{00}})=m_{01},V(\tred{z_{00}})=m_{10}\big)\\[-4mm]
        &\qqq P\big(M(\tred{z_{01}})=m_{01}\big|V(\tred{z_{01}})=m_{11}\big) P\big(V_{10}(\tred{z_{10}})=m_{10}\big) P\big(V_{11}(\tred{z_{11}})=m_{11}\big)\\  
        \eqbb{8mm}{\text{ICC}}{V(z)=V_{10}(z)=V_{11}(z)} &\sum_{m_{01}}\sum_{m_{10}}\sum_{m_{11}} P\big(Y(z_{00})=y\big|M(z_{00})=m_{01},V(z_{00})=m_{10}\big)\\[-4mm]
        &\qqq P\big(M(z_{01})=m_{01}\big|V(z_{01})=m_{11}\big) P\big(\tred{V(z_{10})}=m_{10}\big) P\big(\tred{V(z_{11})}=m_{11}\big)\\      
        \eqbb{0mm}{\text{exchangeability \& consistency}}{\text{\& component consistency}} &\sum_{m_{01}}\sum_{m_{10}}\sum_{m_{11}}P\big(Y=y\big|z_{00},M=m_{01},V=m_{10}\big) P\big(M=m_{01}\big|z_{01},V=m_{11}\big)\\[-4mm]
        &\qqq P\big(V=m_{11}\big|z_{11}\big) P\big(V=m_{10}\big|z_{10}\big).~\label{eq:RTs}
}

\clearpage
\section{Generalized Separable Effects}
  Let $Z$ be an exposure with manipulable components $\big\{Z_i\big\}_{i\in\mathbb{I}}$. Then
  \begin{itemize}
    \item \textit{Component manipulability:} $\{Z_i\}_{i\in\mathbb I}$ are mutually independent.
    \item \textit{Component consistency:} $\forall i \in \mathbb{I},\; Z = Z_i$ in the observed data.
  \end{itemize}
  These properties motivate the \textit{generalized decomposition assumption} in~\cite{stensrud2021generalized}. 
  For a joint intervention $\bzt=(z_i)_{i\in\mathbb I}$ on the exposure components, let $L(\bzt)$ denote the observed outcome under this joint intervention and $\big\{L_k(\bzt)\big\}_{k\in\mathbb{K}}$ its components, $\mathbb{K}\subseteq\mathbb{I}$. Then:
  \begin{itemize}
    \item \textit{Interventional component manipulability: Components corresponding to distinct interventions are mutually independent. That is, for any collection of distinct intervention values $\{z_k\}_{k\in\mathbb K'}$,
\[
\big\{L_k(\bzt)\big\}_{k\in\mathbb K'}
\text{ are mutually independent}.
\]
Components corresponding to the same intervention are not required to be
independent.} 
    \begin{itemize}
        \item This extends component manipulability by stating the independence between the components of the intervened variable.
    \end{itemize}
    \item \textit{Interventional component consistency: If the vector of joint interventions collapses into the same intervention as described in DCCs, i.e., $\bzt=(z)$, then}
    \[
      \forall k \in \mathbb{K},\quad L_k(\bzt)=L(z).
    \]
    \begin{itemize}
        \item This extends component consistency by ensuring alignment between the original variable and its components under the same intervention.
    \end{itemize}
  \end{itemize}

  In the mediated confounder scenario in Figure~\ref{fig:SE3s}, suppose $\bzt=(z_0,z_1,z_2)$ with $z_0,z_1,z_2$ distinct and the manipulable components of $V(\bzt)$ are $\big\{V_0(\bzt),V_1(\bzt),V_2(\bzt)\big\}$, thus $\mathbb{K}=\mathbb{I}$, then ICM implies that $V_0(\bzt),V_1(\bzt),V_2(\bzt)$ are mutually independent. However, when two intervention components coincide, for example $\bzt=(z_0,z_1,z_1)$, then only $V_0(\bzt)\indp \{V_1(\bzt),V_2(\bzt)\}$. In this case, for example, Equation~\eqref{eq:WCsvv} becomes
\eqev{
        P\big(Y(\tred{z_0,z_1,z_1})=y\big)  
        \eqtt{ICM}&\sum_{m_1}\sum_{m_2}\sum_{v_0}\sum_{v_1}\sum_{v_2} P\big(Y(\tred{z_0,z_1,z_1})=y\big|M_1(\tred{z_0,z_1,z_1})=m_1,M_2(\tred{z_0,z_1,z_1})=m_2,\\[-4mm]
        &\qqq\qqq V_0(\tred{z_0,z_1,z_1})=v_0,V_1(\tred{z_0,z_1,z_1})=v_1,V_2(\tred{z_0,z_1,z_1})=v_2\big)\\
        &\quq P\big(M_1(\tred{z_0,z_1,z_1})=m_1\big|M_2(\tred{z_0,z_1,z_1})=m_2,\\
        &\qqq\qqq V_0(\tred{z_0,z_1,z_1})=v_0,V_1(\tred{z_0,z_1,z_1})=v_1,V_2(\tred{z_0,z_1,z_1})=v_2\big)\\
        &\quq P\big(M_2(\tred{z_0,z_1,z_1})=m_2\big|V_0(\tred{z_0,z_1,z_1})=v_0,V_1(\tred{z_0,z_1,z_1})=v_1,V_2(\tred{z_0,z_1,z_1})=v_2\big)\\
        &\quq \tred{P\big(V_0(\tred{z_0,z_1,z_1})=v_0\big)P\big(V_1(\tred{z_0,z_1,z_1})=v_1\big|V_2(\tred{z_0,z_1,z_1})=v_2\big)P\big(V_2(\tred{z_0,z_1,z_1})=v_2\big)}\\
        \eqbb{0mm}{\text{dismissible component condition}}{\text{Figure~\ref{fig:SE3s}(a)}} &\sum_{m_1}\sum_{m_2}\sum_{v_0}\sum_{v_1}\sum_{v_2} P\big(Y(\tred{z_0})=y\big|M_1(\tred{z_0})=m_1,M_2(\tred{z_0})=m_2,\\[-4mm]
        &\qqq\qqq V_0(\tred{z_0})=v_0,V_1(\tred{z_0})=v_1,V_2(\tred{z_0})=v_2\big)\\
        &\quq P\big(M_1(\tred{z_1})=m_1\big|M_2(\tred{z_1})=m_2,V_0(\tred{z_1})=v_0,V_1(\tred{z_1})=v_1,V_2(\tred{z_1})=v_2\big)\\
        &\quq P\big(M_2(\tred{z_1})=m_2\big|V_0(\tred{z_1})=v_0,V_1(\tred{z_1})=v_1,V_2(\tred{z_1})=v_2\big)\\
        &\quq P\big(V_0(\tred{z_0})=v_0\big) P\big(V_1(\tred{z_1})=v_1\big|V_2(\tred{z_1})=v_2\big) P\big(V_2(\tred{z_1})=v_2\big)\\
        \eqbb{12mm}{\text{ICC}}{V(z)=V_0(z)=V_1(z)=V_2(z)}&\sum_{m_1}\sum_{m_2}\sum_{v_0}\sum_{v_1}P\big(Y(z_0)=y\big|M_1(z_0)=m_1,M_2(z_0)=m_2,\tred{V(z_0)=v_0}\big) P\big(\tred{V(z_0)}=v_0\big)\\[-1mm]
        &\quq P\big(M_1(z_1)=m_1\big|M_2(z_1)=m_2,\tred{V(z_1)=v_1}\big) P\big(\tred{V(z_1)}=v_1\big)\\
        &\quq P\big(M_2(z_1)=m_2\big|\tred{V(z_1)=v_1}\big)\\
        \eqbb{0mm}{\text{exchangeability \& consistency}}{\text{\& component consistency}} &\sum_{m_1}\sum_{m_2}\sum_{v_0}\sum_{v_1} P\big(Y=y\big|z_0,M_1=m_1,M_2=m_2,V=v_0\big) P\big(V=v_0|z_0\big)\\[-1mm]
        &\quq P\big(M_1=m_1,M_2=m_2\big|z_1,V=v_1\big) P\big(V=v_1|z_1\big).
}\normalsize

By ICC, setting $z_2=z_1$ implies $V_2(z_0,z_1,z_1)=V_1(z_0,z_1,z_1)$, allowing the identification formula under the finer decomposition to collapse to that under the coarser interventions. 
In other words, finer intervention decompositions provide greater mechanistic resolution at the cost of stronger component-level independence assumptions. Coarser interventions are recovered as special cases through ICC, which collapses component variables corresponding to identical interventions.
When $\bzt=(z_0,z_0,z_0)$, the vector of joint interventions collapses to a single intervention. Therefore, conditioning on $\{V_0(\bzt),V_1(\bzt),V_2(\bzt)\}$ in Figure~\ref{fig:SE3s}(b) corresponds, under ICC, to conditioning
on the original $V(z_0)$ in Figure~\ref{fig:SE3s}(c).
Specifically, ICC implies that
  \[V_0\big[(Z_0,Z_1,Z_2)=(z_0,z_0,z_0)\big]=V_1\big[(Z_0,Z_1,Z_2)=(z_0,z_0,z_0)\big]=V_2\big[(Z_0,Z_1,Z_2)=(z_0,z_0,z_0)\big]=V(Z=z_0)\] or shortly \(V_0(z_0)=V_1(z_0)=V_2(z_0)=V(z_0)\). Together with DCC, this gives
  \eqev{
        &\;P\big(Y\zpd=y\big|M_1\zpd=m_1,M_2\zpd=m_2,V_0\zpd=v_0,V_1\zpd=v_1,V_2\zpd=v_2\big)\\[-1mm]
        \eqtt{DCC}&\;{P\big(Y(z_0)=y\big|M_1(z_0)=m_1,M_2(z_0)=m_2,V_0(z_0)=v_0,V_1(z_0)=v_1,V_2(z_0)=v_2\big)}\\[-1mm]
         \eqtt{ICC}&\;{P\big(Y(z_0)=y\big|M_1(z_0)=m_1,M_2(z_0)=m_2,V(z_0)=v_0\big)},}\normalsize      
  and so for other DCCs in Figure~\ref{fig:SE3s}(a).
  
  Similarly, in the mediator scenario in Figure~\ref{fig:SE4}, $\bzt=(z_{00},z_{01},z_{10},z_{11})$ and the manipulable components of $M_2(\bzt)$ are $\big\{M_2^{10}(\bzt),M_2^{11}(\bzt)\big\}$, thus $\mathbb{K}\subset\mathbb{I}$, then ICM implies that $M_2^{10}(\bzt)\indp M_2^{11}(\bzt)$, and ICC implies that 
  \[M_2^{10}\big[(Z_{00},Z_{01},Z_{10},Z_{11})=(z_{00},z_{00},z_{00},z_{00})\big]=M_2^{11}\big[(Z_{00},Z_{01},Z_{10},Z_{11})=(z_{00},z_{00},z_{00},z_{00})\big]=M_2(Z=z_{00})\] or shortly \(M_2^{10}(z_{00})=M_2^{11}(z_{00})=M_2(z_{00})\). 
  Thus, when multiple intervention components are assigned the same value, ICC again identifies the corresponding mediator components, causing the decomposed representation to collapse to the original mediator. Together with DCC, this yields
  \eqev{
        &\;P\big(Y\zcd=y\big|M_1\zcd=m_{01},M_2^{10}\zcd=m_{10},M_2^{11}\zcd=m_{11}\big)\\
        \eqtt{DCC}&\;{P\big(Y(z_{00})=y\big|M_1(z_{00})=m_{01},M_2^{10}(z_{00})=m_{10},M_2^{11}(z_{00})=m_{11}\big)}\\[-1mm]
        \eqtt{ICC}&\;{P\big(Y(z_{00})=y\big|M_1(z_{00})=m_{01},M_2(z_{00})=m_{10}\big)},}\normalsize
  and so for other DCCs in Figure~\ref{fig:SE4}(a).

  \clearpage

\section{Positivity Conditions of Different Causal Semantics}

We also highlight the positivity conditions required for identification under each causal framework considered in this paper. Positivity ensures that the identification formulas are well-defined by requiring sufficient support in the observed data for all relevant configurations of variables.
In particular, for each conditional density
$P\bigl(V=v \gvn \operatorname{pa}(V)=\mathbf{p}\bigr)$
appearing in the identification formula, positivity requires
\[
P\bigl(\operatorname{pa}(V)=\mathbf{p}\bigr)>0
\]
for the values $(v,\mathbf{p})$ over which the identification formula is evaluated.
This condition ensures that the required conditional distributions are identified from observed data rather than extrapolated to parent configurations outside the observed support.

When no mediated confounder $V$ is present, the identification formulas of the three causal semantics coincide, and so do the corresponding positivity conditions.
For example, from Equation~\eqref{eq:WC},~\eqref{eq:WCi}, and~\eqref{eq:WCs}, positivity requires
\begin{equation*}
P(Z=z_2)>0,\quad
P(Z=z_1,M_2=m_2)>0,\quad
P(Z=z_0,M_1=m_1,M_2=m_2)>0,
\end{equation*}
for the values $(z_0,z_1,z_2,m_1,m_2)$ appearing in the identification formula.
When $V$ is present, differences arise from how each framework incorporates $V$ into the identification formula.
Under the classical assumptions with $V(z)\indp V(z')$, Equation~\eqref{eq:WCv} requires
\begin{align*}
P(Z=z_2,V=v_2)>0,\quad &P(Z=z_2)>0,\\
P(Z=z_1,M_2=m_2,V=v_1)>0,\quad &P(Z=z_1)>0,\\
P(Z=z_0,M_1=m_1,M_2=m_2,V=v_0)>0,\quad &P(Z=z_0)>0.
\end{align*}
Under the interventional approach, Equation~\eqref{eq:WCiv} imposes weaker requirements, with $V$ entering only through the outcome model:
\begin{align*}
P(Z=z_2)>0,\\ 
P(Z=z_1,M_2=m_2)>0,\\
P(Z=z_0,M_1=m_1,M_2=m_2,V=v_0)>0,&\quad P(Z=z_0)>0,
\end{align*}
reflecting the independence of interventional mediators from $V$.

In contrast, separable effects introduce component-specific variables $V_0,V_1,V_2$, leading to the factorization in Equation~\eqref{eq:WCsv}. Positivity then requires support for all parent configurations induced by this factorization:
\begin{align*}
P(Z=z_2,V_0=v_0,V_1=v_1,V_2=v_2)>0,\quad &P(Z=z_2)>0,\\
P(Z=z_1,M_2=m_2,V_0=v_0,V_1=v_1,V_2=v_2)>0,\quad &P(Z=z_1,V_2=v_2)>0,\\
P(Z=z_0,M_1=m_1,M_2=m_2,V_0=v_0,V_1=v_1,V_2=v_2)>0,\quad &P(Z=z_0,V_1=v_1,V_2=v_2)>0.
\end{align*}
If $V_0$, $V_1$, and $V_2$ are interpreted as manipulable components of $V$, then the identification formula reduces to Equation~\eqref{eq:WCv}. Otherwise, positivity requires support over configurations of $(V_0,V_1,V_2)$ induced by the factorization, which may be restrictive unless these variables are distinct.


\clearpage
\bibliographystyle{unsrt}
\bibliography{rci} 

\end{document}

%% file: notation.tex
\tikzstyle{da}=[-{Latex[length=1.8mm,angle'=50]},thick]
\tikzstyle{lo}=[-{Latex[length=1.8mm,angle'=50]},thick]
\tikzstyle{dw}=[{Circle[open,fill=\tcw]}-{Circle[open,fill=black]},thick,dashed]
\tikzstyle{do}=[{Circle[open,fill=\tco]}-{Circle[open,fill=black]},thick,dashed]
\tikzstyle{db}=[{Circle[open,fill=\tcw]}-{Circle[open,fill=\tco]},thick,dashed]
\tikzstyle{te}=[-{Circle[open]},thick]
\tikzstyle{co}=[-,thick]
\tikzstyle{ff}=[circle,thick,inner sep=.7mm,align=center]
\tikzstyle{rc}=[rectangle,thick,inner sep=2mm,align=center]
\tikzstyle{pp}=[fill=red!10,inner sep=1.5mm,rounded corners=2mm]
\tikzstyle{pe}=[red!10,line width=2mm]

\newcommand{\eqev}[1]{\vspace{-4mm}\begin{footnotesize}\begin{align}\begin{split}
    #1\end{split}\end{align}\end{footnotesize}}
\newcommand{\eqevs}[1]{\vspace{-2mm}\begin{scriptsize}\begin{align}\begin{split}
    #1\end{split}\end{align}\end{scriptsize}\vspace{-2mm}}
\newcommand{\eqevsn}[1]{\begin{scriptsize}\begin{align*}\begin{split}
    #1\end{split}\end{align*}\end{scriptsize}}  

\newcolumntype{L}[1]{>{\raggedright\let\newline\\\arraybackslash\hspace{0pt}}p{#1}}
\newcolumntype{C}[1]{>{\centering\let\newline\\\arraybackslash\hspace{0pt}}p{#1}}
\newcolumntype{R}[1]{>{\raggedleft\let\newline\\\arraybackslash\hspace{0pt}}p{#1}}

\newcommand{\eqtt}[1]{\xlongequal{\text{#1}}}
\newcommand{\eqbb}[3]{\underset{#3}{\xlongequal{\hspace{#1}#2\hspace{#1}}}}

\newcommand{\tco}{green!45!blue!90!black}
\newcommand{\tcw}{red!40!yellow!50!black}

\newcommand{\rred}{red!60!black}
\newcommand{\tred}[1]{\textcolor{\rred}{#1}}

\newcommand{\qqq}{\qquad\qquad\qquad}
\newcommand{\quq}{\qquad\qquad}
\newcommand{\indp}{\mathrel{\perp\!\!\!\!\perp}}

\newcommand{\gvn}{\;\mid\;}

\newcommand{\UV}{\mathcal{U}_V}
\newcommand{\zpd}{(z_0,z_1,z_2)}
\newcommand{\zcd}{(z_{00},z_{01},z_{10},z_{11})}

\newcommand{\zi}{\bm{z}}
\newcommand{\bzt}{\mathbf z}

\newcommand{\bn}[1]{{\mathrm{bn}(#1)}}